\documentclass[aps,prd,showpacs,notitlepage,nofootinbib,preprintnumbers,amsmath,amssymb]{revtex4-1}

	\usepackage{natbib}
	\usepackage{amsmath}
	\usepackage{makeidx}
	\usepackage{amsfonts}
	\usepackage[ansinew]{inputenc}
	\usepackage[usenames,dvipsnames]{pstricks}
	\usepackage{subcaption}
	\usepackage{epsfig}
	\usepackage{pst-grad} 
	\usepackage{pst-plot} 
	\usepackage[colorlinks,hyperindex]{hyperref}
	\usepackage{mathrsfs}
	\usepackage{bbold}
	\usepackage[margin=2cm,top=2cm, bottom=2cm]{geometry}
	\usepackage{physics}
	\usepackage{tensor}
	\usepackage{slashed}
	\usepackage{tikz}
	\usepackage{tikz-feynman}
	\usetikzlibrary{positioning}
	\tikzfeynmanset{compat=1.1.0}
	\usepackage{graphicx}
	\usepackage{simplewick}
	\usepackage[normalem]{ulem}
	\usepackage{mathtools}

\makeatletter
\DeclareRobustCommand{\cev}[1]{%
  {\mathpalette\do@cev{#1}}%
}
\newcommand{\do@cev}[2]{%
  \vbox{\offinterlineskip
    \sbox\z@{$\m@th#1 x$}%
    \ialign{##\cr
      \hidewidth\reflectbox{$\m@th#1\vec{}\mkern4mu$}\hidewidth\cr
      \noalign{\kern-\ht\z@}
      $\m@th#1#2$\cr
    }%
  }%
}
\makeatother

\def\sec#1{{Sec.~\ref{#1}}}
\def\app#1{{Appendix~\ref{#1}}}
\def\eq#1{{Eq.~(\ref{#1})}}
\def\fig#1{{Fig.~\ref{#1}}}
\newcommand{\ben}{\begin{eqnarray*}}
\newcommand{\een}{\end{eqnarray*}}
\newcommand{\un}[1]{\underline{#1}}

\newcommand{\pd}{\partial}

\newcommand{\llangle}{\Big\langle \!\! \Big\langle}
\newcommand{\rrangle}{\Big\rangle \!\! \Big\rangle}
\newcommand{\tord}{\textrm{T} \:}
\newcommand{\atord}{\bar{\textrm{T}} \:}

\newcommand{\as}{\alpha_s}

	\hypersetup
	{
		colorlinks,%
		citecolor=black,%
		linkcolor=black,%
		urlcolor=black,%
	}

\begin{document}

\title{Quark Sivers Function at Small $x$: \\ Spin-Dependent Odderon and the Sub-Eikonal Evolution}

\author{Yuri V. Kovchegov} 
         \email[Email: ]{kovchegov.1@osu.edu}
         \affiliation{Department of Physics, The Ohio State
           University, Columbus, OH 43210, USA}

\author{M. Gabriel Santiago}
  \email[Email: ]{santiago.98@buckeyemail.osu.edu}
	\affiliation{Department of Physics, The Ohio State
           University, Columbus, OH 43210, USA}

\begin{abstract}
We apply the formalism developed earlier \cite{Kovchegov:2018znm,Kovchegov:2018zeq} for studying transverse momentum dependent parton distribution functions (TMDs) at small Bjorken $x$ to construct the small-$x$ asymptotics of the quark Sivers function. First, we explicitly construct  the complete fundamental ``polarized Wilson line'' operator to sub-sub-eikonal order: this object can be used to study a variety of quark TMDs at small $x$. We then express the quark Sivers function in terms of dipole scattering amplitudes containing various components of the ``polarized Wilson line" and show that the dominant (eikonal) term which contributes to the quark Sivers function at small $x$ is the spin-dependent odderon, confirming the recent results of Dong, Zheng and Zhou \cite{Dong:2018wsp}. Our conclusion is also similar to the case of the gluon Sivers function derived by Boer, Echevarria, Mulders and Zhou \cite{Boer:2015pni} (see also \cite{Szymanowski:2016mbq}). We also analyze the sub-eikonal corrections to the quark Sivers function using the constructed ``polarized Wilson line'' operator. We derive new small-$x$ evolution equations re-summing double-logarithmic powers of $\as \, \ln^2 (1/x)$ with $\as$ the strong coupling constant. We solve the corresponding novel evolution equations in the large-$N_c$ limit, obtaining a sub-eikonal correction to the spin-dependent odderon contribution. We conclude that the quark Sivers function at small $x$ receives contributions from two terms and is given by
\begin{align}\label{eq_abs}
f_{1 \: T}^{\perp \: q} (x, k_T^2) = C_O (x, k_T^2) \, \frac{1}{x} + C_1 (k_T^2) \, \left( \frac{1}{x} \right)^0 + \ldots
\end{align}
with the function $C_O (x, k_T^2)$ varying slowly with $x$ and the ellipsis denoting the sub-asymptotic and sub-sub-eikonal (order-$x$) corrections. 
\end{abstract}

\maketitle

\tableofcontents

%||||||||||||||||||||||||||||||||||||||||||||||||||||||||||||||||||||||||||||||||||||||||||||||||||||||||||||||||||||||||
%||||||||||||||||||||||||||||||||||||||||||||||||||||||||||||||||||||||||||||||||||||||||||||||||||||||||||||||||||||||||
%||||||||||||||||||||||||||||||||||||||||||||||||||||||||||||||||||||||||||||||||||||||||||||||||||||||||||||||||||||||||

%%%%%%%%%%%%%%%%%%%%%%%%%%%%%%%%%%%%%%%%%%%%%%%%%%%%%%%%%%%%%%%%%

\section{Introduction}
\label{sec:Int}

The Sivers function \cite{Sivers:1989cc,Sivers:1990fh} is crucial for our understanding of the internal structure of the proton (and other hadrons) in terms of its quark and gluon degrees of freedom. It encodes the information about the orbital angular momentum of quarks and gluons in the proton, including spin-orbit coupling effects which lead to an asymmetric distribution of the partonic transverse momentum  \cite{Accardi:2012qut,Boer:2011fh,Aschenauer:2013woa,Aschenauer:2015eha,Proceedings:2020eah,AbdulKhalek:2021gbh}. The quark Sivers function is one of the two quark transverse momentum dependent parton distribution functions (TMDs) which is odd under time reversal, the other being the Boer-Mulders function \cite{Boer:1997nt}. This leads to the celebrated sign change between the quark Sivers function for semi-inclusive deep inelastic scattering (SIDIS) and Drell-Yan (DY) lepton pair production \cite{Brodsky:2002cx,Brodsky:2002rv,Collins:2002kn,Collins:2011zzd,Brodsky:2013oya}
\begin{align}
f_{1T \, \textrm{SIDIS}}^{\perp q} = - f_{1T \, \textrm{DY}}^{\perp q} .
\end{align}
This process dependence shows that while the intrinsic motion of the quarks inside the proton is a universal property of the proton state in QCD and the Sivers function does probe this intrinsic motion, this function also depends on how the proton is probed. (See \cite{Kovchegov:2013cva} for a simple and intuitive illustration of this statement.) Thus, the Sivers function is a process-dependent TMD, with the process dependence being under theoretical control in SIDIS and DY.

The small-Bjorken $x$ evolution of the Sivers function lies at the intersection of several facets of QCD research. It probes the spin and angular momentum structure of protons in the $x$-range where sea quarks and gluons are the dominant constituents of the proton.  The intersection of spin physics and small-$x$ physics has received extensive attention lately \cite{Kovchegov:2012ga,Kovchegov:2013cva,Zhou:2013gsa,Kovchegov:2015zha,Kovchegov:2015pbl, Altinoluk:2014oxa, Hatta:2016aoc, Hatta2016a,Hatta:2016khv,Boer:2016bfj,Balitsky:2016dgz, Kovchegov:2016zex, Kovchegov:2016weo, Kovchegov:2017jxc, Kovchegov:2017lsr, Kovchegov:2018zeq, Kovchegov:2018znm, Chirilli:2018kkw, Altinoluk:2019wyu,Kovchegov:2019rrz, Boussarie:2019icw, Cougoulic:2019aja, Kovchegov:2020hgb, Cougoulic:2020tbc, Altinoluk:2020oyd, Kovchegov:2020kxg, Chirilli:2021lif, Adamiak:2021ppq, Kovchegov:2021lvz,Bondarenko:2021rbp,Abir:2021kma} with small-$x$ evolution equations constructed for the various quark and gluon TMDs, partially in anticipation of the future Electron-Ion Collider (EIC) \cite{Accardi:2012qut,Boer:2011fh,Proceedings:2020eah,AbdulKhalek:2021gbh} where many of the TMDs will be measured at small $x$ with high precision. The small-$x$ asymptotics of the Sivers function also intersect with research on the QCD odderon, the $\mathcal{C}$-odd $t$-channel gluon exchange which was originally proposed as a mechanism for generating the asymmetry between $pp$ and $p\bar{p}$ cross sections at very high energies \cite{Lukaszuk:1973nt}. At leading order in $\alpha_s$, the odderon is given by a three-gluon exchange in the $t$-channel with the gluons in the symmetric $d^{abc}$ color configuration ($d^{abc} = 2 \tr [t^a \{ t^b, t^c \} ]$ with $t^a$ the fundamental generators of SU($N_c$) and $N_c$ the number of colors). The odderon has its own extensive body of research at small-$x$ \cite{Bartels:1980pe,Kwiecinski:1980wb,Nicolescu:1990ii,Janik:1998xj,Bartels:1999yt,Korchemsky:2001nx,Kovchegov:2003dm,Ewerz:2003xi,Hatta:2005as,Kovner:2005qj,Jeon:2005cf,Hagiwara:2020mqb}, as well as an exciting recent announcement of the odderon detection in $pp$ and $p\bar{p}$ collisions by the D0 and TOTEM collaborations \cite{TOTEM:2020zzr} (see also \cite{Antchev:2017dia,Martynov:2017zjz,Contreras:2020lrh,Braun:2020vmd,Csorgo:2020wmw}). It was shown in \cite{Boer:2015pni} that the gluon Sivers function at small $x$ is mainly generated by the spin-dependent odderon, arising from a fundamental-representation Wilson loop in the gauge link of the operator. One might expect a similar result in the case of the quark (SIDIS or DY) Sivers function as it also comes in with a fundamental-representation gauge link. Indeed, in \cite{Szymanowski:2016mbq} the odderon was shown to generate single transverse spin asymmetry for the quark jet production in scattering of a quark on a transversely polarized nucleon, further suggesting the odderon's connection with the quark Sivers function. Moreover, recently, in \cite{Dong:2018wsp} it was shown that the spin-dependent odderon does contribute to the quark Sivers function. In this work we study the intersection of these diverse areas of QCD by constructing the small-$x$ asymptotics of the quark Sivers function. Our approach is different from \cite{Boer:2015pni,Dong:2018wsp} since we do not limit our analysis to the eikonal contribution and study the sub-eikonal corrections to the quark Sivers function as well.

Small-$x$ evolution for the helicity TMD and quark transversity TMD were constructed in  \cite{Kovchegov:2015pbl,Kovchegov:2016zex,Kovchegov:2018znm} and in \cite{Kovchegov:2018zeq}, respectively, using the saturation/color glass condensate (CGC) framework \cite{Iancu:2003xm,Weigert:2005us,JalilianMarian:2005jf,Gelis:2010nm,Albacete:2014fwa,Kovchegov:2012mbw} where the operator definitions of the TMDs can be rewritten in terms of the so-called polarized dipole scattering amplitudes defined in terms of Wilson line correlators. For the helicity TMDs, a new `longitudinally polarized Wilson line' operator had to be constructed in \cite{Kovchegov:2017lsr,Kovchegov:2018znm}, where a sub-eikonal operator (or two operators) were inserted into the usual eikonal light-cone Wilson lines in order to couple the proton's helicity to the helicity of the quark or anti-quark in the dipole. (In our notation, sub-eikonal refers to the object suppressed by one power of $x$ compared to the eikonal scattering, sub-sub-eikonal refers to suppression by $x^2$, etc.) Similarly, for the quark transversity TMD a new `transversely polarized Wilson line' operator was constructed in \cite{Kovchegov:2018zeq}, containing sub-sub-eikonal operators needed to couple the proton's transverse spin to the transverse spin of the quarks in the dipole. In order to study the small-$x$ asymptotics of the leading-twist quark TMDs dependent on the proton's spin, one needs to couple the proton's spin to the quarks in the dipole. To evaluate such coupling, one needs {\sl a priori} to find all the other terms in the scattering of a quark on a polarized proton, including all the eikonal, sub-eikonal, and sub-sub-eikonal operators which were not included in the `polarized Wilson line' operators constructed in \cite{Kovchegov:2017lsr,Kovchegov:2018znm,Kovchegov:2018zeq}. Here we will perform this construction, thus completing the calculations started in \cite{Kovchegov:2017lsr,Kovchegov:2018znm,Kovchegov:2018zeq}. The result of our calculation is a fundamental polarized Wilson line operator up to and including the sub-sub-eikonal order: it is given in \eq{Full} below. This is our main formal result: it can be used for the future analyses of different quark (and gluon) TMDs at small $x$. Here we apply this result to obtain the small-$x$ asymptotics of the quark Sivers function. We find two contributions: one coming directly from the exchange of a spin-dependent odderon and a novel sub-eikonal one with the double-logarithmic evolution.

The structure of this paper is as follows. In \sec{sec:polwline} we construct the full  `polarized Wilson line' operator in \eq{Full} working to sub-sub-eikonal order. One can also think of this object as 
the full quark $S$-matrix in the background field of a target calculated up to (and including) the sub-sub-eikonal order and expressed in the transverse spin basis. (A conversion to helicity basis can be easily accomplished using Table~\ref{table:conversion} below). 

In \sec{sec:sivop} we take the operator definition of the quark Sivers TMD and rewrite it in the small-$x$/saturation formalism in terms of dipoles containing different parts of the new `Wilson line' operator. We separately investigate the eikonal and sub-eikonal contributions. In \sec{sec:gen_exp}, working at the eikonal level, we show that the leading contribution to the quark Sivers function comes from the eikonal odderon exchange which has a known small-$x$ evolution \cite{Bartels:1980pe,Kwiecinski:1980wb,Kovchegov:2003dm,Hatta:2005as} and asymptotics \cite{Bartels:1999yt,Kovchegov:2003dm}. The odderon intercept is known to be exactly zero at the leading \cite{Bartels:1999yt,Kovchegov:2003dm} and next-to-leading \cite{Kovchegov:2012rz} logarithmic approximation in $1/x$, and also at any order in the strong coupling $\as$ in the large-$N_c$ limit \cite{Caron-Huot:2013fea}. In addition, the odderon intercept appears to be zero in the strong-coupling calculations based on the anti-de Sitter/Conformal Field Theory (AdS/CFT) correspondence \cite{Brower:2008cy,Avsar:2009hc,Brower:2014wha}. We thus conclude, with the perturbative and non-perturbative accuracy of the odderon calculations \cite{Bartels:1999yt,Kovchegov:2003dm,Kovchegov:2012rz,Caron-Huot:2013fea,Brower:2008cy,Avsar:2009hc,Brower:2014wha}, that the leading eikonal small-$x$ asymptotics of the quark Sivers function is $f_{1T}^{\perp q} (x, k_T^2) \sim 1/x$ for both the SIDIS and DY cases. 

In \sec{sec:spinodd} we explicitly calculate the odderon-generated quark Sivers function in the scalar diquark model of the proton. This contribution dominates at small $x$ as it is proportional to $1/x$ (see \eq{oddsiv}) and receives no corrections to this power of $x$  from evolution, as described above. It can be compared to the lowest fixed-order (one-loop) contribution to the quark Sivers function in the scalar diquark model of the proton, which, for massless quarks, gives $f_{1T}^{\perp q} (x, k_T^2) \sim x$ and, therefore, falls off rather rapidly at small $x$ \cite{Meissner:2007rx}. 

The sub-eikonal small-$x$ asymptotics of the quark Sivers function is studied in \sec{sec:sub-eik}. There we identify the sub-eikonal part of the polarized Wilson line operator \eqref{Vi} responsible for the Sivers function and construct a new evolution equations \eqref{Fi_evol} (or, at large $N_c$, Eqs.~\eqref{Fi_evol5} or \eqref{Fi_evol6}). These evolution equations re-sum powers of $\as \, \ln^2 (1/x)$: we will refer to this as the double-logarithmic approximation (DLA) (cf., e.g., \cite{Kovchegov:2015pbl,Kovchegov:2016zex,Kovchegov:2018znm,Kovchegov:2018zeq}). Solving these novel evolution equations in the large-$N_c$ limit we arrive at the sub-eikonal small-$x$ asymptotic term \eqref{Sivers_sub_eik8} for the quark Sivers function. Combining this with the spin-dependent odderon we arrive at the eikonal and sub-eikonal small-$x$ asymptotics of the quark Sivers function given in \eq{eq_abs}, which we will repeat here for completeness,
\begin{align}\label{Siv_asympt}
f_{1 \: T}^{\perp \: q} (x, k_T^2) = C_O (x, k_T^2) \, \frac{1}{x} + C_1 (k_T^2) \, \left( \frac{1}{x} \right)^0 .
\end{align}
This is the main physical result of this work. The function $C_O (x, k_T^2)$ is slowly-varying with $x$ and can be exactly determined by the odderon evolution equation (see \cite{Contreras:2020lrh}). 

We conclude in \sec{sec:con} by summarizing our results and by describing possible future applications of our technique, such as the determination of the small-$x$ asymptotics for various other TMDs.

%%%%%%%%%%%%%%%%%%%%%%%%%%%%%%%%%%%%%%%%%%%%%%%%%%%%%%%%%%%%%%%%%%%%%

\section{Quark $S$-Matrix in the Background Field}
\label{sec:polwline}

At small $x$, the quark and gluon helicity TMD \cite{Kovchegov:2018znm} and the quark transversity TMD \cite{Kovchegov:2018zeq} can be expressed in terms of the `polarized dipole operators' given by correlators of polarized Wilson lines with the usual eikonal Wilson lines. These polarized dipole operators obey small-$x$ evolution equations, which differ for different TMDs as dictated by the structure of the corresponding part of the polarized Wilson line operator. The polarized fundamental Wilson line up to sub-sub-eikonal order was partially constructed in \cite{Kovchegov:2018zeq}, where, for the transversity calculation, the only contributing portion of the Wilson line operator was the term which coupled the transverse spin of the proton to the transverse spin of a probed quark. Here we want to construct the full polarized Wilson line (quark $S$-matrix) which applies for scattering of any-polarization quark on any-polarization nucleon, including terms up to (and including) the sub-sub-eikonal order. The calculation will be carried out in the transverse spin basis, with the conversion of our results into the helicity basis being easy to accomplish with the help of Table~\ref{table:conversion} below.

We consider high-energy scattering of a quark moving in the $x^-$ light-cone direction with the large `minus' momentum $p_2^-$ on a proton moving in the $x^+$ light-cone direction with the large `plus' momentum $p_1^+$. The light-cone coordinates are defined by
\begin{align}\label{LC_coord}
x^+ = t + z,  \ \ \  x^- = t - z, \ \ \  \un{x} = (x,y),
\end{align}
such that a space-time 4-vector is written in terms of light-cone coordinates as $x^{\mu} = (x^+, x^-, \un{x})$. The inner product of two 4-vectors is 
\begin{align}
a_{\mu}b^{\mu} = \frac{a^+ b^- + a^- b^+}{2} - \un{a} \vdot \un{b}.
\end{align}
The transverse portion of 4-vectors is denoted by $\un{x}= (x,y)$ while its magnitude we is labeled by $|\un{x}|=x_{\perp}$. We will make one exception for the transverse momentum $\un{k}$, whose magnitude will be denoted by $k_T = |\un{k}|$. 

The fundamental Wilson line which sums up the eikonal scattering of the quark at transverse position $\un{x}$ moving in the $x^-$ direction on the gluon field of the proton is \cite{Balitsky:1995ub}
\begin{align}\label{Wline}
V_{\un{x}} [x^-_f,x^-_i] = \mathcal{P} \exp \left[ \frac{ig}{2} \int\limits_{x^-_i}^{x^-_f} \dd{x}^- A^+ (0^+, x^-, \un{x}) \right],
\end{align}
with $\mathcal{P}$ the path ordering operator, $A^{\mu} = \sum_a A^{a \, \mu} \, t^a$ the background gluon field, $g$ the strong coupling constant, and the integral running over the light-cone path of the quark. In the case of an infinite path we will abbreviate the Wilson line as $V_{\un{x}} \equiv V_{\un{x}} [\infty,-\infty]$.

We are interested in constructing the generalization of the Wilson line operator which includes all possible sub-eikonal and sub-sub-eikonal interactions of a quark with the target irrespective of the quark and the target polarizations. To this end, we define an $S$-matrix for the quark--target scattering by 
\begin{align}\label{Vxy}
V_{\un{x}, \un{y}; \sigma', \sigma}  \equiv \int \frac{d^2 p_{in}}{(2\pi)^2} \, \frac{d^2 p_{out}}{(2\pi)^2} \, e^{i \un{p}_{out} \cdot \un{x} - i \un{p}_{in} \cdot \un{y}} \ \left[ \delta_{\sigma, \sigma'} \, (2 \pi)^2 \, \delta^2 \left( \un{p}_{out} - \un{p}_{in} \right) + i \, A_{\sigma', \sigma} (\un{p}_{out}, \un{p}_{in}) \right],
\end{align}
where $A (\un{p}_{out}, \un{p}_{in})$ is the scattering amplitude for a quark on a target with $\un{p}_{in}$ and $\un{p}_{out}$ the incoming and outgoing quark transverse momenta, respectively, while $\sigma'$ and $\sigma$ are the outgoing and incoming quark polarizations in helicity basis. The amplitude $A$ is normalized such that $A = M/(2 s)$ \cite{Kovchegov:2012mbw}, where $M$ is the standard textbook scattering amplitude and $s$ is the center-of-mass energy squared. Extending the notation of \cite{Kovchegov:2015pbl, Kovchegov:2018znm, Kovchegov:2018zeq}, we will denote by $V^{\textrm{pol} }_{\un{x}, \un{y}; \sigma', \sigma}$ the entire non-eikonal part of the quark scattering $S$-matrix \eqref{Vxy}, independent of whether the terms it contains depend on quark or target polarizations or not:
\begin{align}\label{Vxy_pol}
V_{\un{x}, \un{y}; \sigma', \sigma} = V_{\un{x}} \, \delta^2 (\un{x} - \un{y}) \, \delta_{\sigma, \sigma'} + V^{\textrm{pol}}_{\un{x}, \un{y}; \sigma', \sigma}  .
\end{align}

From the earlier calculations \cite{Kovchegov:2017lsr, Kovchegov:2018znm, Kovchegov:2018zeq, Chirilli:2018kkw, Altinoluk:2020oyd} we know that sub-eikonal and sub-sub-eikonal corrections come in as insertions of one or more sub- and/or sub-sub-eikonal operators anywhere along the $x^-$ path of the quark. For an insertion of a local operator we have
\begin{align}
\label{polwilgen}
V^{\textrm{pol}}_{\un{x}, \un{y}; \sigma', \sigma} = \int\limits_{-\infty}^{\infty} \dd{z}^- d^2 z \ V_{\un{x}} [ \infty, z^-] \, \delta^2 (\un{x} - \un{z}) \, \mathcal{O}^{\textrm{pol}}_{\sigma', \sigma} (z^-, \un{z}) \, V_{\un{y}} [ z^-, -\infty] \, \delta^2 (\un{y} - \un{z}) ,
\end{align}
while for an insertion of a bi-local operator (cf.  \cite{Kovchegov:2018znm,Kovchegov:2018zeq}) we write
\begin{align}
\label{polwilgendoub}
V^{\textrm{pol}}_{\un{x}, \un{y}; \sigma', \sigma} = \int\limits_{-\infty}^{\infty} \dd{z}_1^- d^2 z_1 \ \int\limits_{z_1^-}^{\infty} \dd{z}_2^- d^2 z_2 \ V_{\un{x}} [ \infty, z_2^-] \, \delta^2 (\un{x} - \un{z}_2) \, \mathcal{O}^{\textrm{pol}}_{\sigma', \sigma} (z_2^-, z_1^-;  \un{z}_2, \un{z}_1) \, V_{\un{y}} [ z_1^-, -\infty]  \, \delta^2 (\un{y} - \un{z}_1). 
\end{align}
Here and below, the two-dimensional integrals denote integration over transverse components of the vector: for instance, $d^2 z = d z^x \, d z^y$. When space allows we will also refer to such integration measure as $d^2 z_\perp$. Note that, in general, the operators $\mathcal{O}^{\textrm{pol}}_{\sigma', \sigma} (z^-, \un{z})$ and $\mathcal{O}^{\textrm{pol}}_{\sigma', \sigma} (z_2^-, z_1^-;  \un{z}_2, \un{z}_1)$ may contain derivatives with respect to $\un{z}$ and $\un{z}_1, \un{z}_2$, respectively, acting on the delta-functions in Eqs.~\eqref{polwilgen} and \eqref{polwilgendoub}. For the helicity and transversity `polarized Wilson lines' such derivatives were absent in \cite{Kovchegov:2018znm,Kovchegov:2018zeq} and one also had $\mathcal{O}^{\textrm{pol}} (z_2^-, z_1^-;  \un{z}_2, \un{z}_1) = \delta^2 (\un{z}_1 - \un{z}_2) \, \mathcal{O}^{\textrm{pol}}_{\sigma', \sigma} (z_2^-, z_1^-, \un{z}_1)$, resulting in the simplification $V^{\textrm{pol}}_{\un{x}, \un{y}; \sigma', \sigma} = V^{\textrm{pol}}_{\un{x}; \sigma', \sigma} \, \delta^2 (\un{x} - \un{y})$ with
\begin{align}
\label{polwilgen1}
V^{\textrm{pol}}_{\un{x}} = \int\limits_{-\infty}^{\infty} \dd{x}^- \, V_{\un{x}} [ \infty, x^-] \,  \mathcal{O}^{\textrm{pol}} (x^-, \un{x}) \, V_{\un{x}} [ x^-, -\infty] ,
\end{align}
and
\begin{align}
\label{polwilgendoub1}
V^{\textrm{pol}}_{\un{x}} = \int\limits_{-\infty}^{\infty} \dd{x}_1^- \int\limits_{x_1^-}^{\infty} \dd{x}_2^- V_{\un{x}} [ \infty, x_2^-] \, \mathcal{O}^{\textrm{pol}} (x_2^-, x_1^-, \un{x}) \, V_{\un{x}} [ x_1^-, -\infty] ,
\end{align}
for the local and bi-local operators, respectively. For brevity we have suppressed the polarization indices in Eqs.~\eqref{polwilgen1} and \eqref{polwilgendoub1}. 

From the analysis in \cite{Kovchegov:2018znm,Cougoulic:2020tbc,Kovchegov:2018zeq} we know that the local operator $\mathcal{O}^{\textrm{pol}}_{\sigma', \sigma} (z^-, \un{z})$ in \eq{polwilgen} can be obtained by calculating the diagrams containing the sub-eikonal and/or sub-sub-eikonal gluon field insertion shown in \fig{FIG:opoldiagglue} below, while the non-local operator $\mathcal{O}^{\textrm{pol}}_{\sigma', \sigma} (z_2^-, z_1^-;  \un{z}_2, \un{z}_1)$ arises due to two quark field insertions with the adjoint Wilson line connecting them, as illustrated diagrammatically in \fig{FIG:opoldiagquark}. Note that at the sub-sub-eikonal level the term with two insertions of $\mathcal{O}^{\textrm{pol}}_{\sigma', \sigma}  (z^-, \un{z})$ needs to be included as well.  Below we will explicitly calculate the non-eikonal operators $\mathcal{O}^{\textrm{pol}}_{\sigma', \sigma}  (z^-, \un{z})$ and $\mathcal{O}^{\textrm{pol}}_{\sigma', \sigma}  (z_2^-, z_1^-;  \un{z}_2, \un{z}_1)$.

%%%%%%%%%%%%%%%%%%%%%%%%%%%%%%%%%%%%%%%%%%%%%%%%%%%%%%%%%%%%%%%%%%%%%%%%%%%%%%%%%%%%%%

\subsection{Gluon insertion operator}

\begin{figure}[ht]
\centering
\includegraphics[width=0.6\linewidth]{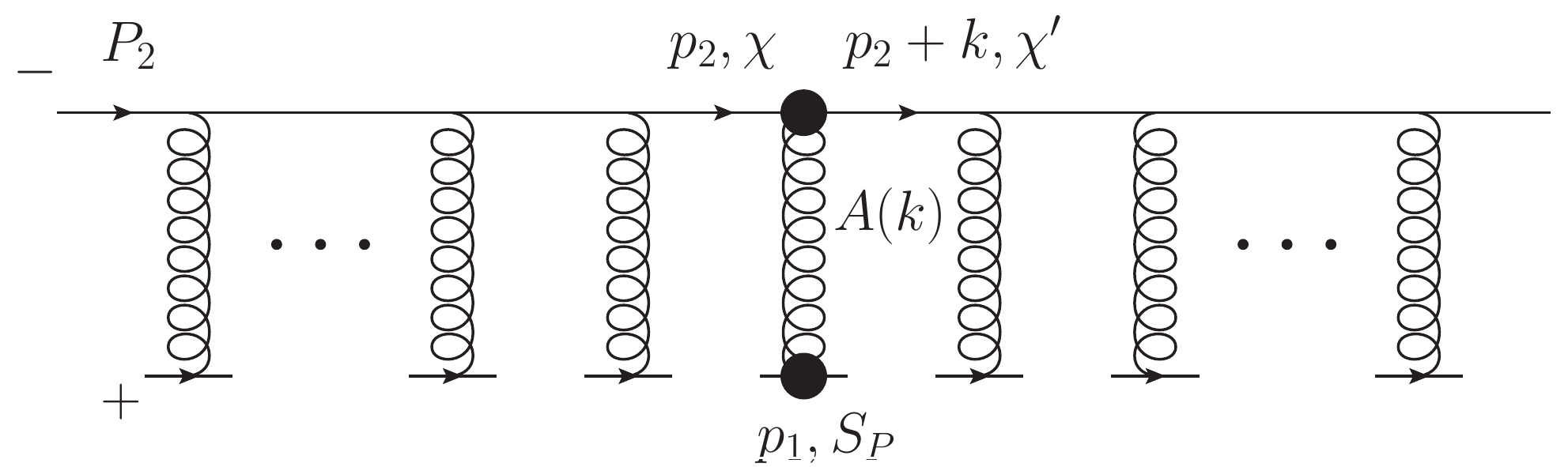}
\caption{The diagram representing the gluon field contribution to the non-eikonal operator. The black circles on the vertices designate potentially spin-dependent sub- and sub-sub-eikonal scattering.}
\label{FIG:opoldiagglue}
\end{figure}

Here we calculate the gluon exchange operator $\mathcal{O}^{\textrm{pol G}}$ in Feynman gauge ($\partial_{\mu} A^{\mu}=0$). The non-eikonal scattering in \fig{FIG:opoldiagglue} is shown by a gluon coupling to the quark line at the top via a black-circle vertex, where the background field transfers a momentum $k$ to the $x^-$-direction moving quark. Following the approach of \cite{Kovchegov:2017lsr, Kovchegov:2018znm, Kovchegov:2018zeq}, which involves replacing the numerators of quark propagators by the polarization sums, we obtain the (potentially non-eikonal) operator in momentum space
\begin{align}
\label{ogunsimp}
\mathcal{O}^{\textrm{G}}_{\chi', \chi} (k) = \frac{1}{2 \sqrt{p_2^- \, (p_2^- + k^-)}} \, i g \, \bar{u}_{\chi'} (p_2 + k) \slashed{A} (k) u_{\chi} (p_2)	.
\end{align}
The normalizing factor of $1/\left[2 \sqrt{p_2^- \, (p_2^- + k^-)} \right]$ results from the square roots of the denominators of the residues for the two quark propagators adjacent to the non-eikonal vertex integrated over their light-cone 'plus' momentum components, see e.g. \eq{phase} below. The normalization for the operators in \cite{Kovchegov:2017lsr, Kovchegov:2018znm, Kovchegov:2018zeq} was chosen to be $1/2 p_2^-$, with the difference between that and the normalization in \eq{ogunsimp} not affecting the helicity and transversity operators considered in those references. 

%%%%%%%%%%%%%%%%%%%%%%%%%%%%%%%%%%%%%%%%%%%%%%%%%%%%%%%%%%%%%%%%%%%%%%%%%%%%%%%%%%%%%%
%%%%%%%%%%%%%%%%%%%%%%%%%%%%%%%%%%%%%%%%%%%%%%%%%%%%%%%%%%%%%%%%%%%%%%%%%%%%%%%%%%%%%%

We take the spinors here for the $x^-$ moving quark to be the $\pm$ reversed, transversely polarized Brodsky-Lepage (BL) spinors \cite{Lepage:1980fj} used in \cite{Kovchegov:2018zeq}, with spin quantized along the $x$-direction. They are related to the helicity-basis $\pm$ reversed BL spinors by \cite{Kovchegov:2012ga}
\begin{align}\label{chi_def}
u_\chi \equiv \frac{1}{\sqrt{2}} \, \left[ u_+ + \chi \, u_- \right], \ \ \ v_\chi \equiv \frac{1}{\sqrt{2}} \, \left[ v_+ + \chi \, v_- \right],
\end{align}
where $\chi = \pm 1$ and  the helicity-basis $\pm$-reversed BL spinors are \cite{Kovchegov:2018znm,Kovchegov:2018zeq}
\begin{align}\label{anti-BLspinors}
u_\sigma (p) = \frac{1}{\sqrt{p^-}} \, [p^- + m \, \gamma^0 +  \gamma^0 \, {\un \gamma} \cdot {\un p} ] \,  \rho (\sigma), \ \ \ v_\sigma (p) = \frac{1}{\sqrt{p^-}} \, [p^- - m \, \gamma^0 +  \gamma^0 \, {\un \gamma} \cdot {\un p} ] \,  \rho (-\sigma),
\end{align}
with $p^\mu = \left( \frac{{\un p}^2+ m^2}{p^-}, p^-, {\un p} \right)$ and
\begin{align}
  \rho (+1) \, = \, \frac{1}{\sqrt{2}} \, \left(
  \begin{array}{c}
      1 \\ 0 \\ -1 \\ 0
  \end{array}
\right), \ \ \ \rho (-1) \, = \, \frac{1}{\sqrt{2}} \, \left(
  \begin{array}{c}
        0 \\ 1 \\ 0 \\ 1
  \end{array}
\right) .
\end{align}

Employing the spinors from \eq{chi_def} and performing a direct calculation, one can show that the Dirac matrix elements in \eq{ogunsimp} are 
\begin{subequations}\label{ubaru}
\begin{align}
\bar{u}_{\chi'} (p_2 + k) \gamma^+ u_{\chi} (p_2) =& \frac{2}{\sqrt{(p_2^- + k^-) p_2^-}} \Bigg( \delta_{\chi,\chi'} \,  \left[ (\underline{p}_2 + \underline{k} ) \vdot \underline{p}_2 + m^2 +
i \, m \, \chi \, \underline{S} \cross \underline{k} \right] \notag \\
& + \delta_{\chi,-\chi'} \, \left[ i (\underline{k} \cross \underline{p}_2 ) - m \, \chi \, \underline{S} \vdot \underline{k} \right] \Bigg) ,	\\
\bar{u}_{\chi'} (p_2 + k) \gamma^- u_{\chi} (p_2) =& \sqrt{(p_2^- + k^-) p_2^-} \, 2 \,  \delta_{\chi,\chi'} , \\
 \bar{u}_{\chi'} (p_2 + k) \gamma_{\perp}^1 u_{\chi} (p_2) =& \sqrt{(p_2^- + k^-) p_2^-} \Bigg( \delta_{\chi,\chi'} \Bigg[ \frac{\underline{S} \vdot \underline{p}_2 }{p_2^-} + \frac{\underline{S} \vdot (\underline{p}_2 + \underline{k})}{p_2^- + k^-} \Bigg] \notag \\
&+ \delta_{\chi,-\chi'} \Bigg[ \frac{i \, \underline{S} \cross \underline{p}_2 - m \, \chi }{p_2^-} + \frac{-i \, \underline{S} \cross (\underline{p}_2 + \underline{k} )  + m \, \chi}{p_2^- + k^-} \Bigg] \Bigg) , \\
\bar{u}_{\chi'} (p_2 + k) \gamma_{\perp}^2 u_{\chi} (p_2) =& \sqrt{(p_2^- + k^-) p_2^-} \Bigg( \delta_{\chi,\chi'} \Bigg[ \frac{\underline{S} \cross \underline{p}_2 + i \, m \, \chi}{p_2^-} + \frac{\underline{S} \cross (\underline{p}_2 + \underline{k}) - i \, m \, \chi}{p_2^- + k^-} \Bigg]  \notag \\
&+ \delta_{\chi,-\chi'} \Bigg[ \frac{-i \, \underline{S} \vdot \underline{p}_2 }{p_2^-} + \frac{i \, \underline{S} \vdot (\underline{p}_2 + \underline{k} )}{p_2^- + k^-} \Bigg] \Bigg) ,
\end{align}
\end{subequations}
with $\un{S}$ the unit vector in the direction of transverse spin quantization (for the proton polarized along  the $x$-axis we have  $\un{S}=\hat{x}$). The cross-product is defined by $\un{a} \times \un{b} = a^x b^y - a^y b^x = \epsilon^{ij} a^i b^j$ with $\epsilon^{ij}$ the 2-dimensional Levi-Civita symbol, $\epsilon^{12} = - \epsilon^{21} = 1, \epsilon^{11} = \epsilon^{22} =0$. 

Plugging the matrix elements \eqref{ubaru} into \eq{ogunsimp} we find (for $\un{S}=\hat{x}$)
\begin{align}\label{O1}
& \mathcal{O}^{\textrm{G}}_{\chi', \chi} (k) = \frac{ig}{2} \Bigg\{ \delta_{\chi,\chi'}  \Bigg( A^+ +   \frac{(\underline{p}_2 + \underline{k} ) \vdot \underline{p}_2  + m^2 + i m \chi \, \underline{S} \cross \underline{k} }{(p_2^- + k^-) p_2^-} \, A^- \! \Bigg) \! + \delta_{\chi, -\chi'}  \frac{i ( \underline{k} \cross \underline{p}_2 ) - m \chi \underline{S} \vdot \underline{k}}{(p_2^- + k^-) p_2^-} A^- \notag	\\
 & - \delta_{\chi,\chi'}  \left[ \Bigg( \frac{\underline{S} \cross \underline{p}_2 + im\chi}{p_2^-} + \frac{\underline{S} \cross (\underline{p}_2 + \underline{k} ) -
im\chi}{p_2^- + k^-} \Bigg)  \underline{S} \times \underline{A} + \Bigg( \frac{\underline{S} \vdot \underline{p}_2 }{p_2^-} + \frac{\underline{S} \vdot (\underline{p}_2 + \underline{k})}{p_2^- + k^-} \Bigg) \underline{S} \cdot \underline{A} \right]  	\\
 &  - \delta_{\chi, -\chi'} \left[ \Bigg( \frac{-i \underline{S} \vdot \underline{p}_2}{p_2^-} + \frac{i \underline{S} \vdot (\underline{p}_2 + \underline{k})}{p_2^- + k^-} \Bigg) \underline{S} \times \underline{A} + \Bigg( \frac{i \underline{S} \cross \underline{p}_2 - m \chi }{p_2^-} + \frac{-i \underline{S} \cross (\underline{p}_2 + \underline{k} )  + m \chi}{p_2^- + k^-} \Bigg) \underline{S} \cdot \underline{A}  \right]   \Bigg\}. \notag
\end{align}
Since $k^- \sim 1/p_1^+$ is small (with $p_1^+$ the large light-cone momentum component of the parton in the target generating the gluon field), $k^- \ll p_2^-$, we expand some of the terms in the powers of $k^-/p_2^-$, obtaining
\begin{align}\label{O12}
\mathcal{O}^{\textrm{G}}_{\chi', \chi} (k) = \ &  \frac{ig}{2} \Bigg\{ \delta_{\chi,\chi'}  \Bigg[  A^+ +   \frac{(\underline{p}_2 + \underline{k} ) \vdot \underline{p}_2  + m^2 }{(p_2^-)^2} \, A^- \! - \frac{\underline{p}_2 \cdot \underline{A}}{p_2^-}   - \frac{(\underline{p}_2 +\underline{k} ) \cdot \underline{A} }{p_2^- + k^-}  \Bigg]  \\
 &  - \delta_{\chi, -\chi'} \, \frac{i }{p_2^-} \left[ \underline{k} \times \underline{A}  - \frac{1}{p_2^-} \left( (\underline{k} \cross \underline{p}_2 ) \, A^- - k^- \, \un{A} \times (\un{p}_2 + \un{k}) \right)  \right] \notag  \\ 
 & + \chi \, \delta_{\chi,\chi'} \,  \frac{i m}{(p_2^-)^2} \left[ ( \underline{S} \cross \underline{k}) \, A^- - ( \underline{S} \cross \underline{A}) \, k^- \right] - \chi \, \delta_{\chi, -\chi'}  \frac{m }{(p_2^-)^2} \left[ (\underline{S} \vdot \underline{k}) \, A^- -  (\underline{S} \vdot \underline{A}) \, k^- \right] \Bigg\}  + {\cal O} \left( \frac{1}{(p_2^-)^3} \right) . \notag
\end{align}
Note that the ``eikonality" of different terms in \eq{O12} is manifest: there is the eikonal $A^+$ term, not suppressed by powers of $p_2^-$, there are the sub-eikonal ${\cal O} (1/p_2^-)$ terms, and the sub-sub-eikonal ${\cal O} (1/(p_2^-)^2)$ terms.

There is one caveat about the momentum $p_2^-$ in the sub-eikonal (order-$1/p_2^-$) terms in \eq{O12}: while in sub-eikonal calculations one does not need to distinguish the minus momentum component of different quark propagators in \fig{FIG:opoldiagglue}, at the sub-sub-eikonal order considered here one has to take into account that the interactions of the quark with the background gluon field result in transfer of a small momentum $q^-$ to the quark. If we label $P_2^-$ the momentum of the incoming quark (see \fig{FIG:opoldiagglue}), before any interaction, then we can write $p_2^- = P_2^- + q^-$. The difference between $1/p_2^-$ and $1/P_2^-$ is of the sub-sub-eikonal order. The difference between $1/(p_2^-)^2$ and $1/(P_2^-)^2$ is of the sub-sub-sub-eikonal order, and is discarded in our calculation. In the coordinate space we replace $q^- \to i \pd^-$ with the derivative acting on everything to the right of (and, hence, earlier than) the insertion of the non-eikonal operator at hand, such that $q^-$ is really a total minus momentum transferred to the quark by the target at the light-cone time $x^-$. The order of the operators also matters: we will replace $1/p_2^- \to 1/(P_2^- + i \pd^-)$ with the operator placed to the right of the non-eikonal gluon field, while a similar replacement $1/(p_2^- + k^-) \to 1/(P_2^- + i \pd^-)$ is different from the former by the placement of the operator to the left of non-eikonal field. Strictly-speaking, for consistency we need to expand $1/(P_2^- + i \pd^-)$ in the powers of $i \pd^-/P_2^-$: we chose to keep the term as is, with the understanding that the overall expression along with this term are correct only up to sub-sub-eikonal accuracy. While the gluon field of the target is often taken to be a function of $x^-$ and $\un{x}$ only, see e.g. \cite{Altinoluk:2020oyd}, below, in Eqs.~\eqref{AplusAperp}, we will see that the $x^+$-dependence comes in through the phase of the field, which usually constitutes a higher-order (in eikonality) correction to the leading-order observables generated by the field. Hence, the partial derivative $\pd^-$ with respect to $x^+$ is non-trivial, even when applying to a regular light-cone Wilson line \eqref{Wline}, if the sub-eikonal $x^+$-dependent phase is included in the otherwise eikonal gluon field $A^+$.

To Fourier-transform the expression \eqref{O12} into coordinate space we also replace $\un{k} \to - i \, \un{\nabla}$, $k^- \to i \pd^-$ (both derivatives acting on the gluon field), $\underline{p}_2 \to -i \, \vec{\un{\nabla}}$, and $(\underline{p}_2 + \un{k}) \to i \, \cev{\un{\nabla}}$, where the operator $\vec{\un{\nabla}}$ acts only on the objects to the right of the operator $\mathcal{O}^{\textrm{G}}_{\chi', \chi}$, while $\cev{\un{\nabla}}$ acts only to the left, and neither of them acts directly on $A^\mu$, somewhat similar to the notation in \cite{Altinoluk:2020oyd}. (The $\un{\nabla}$ operator is defined by $\nabla^i = - \pd^i = \pd_i$.) This yields, with sub-sub-eikonal accuracy,
\begin{align}\label{O22}
\mathcal{O}^{\textrm{G}}_{\chi', \chi} (x) = \ &  \frac{ig}{2} \Bigg\{ \delta_{\chi,\chi'} \! \Bigg[ \!  A^+   +   \frac{\cev{\nabla}^i \,  A^- \, \vec{\nabla}^i  + m^2 A^-}{(P_2^-)^2} +  \underline{A}  \cdot \vec{\un{\nabla}} \, \frac{i}{P_2^- + i \pd^-}  - \frac{i}{P_2^- + i \pd^-} \, \cev{\un{\nabla}} \cdot \underline{A}  \Bigg] \\
 &  - \delta_{\chi, -\chi'} \, \left[ (\underline{\nabla} \times \underline{A}) \, \frac{1}{P_2^- + i \pd^-} + \frac{i}{(P_2^-)^2} \, \cev{\un{\nabla}}  \cross ( \un{\nabla} \, A^- + \pd^- \un{A})  \right]   \notag \\ 
 & + \chi \, \delta_{\chi,\chi'} \,  \frac{m}{(P_2^-)^2} \, \underline{S} \cross (\un{\nabla} \, A^- + \pd^- \un{A}) + \chi \, \delta_{\chi, -\chi'}  \frac{i m }{(P_2^-)^2} \, \underline{S} \vdot (\underline{\nabla} \, A^- + \pd^- \un{A}) \Bigg\}   , \notag
\end{align}
where now $A^\mu = A^\mu (x^+, x^-, \un{x})$. While the operator in \eq{O22} is also a function of $x^\mu = (x^+, x^-, \un{x})$, in the future, by analogy to the light-cone Wilson line \eqref{Wline}, we will use $\mathcal{O}^{\textrm{G}}_{\chi', \chi} (x^-, \un{x}) \equiv \mathcal{O}^{\textrm{G}}_{\chi', \chi} (x^+ =0, x^-, \un{x})$, that is, we will just put $x^+ =0$ in all expressions, while implying that the $\pd^-$-derivatives are applied before putting $x^+$ to zero. 

The $\delta_{\chi, -\chi'}$ term in \eq{O22} may appear left-right asymmetric at the sub-sub-eikonal order. Note, however, that the order of $(\underline{\nabla} \times \underline{A})$ and $1/(P_2^- + i \pd^-)$ in the first term multiplying $\delta_{\chi, -\chi'}$ can be interchanged simultaneously with replacing $\cev{\un{\nabla}} \to - \vec{\un{\nabla}}$ in the second term multiplying $\delta_{\chi, -\chi'}$: taking a half-sum of the original and interchanged $\delta_{\chi, -\chi'}$ terms one can left-right symmetrize the coefficient of $\delta_{\chi, -\chi'}$, if needed. 

The expression \eqref{O22}, while including all the eikonal, sub-eikonal and sub-sub-eikonal corrections to the quark scattering due to an insertion of the background gluon field, is missing the non-eikonal corrections to the free quark propagator. Consider the propagator of the $p_2$ quark line in \fig{FIG:opoldiagglue}, concentrating on the denominator of the propagator. (The numerators in our calculation are written as polarization sums plus the instantaneous terms, just like in the light-cone perturbation theory (LCPT) \cite{Lepage:1980fj,Brodsky:1997de}.) The integral over $p_2^+$ yields (for $p_2^- >0$)
\begin{align}\label{phase}
& \int\limits_{-\infty}^\infty \frac{d p_2^+}{4 \pi} e^{- \frac{i}{2} p_2^+ (x^- - y^-)} \, \frac{i}{p_2^2 - m^2 + i \epsilon} = \frac{1}{2 p_2^-} \, e^{- i \frac{{\un p}_2^2 + m^2}{2 p_2^-} (x^- - y^-)} \\ & = \frac{1}{2 p_2^-} \, \exp \left\{ - i \, \frac{{\un p}_2^2 + m^2}{2 p_2^-} \int\limits^{x^-}_{y^-} d z^- \right\} =  \frac{1}{2 p_2^-} \, \left[ 1 -  i \, \frac{{\un p}_2^2 + m^2}{2 p_2^-} \int\limits^{x^-}_{y^-} d z^- + \ldots \right] \notag
\end{align}
where the ellipsis denote the higher-order (sub-sub-eikonal and beyond) corrections, which are obtained by further expanding the exponential in the powers of the phase. Those corrections correspond to multiple insertions of the sub-eikonal operator $-  i ({\un p}_2^2 + m^2)/(2 p_2^-)$ in \eq{phase}. The sub-eikonal phase corrections were employed two decades ago in the medium jet quenching calculations for heavy ion collisions \cite{Baier:1996sk,Baier:1996kr,Zakharov:1996fv,Zakharov:1997uu,Gyulassy:2000er}.

The above considerations for the sub-sub-eikonal minus momentum transfer apply here as well: therefore, we need to replace $1/p_2^- \to 1/(P_2^- + i \pd^-)$ in the phase operator from \eq{phase}. Free quark propagation preserves the quark polarization, and hence the phase operator $-  i ({\un p}_2^2 + m^2)/[2 (P_2^- + i \pd^-)]$ enters \eq{O22} with $\delta_{\chi,\chi'}$. Adding the $-  i \delta_{\chi,\chi'} ({\un p}_2^2 + m^2)/[2 (P_2^- + i \pd^-)]$ operator into \eq{O22} while replacing ${\un p}_2^2 \to \cev{\un{\nabla}} \cdot \vec{\un{\nabla}}$ we obtain\footnote{The sub-eikonal phase in \eq{phase} does not constitute an interaction with the target. Moreover, the phase should not be included on the external (incoming and outgoing) quark legs. Therefore, strictly-speaking, by including this phase into the amplitude $A_{\sigma', \sigma}$ from \eq{Vxy}, we are also including the contributions where the phase originates on the external legs: such contributions have to be subtracted out if the incoming and outgoing quark lines are truly the external lines of the scattering process at hand. If the lines are internal, e.g., for small-$x$ evolution step with the soft quark emission, then the phase can be kept. For brevity, and due to these stated reasons, we will not include the external-line phase subtraction in our expressions. }
\begin{align}\label{O23}
& \mathcal{O}^{\textrm{G}}_{\chi', \chi} (x^-, \un{x}) = -  i \, \delta_{\chi,\chi'} \left[\cev{D}^i \frac{1}{2 (P_2^- + i D^-)}  \vec{D}^i + \frac{m^2}{2 (P_2^- + i D^-)}  \right]   \\ 
 & +  \frac{ig}{2} \Bigg\{ \delta_{\chi,\chi'} \,  A^+  + \delta_{\chi, -\chi'} \left[ F^{12}  \, \frac{1}{P_2^- + i D^-} - \frac{i}{(P_2^-)^2}  \epsilon^{ij} \, \cev{\nabla}^i \, F^{-j}   \right]  + \chi \, \delta_{\chi,\chi'}  \frac{m}{(P_2^-)^2} \, \epsilon^{ij} S^i F^{-j} + \chi \, \delta_{\chi, -\chi'}  \frac{i m }{(P_2^-)^2} \, S^i  \, F^{-i} \Bigg\}   , \notag
\end{align}
where $\vec{D}^i = \pd^i - i g A^i$, $\cev{D}^i = \cev{\pd}^i + i g A^i$ (cf. \cite{Altinoluk:2020oyd}). The $\cev{D}^i \cdot \vec{D}^i$ term contains the $A^i A^i$ product, which diagrammatically arises from the insertion of two non-eikonal gluon fields, with the instantaneous part of the quark propagator inserted between them \cite{Altinoluk:2020oyd}. Here we have restored this term by simply requiring gauge-covariance of the operator $\mathcal{O}^{\textrm{G}}_{\chi', \chi}$, which would lead to gauge-invariance of the quark scattering amplitude. The same philosophy allowed us to replace $\pd^- \to D^-$ in the denominators, with $D^\mu = \pd^\mu - i g A^\mu$ the covariant derivative. While the terms linear in the field $A^\mu$ and the terms independent of $A^\mu$ coming from the first term on the right of \eq{O23} are explicitly present in \eq{O22} already, the remaining terms were again restored by requiring gauge-covariance of the expression. Note that the $A^i A^- A^i$ term appears to require a calculation of a diagram with two consecutive insertions of the instantaneous term in the quark propagator.

In arriving at \eq{O23} we have also completed $\underline{\nabla} \times \underline{A} = - (\pd^x A^y - \pd^y A^x) \to - F^{12}$, with $F^{\mu\nu}$ the full non-Abelian gluon field strength tensor (cf. \cite{Kovchegov:2017lsr, Kovchegov:2018znm}). The $F^{12}$-terms corresponds to the gluon helicity operator found in \cite{Kovchegov:2017lsr, Kovchegov:2018znm}, since the $\delta_{\chi, -\chi'}$ structure in the transverse spinor basis corresponds to $\sigma \delta_{\sigma \sigma'}$ in the helicity spinor basis. Similarly, we have completed $\underline{S} \cross (\un{\nabla} \, A^- + \pd^- \un{A}) \to \epsilon^{ij} S^i F^{-j} $ for the transversity term $\sim \chi \, \delta_{\chi,\chi'}$ \cite{Kovchegov:2018zeq}, along with replacing $\un{\nabla}^i \, A^- + \pd^- \un{A}^i = - \pd^i A^- + \pd^- A^i \to F^{-i}$ elsewhere in \eq{O23}. The reason we can make this replacement is due to the commutator $[A^-, A^i]$ being further energy-suppressed, since the transverse components of the gluon field $A^i$ are at most sub-eikonal, while $A^-$ is sub-sub-eikonal. (This is the case only if the gluon field is emitted by a source, and does not represent the incoming source gluon \cite{Cougoulic:2020tbc}. In the latter case one needs to perform a calculation with two gluon field insertions and the instantaneous term in the quark propagator we mentioned above, similar to \cite{Altinoluk:2020oyd}).

The $A^+$-term in \eq{O23} will just yield the usual eikonal Wilson line: we will remove it from the operator $\mathcal{O}^{\textrm{G}}_{\chi', \chi}$, labeling the remainder of the operator by $\mathcal{O}^{\textrm{pol G}}_{\chi', \chi}$. Finally, replacing all the remaining $\cev{\nabla}^i \to \cev{D}^i$ with the sub-sub-eikonal accuracy of our calculation, we arrive at a completely gauge-covariant operator
\begin{align}\label{O24}
& \mathcal{O}^{\textrm{pol G}}_{\chi', \chi} (x^-, \un{x}) = -  i \, \delta_{\chi,\chi'} \left[\cev{D}^i \frac{1}{2 (P_2^- + i D^-)}  \vec{D}^i + \frac{m^2}{2 (P_2^- + i D^-)}  \right]   \\ 
 & +  \frac{ig}{2} \Bigg\{\delta_{\chi, -\chi'} \left[ F^{12}  \, \frac{1}{P_2^- + i D^-} - \frac{i}{(P_2^-)^2}  \epsilon^{ij} \, \cev{D}^i \, F^{-j}   \right]  + \chi \, \delta_{\chi,\chi'}  \frac{m}{(P_2^-)^2} \, \epsilon^{ij} S^i F^{-j} + \chi \, \delta_{\chi, -\chi'}  \frac{i m }{(P_2^-)^2} \, S^i  \, F^{-i} \Bigg\}   . \notag
\end{align}
Once again let us point out that the second-to-last term in the curly brackets of \eq{O24} corresponds to transversity \cite{Kovchegov:2018zeq}, while the $F^{12}$-term corresponds to helicity \cite{Kovchegov:2017lsr, Kovchegov:2018znm}. 

Equation \eqref{O24} is our final general result for the sub- and sub-sub-eikonal gluon insertion operator. The corresponding $S$-matrix from \eq{Vxy} is\footnote{Note, again, that subtraction of the sub-eikonal and sub-sub-eikonal phase corrections on the external quark legs is implied, but not shown explicitly.}
\begin{align}
\label{polwilgen2}
V^{\textrm{G}}_{\un{x}, \un{y}; \chi', \chi} = & V_{\un{x}} \, \delta^2 (\un{x} - \un{y}) \, \delta_{\chi,\chi'}  + \int\limits_{-\infty}^{\infty} \dd{z}^- d^2 z \ V_{\un{x}} [ \infty, z^-] \, \delta^2 (\un{x} - \un{z}) \, \mathcal{O}_{\chi', \chi}^{\textrm{pol G}} (z^-, \un{z}) \, V_{\un{y}} [ z^-, -\infty] \, \delta^2 (\un{y} - \un{z}) \\ & + \int\limits_{-\infty}^{\infty} \dd{z}_1^- d^2 z_1 \int\limits_{z_1^-}^{\infty} \dd{z}_2^- d^2 z_2 \sum_{\chi'' = \pm 1} \ V_{\un{x}} [ \infty, z_2^-] \, \delta^2 (\un{x} - \un{z}_2) \, \mathcal{O}_{\chi', \chi''}^{\textrm{pol G}} (z_2^-, \un{z}_2) \, V_{\un{z}_1} [z_2^-,  z_1^-] \, \delta^2 (\un{z}_2 - \un{z}_1) \notag \\ & \times \, \mathcal{O}_{\chi'', \chi}^{\textrm{pol G}} (z_1^-, \un{z}_1) \,   V_{\un{y}} [ z_1^-, -\infty] \, \delta^2 (\un{y} - \un{z}_1), \notag
\end{align}
where $\mathcal{O}_{\chi', \chi}^{\textrm{pol G}}$ is given by \eq{O24}. While the second term on the right of \eq{polwilgen2} contains the entire $\mathcal{O}_{\chi', \chi}^{\textrm{pol G}}$ from \eq{O24}, only sub-eikonal terms in each $\mathcal{O}_{\chi', \chi}^{\textrm{pol G}}$ need to be kept in the last term on the right in \eq{polwilgen2}. Equation \eqref{polwilgen2} contains the quark scattering amplitude on any target up to the sub-sub-eikonal order in the gluon exchange channel. The expression \eqref{polwilgen2} is valid in the gauges where the transverse links at $x^- \to \pm \infty$ do not contribute (e.g., to the dipole amplitude). 

%%%%%%%%%%%%%%%%%%%%%%%%%%%%%%%%%%%%%%%%%
\begin{table}[h!]
\begin{tabular}{|c|c|}
\hline
Helicity basis & Transverse basis \\
\hline
$\delta_{\sigma, \sigma'}$ & $\delta_{\chi, \chi'}$ \\
$\sigma \, \delta_{\sigma, \sigma'}$ & $\delta_{\chi, - \chi'}$ \\
$\delta_{\sigma, - \sigma'}$ & $\chi \, \delta_{\chi, \chi'}$ \\
$\sigma \, \delta_{\sigma, - \sigma'}$ & $-\chi \, \delta_{\chi, - \chi'}$ \\
\hline
\end{tabular}
\caption{A conversion table between the helicity and transverse spinor bases.}
\label{table:conversion}
\end{table}
%%%%%%%%%%%%%%%%%%%%%%%%%%%%%%%%%%%%%%%%%

In the helicity basis for spinors the operator \eqref{O24} can be re-written with the help of the conversion Table~\ref{table:conversion}. This yields
\begin{align}\label{O25}
& \mathcal{O}^{\textrm{pol G}}_{\sigma', \sigma} (x^-, \un{x}) = -  i \, \delta_{\sigma,\sigma'}  \left[\cev{D}^i \frac{1}{2 (P_2^- + i D^-)}  \vec{D}^i + \frac{m^2}{2 (P_2^- + i D^-)}  \right]   \\ 
 & +  \frac{ig}{2} \Bigg\{ \sigma \, \delta_{\sigma, \sigma'} \left[ F^{12}  \, \frac{1}{P_2^- + i D^-} - \frac{i}{(P_2^-)^2}  \epsilon^{ij} \, \cev{D}^i \, F^{-j}   \right]  + \delta_{\sigma , -\sigma'}  \frac{m}{(P_2^-)^2} \, \epsilon^{ij} S^i F^{-j} - \sigma \delta_{\sigma , -\sigma'}   \frac{i m }{(P_2^-)^2} \, S^i  \, F^{-i} \Bigg\}   . \notag
\end{align}
In the helicity basis \eq{polwilgen2} becomes
\begin{align}
\label{polwilgen3}
V^{\textrm{G}}_{\un{x}, \un{y}; \sigma', \sigma} = & V_{\un{x}} \, \delta^2 (\un{x} - \un{y}) \, \delta_{\sigma,\sigma'}  + \int\limits_{-\infty}^{\infty} \dd{z}^- d^2 z \ V_{\un{x}} [ \infty, z^-] \, \delta^2 (\un{x} - \un{z}) \, \mathcal{O}_{\sigma', \sigma}^{\textrm{pol G}} (z^-, \un{z}) \, V_{\un{y}} [ z^-, -\infty] \, \delta^2 (\un{y} - \un{z}) \\ & + \int\limits_{-\infty}^{\infty} \dd{z}_1^- d^2 z_1 \int\limits_{z_1^-}^{\infty} \dd{z}_2^- d^2 z_2 \sum_{\sigma'' = \pm 1} \ V_{\un{x}} [ \infty, z_2^-] \, \delta^2 (\un{x} - \un{z}_2) \, \mathcal{O}_{\sigma', \sigma''}^{\textrm{pol G}} (z_2^-, \un{z}_2) \, V_{\un{z}_1} [z_2^-,  z_1^-] \, \delta^2 (\un{z}_2 - \un{z}_1) \notag \\ & \times \, \mathcal{O}_{\sigma'', \sigma}^{\textrm{pol G}} (z_1^-, \un{z}_1) \,   V_{\un{y}} [ z_1^-, -\infty] \, \delta^2 (\un{y} - \un{z}_1). \notag
\end{align}

%%%%%%%%%%%%%%%%%%%%%%%%%%%%%%%%%%%%%%%%%%%%%%%%%%%%%%%%%%%%%%%%%%%%%%%%%%%%%%%%%%

\subsection{Quark insertion operator}

%%%%%%%%%%%%%%%%%%%%%%%%%%%%%%%%%%%%%%%%%%%%%%%%%%%%%%%%%%%%%%%%%%%%%%%%%%%%%%%%%%%%%%
\begin{figure}[h]
\centering
\includegraphics[width=0.7\linewidth]{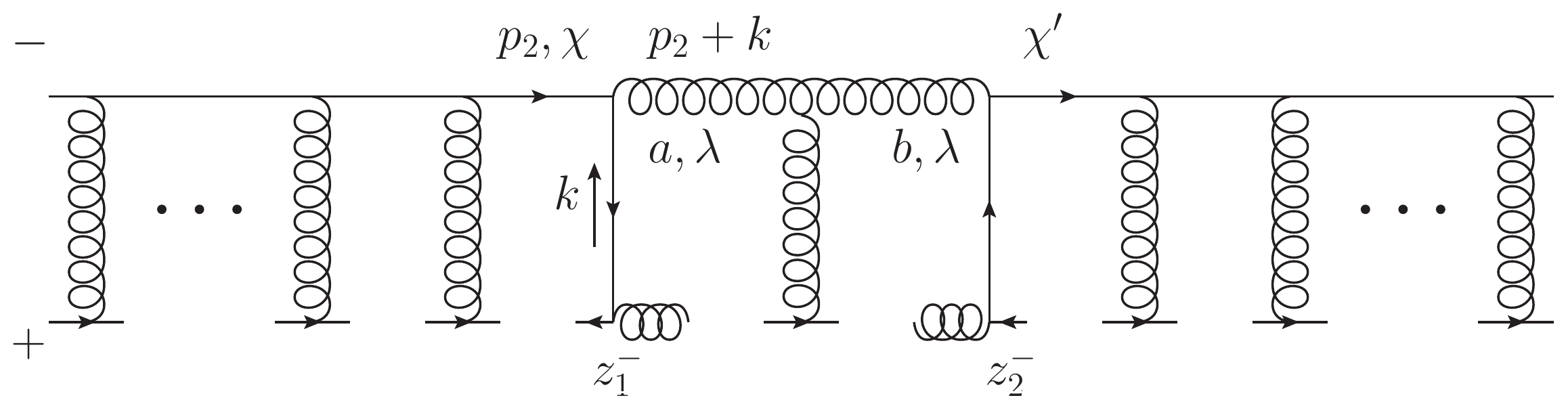}  
\caption{The diagram representing the quark field contribution to the sub-eikonal and sub-sub-eikonal operators. }
\label{FIG:opoldiagquark}
\end{figure}
%%%%%%%%%%%%%%%%%%%%%%%%%%%%%%%%%%%%%%%%%%%%%%%%%%%%%%%%%%%%%%%%%%%%%%%%%%%%%%%%%%%%%%

Now we turn to the two quark exchanges shown in \fig{FIG:opoldiagquark}. The part of this operator related to transversity at the sub-sub-eikonal order was calculated in \cite{Kovchegov:2018zeq}. Here we follow similar steps, but this time retaining all the other terms at the sub-eikonal and sub-sub-eikonal orders. We begin with the left quark exchange in \fig{FIG:opoldiagquark}, which, in the $A^- =0$ gauge, after expansion in $k^-/p_2^-$ in some of the terms, gives
\begin{align}\label{op1}
& \frac{1}{2 \sqrt{p_2^- \, (p_2^- + k^-)}} \, \epsilon_\lambda^{\mu \, *} (p_2+k) \, (i g) \, {\bar \psi} (k) t^a \gamma_\mu u_{\chi} (p_2) = - \frac{i g \, \chi}{2 \,  \sqrt{p_2^- + k^-}} \, {\bar \psi} (k) \, t^a  \Bigg[  \delta_{\lambda, -1} \, \rho (+1) + \chi \, \delta_{\lambda, 1} \, \rho(-1) \notag \\ & - \frac{m}{p_2^-} \, \left[ \chi \, \delta_{\lambda,  1} \, \gamma^0 \, \rho (-1) + \delta_{\lambda, -1} \, \gamma^0 \, \rho (+1) \right]  - \sqrt{2} \, \frac{{\un \epsilon}_\lambda^* \cdot {\un k}}{p_2^-} \, \left[ \chi \, \gamma^0 \, \rho (+1) + \gamma^0 \, \rho (-1) \right] \notag \\ & - \sqrt{2} \, \frac{{\un \epsilon}_\lambda^* \cdot {\un p}_2}{p_2^-} \, \left[ \chi \, \delta_{\lambda,  1} \, \gamma^0 \, \rho (+1) + \delta_{\lambda, -1} \, \gamma^0 \, \rho (-1) \right] + {\cal O} \left( \frac{1}{(p_2^-)^2} \right) \Bigg].
\end{align}
The gluon polarization 4-vector in the $A^- =0$ gauge is $\epsilon^\mu_\lambda (p_2 + k) = \left( \frac{2 \, {\un \epsilon}_\lambda \cdot ({\un p}_2 + {\un k})}{p_2^- + k^-}, 0, {\un \epsilon}_\lambda \right)$ with ${\un \epsilon}_\lambda = (-1/\sqrt{2}) \, (\lambda, i)$ \cite{Lepage:1980fj}. The expression \eqref{op1} is consistent with Eq.~(12) in \cite{Kovchegov:2018zeq}, if we take into account that our definition of the light-cone coordinates \eqref{LC_coord} is different from that in \cite{Kovchegov:2018zeq} and that now we do not put $\un{p}_2 =0$, as was assumed in \cite{Kovchegov:2018zeq}. 

In the coordinate space \eq{op1} gives
\begin{align}\label{op2}
& - \frac{i g}{2 \,  \sqrt{P_2^- + i \pd^-}} \, {\bar \psi} (z_1^-, {\un z}) \, t^a \, \cev{M} (\lambda, \chi) \equiv - \frac{i g \, \chi}{2 \,  \sqrt{P_2^- + i \pd^-_{z_1}}} \, {\bar \psi} (z_1^-,\un{z}_1) \, t^a \,  \Bigg[  \delta_{\lambda, -1} \, \rho (+1) + \chi \, \delta_{\lambda, 1} \, \rho(-1) \notag \\ & - \frac{m}{p_2^-} \, \left[ \chi \, \delta_{\lambda,  1} \, \gamma^0 \, \rho (-1) + \delta_{\lambda, -1} \, \gamma^0 \, \rho (+1) \right] + i \sqrt{2} \, \frac{{\un \epsilon}_\lambda^* \cdot \cev{\un \nabla}_{z_1}}{p_2^-} \, \left[ \chi \, \gamma^0 \, \rho (+1) + \gamma^0 \, \rho (-1) \right] \notag \\ & + i \sqrt{2} \, \frac{{\un \epsilon}_\lambda^* \cdot {\un \nabla}_{z_1}}{p_2^-} \, \left[ \chi \, \delta_{\lambda,  1} \, \gamma^0 \, \rho (+1) + \delta_{\lambda, -1} \, \gamma^0 \, \rho (-1) \right] + {\cal O} \left( \frac{1}{(p_2^-)^2} \right) \Bigg],
\end{align}
where $\cev{\un \nabla}_{z_1}$ acts only on $\bar \psi$, while ${\un \nabla}_{z_1}$ will act on the objects to the right of the operator $\mathcal{O}^{\textrm{pol q}\overline{\textrm{q}}}_{\chi', \chi}$ (that is, earlier in $z^-$)  when we assemble everything together.  In \eq{op2} we have also defined a new object $\cev{M} (\lambda, \chi)$. The two-quark exchange contribution to the polarized Wilson line/quark $S$-matrix operator, as depicted in \fig{FIG:opoldiagquark}, can be calculated in terms of $\cev{M} (\lambda, \chi)$ as \cite{Kovchegov:2018zeq}
\begin{align}\label{2q_1}
& V^{\textrm{pol q}\overline{\textrm{q}}}_{\un{x}, \un{y}; \chi', \chi} \supset - \frac{g^2 \, p_1^+ }{4 \, s}
  \int\limits_{-\infty}^\infty  d z_1^- \, d^2 z_1  \int\limits_{z_1^-}^\infty d z_2^- \, d^2 z_2 \, \sum_\lambda \, V_{\un x} [\infty, z_2^-] \: \delta^2 (\un{x} - \un{z}_2) \, t^b  \, \left[ {\overrightarrow M}^\dagger (\lambda, \chi') \, \gamma^0 \, {\psi} (z_2^-, {\un z}_2)  \right] \\ 
 & \times \,   \frac{1}{\sqrt{1 + (i \pd^-_{z_2}/P_2^-)}} \, U_{{\un z}_2}^{ba} [z_2^-,  z_1^-] \, \delta^2 (\un{z}_2 - \un{z}_1)  \,  \frac{1}{\sqrt{1 + (i \pd^-_{z_1}/P_2^-)}} \left[ {\bar \psi} (z_1^-, {\un z}_1) \, \overleftarrow{M} (\lambda, \chi) \right] \: t^a \,V_{\un y} [z_1^- , -\infty] \, \delta^2 (\un{y} - \un{z}_1),  \notag
\end{align}
where $s = p_1^+ P_2^-$ with the quark in the target carrying large `plus' momentum component $p_1^+$. The adjoint light-cone Wilson line $U$ is defined analogously to the fundamental one in \eq{Wline}, 
\begin{align}\label{Uline}
U_{\un{x}} [x^-_f,x^-_i] = \mathcal{P} \exp \left[ \frac{ig}{2} \int\limits_{x^-_i}^{x^-_f} \dd{x}^- {\cal A}^+ (0^+, x^-, \un{x}) \right],
\end{align}
with ${\cal A}^\mu = \sum_a A^{a \, \mu} \, T^a$ and with the adjoint SU($N_c$) generators $T^a$ defined by $(T^a)_{bc} = - i f^{abc}$.

Employing \eq{op2} in \eq{2q_1} and comparing the result to \eq{polwilgendoub}, yields, after some algebra,
\begin{align}\label{Oq1}
& \mathcal{O}^{\textrm{pol q}\overline{\textrm{q}}}_{\chi', \chi} (z_2^-, z_1^-;  \un{z}_2, \un{z}_1) = - \frac{g^2 \, p_1^+}{8 \, s} \, t^b \, \psi_{\beta} (z_2^-,\un{z}_2) \,  \left[ 1 - \frac{i \, p_1^+ \, \pd^-_{z_2}}{2 s} \right] \, U_{\un{z}_2}^{ba} [z_2^-,z_1^-] \, \delta^2 (\un{z}_2 - \un{z}_1)  \left[ 1 - \frac{i \, p_1^+ \, \pd^-_{z_1}}{2 s} \right] \notag \\
& \times \, \Bigg\{ \gamma^+ \, \delta_{\chi, \chi'} - \gamma^+ \, \gamma^5 \, \delta_{\chi, -\chi'} - \frac{2 m p_1^+}{s} \left[ \delta_{\chi, \chi'} - \delta_{\chi, -\chi'} \, i \, \gamma^1 \, \gamma^2 \right]  \notag \\ 
& + \frac{p_1^+}{2 s} \Bigg[ - \chi \chi' (1-\gamma^5) \, \Big( [i \gamma^2 + \gamma^1 ]  [i \underline{S} \vdot \underline{{\pd}}_{z_2} + \underline{S} \cross \underline{\pd}_{z_2} ] + [i \gamma^2 - \gamma^1 ]  [i \underline{S} \vdot \underline{{\pd}}_{z_1} - \underline{S} \cross \underline{\pd}_{z_1} ]  \Big) \notag \\ 
& + (1 + \gamma^5) \, \Big( [i \gamma^2 - \gamma^1 ]  [i \underline{S} \vdot \underline{{\pd}}_{z_2} - \underline{S} \cross \underline{\pd}_{z_2} ] + [i \gamma^2 + \gamma^1 ]  [i \underline{S} \vdot \underline{{\pd}}_{z_1} + \underline{S} \cross \underline{\pd}_{z_1} ]  \Big) \notag \\ 
& + \chi \, (1-\gamma^5) \, \Big( (1 + i \gamma^1 \gamma^2)  [i \underline{S} \vdot \underline{\cev{\pd}}_{z_2} + \underline{S} \cross \underline{\cev{\pd}}_{z_2} ] + (1 - i \gamma^1 \gamma^2)  [i \underline{S} \vdot \underline{\vec{\pd}}_{z_1} + \underline{S} \cross \underline{\vec{\pd}}_{z_1} ]  \Big) \notag \\ 
& - \chi' \, (1+\gamma^5) \, \Big( (1 - i \gamma^1 \gamma^2)  [i \underline{S} \vdot \underline{\cev{\pd}}_{z_2} - \underline{S} \cross \underline{\cev{\pd}}_{z_2} ] + (1 + i \gamma^1 \gamma^2)  [i \underline{S} \vdot \underline{\vec{\pd}}_{z_1} - \underline{S} \cross \underline{\vec{\pd}}_{z_1} ]  \Big) \Bigg] \Bigg{\}}_{\alpha \beta} \notag \\
&\times \bar{\psi}_\alpha (z_1^-,\un{z}_1) \, t^a + {\cal O} \left( \frac{1}{s^3} \right)  . 
\end{align}
Here the partial derivatives $\underline{{\pd}}_{z_1}$ and $\underline{{\pd}}_{z_2}$ are acting on $U_{\un{z}_2}^{ba} [z_2^-,z_1^-] \, \delta^2 (\un{z}_2 - \un{z}_1)$ only, while the partial derivatives with arrows, $\underline{\vec{\pd}}_{z_1}$ and $\underline{\cev{\pd}}_{z_2}$, are only acting on $\bar \psi (z_1^-,\un{z}_1)$ and $\psi (z_2^-,\un{z}_2)$, respectively (as indicated by the direction of the arrows). Additionally, $\alpha, \beta = 1, 2, 3, 4$ are the Dirac spinor indices. 

When comparing \eq{Oq1} to Eq.~(16) in \cite{Kovchegov:2018zeq} one has to take into account a different definition of the light-cone coordinates \eqref{LC_coord} used here and that now we do not put $\un{p}_2 =0$: even then, \eq{Oq1} differs from Eq.~(16) in \cite{Kovchegov:2018zeq} by several minus signs. In addition, Eq.~(16) in \cite{Kovchegov:2018zeq} was partially projected onto $\chi = \chi'$, with some transverse polarization-independent terms discarded, as not contributing to transversity, calculating which was the main goal of \cite{Kovchegov:2018zeq}. Hence it is difficult to fully compare \eq{Oq1} to Eq.~(16) in \cite{Kovchegov:2018zeq}. However, the transverse spin-dependent parts ($\sim \chi \delta_{\chi, \chi'}$) of the two expressions are the same: we, therefore, confirm the polarized Wilson line operator used in the transversity calculation of \cite{Kovchegov:2018zeq}. 

The next step is to rewrite the expression \eqref{Oq1}, derived in the $A^- =0$ gauge, in a gauge-covariant form, similar to \eq{O24}. Following \cite{Kovchegov:2018zeq}, this is easily accomplished by ``promoting" the partial derivatives into covariant derivatives, except now the derivatives acting on $U_{\un{z}_2}^{ba} [z_2^-,z_1^-] \, \delta^2 (\un{z}_2 - \un{z}_1)$ will become adjoint covariant derivatives. (In the $A^- =0$ gauge, the $A^i$ field of the target is energy-suppressed: therefore, replacing transverse derivatives in \eq{Oq1} by covariant derivatives constitutes adding sub-sub-sub-eikonal corrections and is allowed by the accuracy of our sub-sub-eikonal approximation.) We thus write
\begin{align}\label{Oq2}
  &\mathcal{O}^{\textrm{pol q}\overline{\textrm{q}}}_{\chi', \chi} (z_2^-, z_1^-;  \un{z}_2, \un{z}_1) = - \frac{g^2 \, p_1^+}{8 \, s} \, t^b \, \psi_{\beta} (z_2^-,\un{z}_2) \,  \left[ \delta^{b'b''} - \frac{i \, p_1^+ \, \mathscr{D}^{b'b''-}_{z_2}}{2 s} \right] \, U_{\un{z}_2}^{b''a''} [z_2^-,z_1^-] \, \delta^2 (\un{z}_2 - \un{z}_1) \\ 
 & \times  \left[ \delta^{a''a'} - \frac{i p_1^+ \mathscr{D}^{a''a'-}_{z_1}}{2 s} \right]  \Bigg\{ \gamma^+ \delta_{\chi, \chi'} \delta^{a'a} \, \delta^{bb'} - \gamma^+ \gamma^5 \delta_{\chi, -\chi'} \delta^{a'a} \delta^{bb'} - \frac{2 m p_1^+}{s} \left[ \delta_{\chi, \chi'} - \delta_{\chi, -\chi'}  i  \gamma^1 \gamma^2 \right] \delta^{a'a} \delta^{bb'} \notag \\ 
 & + \frac{p_1^+}{2 s} \Bigg[ - \chi \chi' (1-\gamma^5) \, \Big( [i \gamma^2 + \gamma^1 ]  [i \underline{S} \vdot \cev{\underline{\mathscr{D}}}_{z_2}^{bb'} + \underline{S} \cross \cev{\underline{\mathscr{D}}}_{z_2}^{bb'}  ] \, \delta^{a'a} + [i \gamma^2 - \gamma^1 ]  [i \underline{S} \vdot \underline{\mathscr{D}}_{z_1}^{a'a} - \underline{S} \cross \underline{\mathscr{D}}_{z_1}^{a'a}  ] \, \delta^{bb'}  \Big) \notag \\ 
 & + (1 + \gamma^5) \, \Big( [i \gamma^2 - \gamma^1 ]  [i \underline{S} \vdot \cev{\underline{\mathscr{D}}}_{z_2}^{bb'} - \underline{S} \cross \cev{\underline{\mathscr{D}}}_{z_2}^{bb'}  ] \, \delta^{a'a} + [i \gamma^2 + \gamma^1 ]  [i \underline{S} \vdot \underline{\mathscr{D}}_{z_1}^{a'a} + \underline{S} \cross \underline{\mathscr{D}}_{z_1}^{a'a}   ]  \, \delta^{bb'} \Big)   \notag \\ 
 & + \chi \, (1-\gamma^5) \, \Big( (1 + i \gamma^1 \gamma^2)  \left[ i \underline{S} \vdot \cev{\underline{D}}_{z_2} + \underline{S} \cross \cev{\underline{D}}_{z_2} \right]  + (1 - i \gamma^1 \gamma^2)  \left[ i \underline{S} \vdot \underline{D}_{z_1}  + \underline{S} \cross \underline{D}_{z_1} \right] \Big) \, \delta^{a'a} \, \delta^{bb'} \notag \\ 
 & - \chi' \, (1+\gamma^5) \, \Big( (1 - i \gamma^1 \gamma^2)  \left[ i \underline{S} \vdot \cev{\underline{D}}_{z_2} - \underline{S} \cross \cev{\underline{D}}_{z_2} \right]  + (1 + i \gamma^1 \gamma^2)  \left[ i \underline{S} \vdot \underline{D}_{z_1} - \underline{S} \cross \underline{D}_{z_1}\right]  \Big) \, \delta^{a'a} \, \delta^{bb'} \Bigg] \Bigg{\}}_{\alpha \beta} \notag \\
 &\times \bar{\psi}_\alpha (z_1^-,\un{z}_1) \, t^a  + {\cal O} \left( \frac{1}{s^3} \right) , \notag
\end{align}
with $D^i_{z_1} = \pd^i_{z_1} - \, i g A^i (z_1^-,\un{z}_1)$ and $\cev{D}^i_{z_2} = \cev{\pd}^i_{z_2} + \, i g A^i (z_2^-,\un{z}_2)$ acting on spinors $\bar \psi$ and $\psi$ with the color matrix in $A^i$ inserted between $\bar{\psi}_\alpha t^a$ and between $t^b \, \psi_{\beta}$, respectively,  while $\cev{\underline{\mathscr{D}}}_{z_2}^{ab} = \cev{\un{\pd}}_{z_2} \delta^{ab} - g f^{acb} \un{A}^c (z_2^-,\un{z}_2)$  and  $\underline{\mathscr{D}}_{z_1}^{ab} = \un{\pd}_{z_1} \delta^{ab} + g f^{acb} \un{A}^c (z_1^-,\un{z}_1)$ both act on $U_{\un{z}_2}^{ba} [z_2^-,z_1^-] \, \delta^2 (\un{z}_2 - \un{z}_1)$. The adjoint covariant derivatives $\mathscr{D}^{b'b''-}_{z_2}$ and $\mathscr{D}^{a''a'-}_{z_1}$ act on everything to their right (with $\mathscr{D}^{ab-}_{z}= \pd^-_{z} \delta^{ab} + g f^{acb} A^{c -}$).

Employing 
\begin{align}
1 = \delta_{\chi, \chi'} + \delta_{\chi, -\chi'} , \ \ \ \chi \, \chi' = \delta_{\chi, \chi'} - \delta_{\chi, -\chi'} , \ \ \ \chi = \chi \delta_{\chi, \chi'} + \chi \delta_{\chi, -\chi'} , \ \ \ \chi' = \chi \delta_{\chi, \chi'} - \chi \delta_{\chi, -\chi'},
\end{align}
we recast \eq{Oq2} as
\begin{align}\label{Oq3}
& \mathcal{O}^{\textrm{pol q}\overline{\textrm{q}}}_{\chi', \chi} (z_2^-, z_1^-;  \un{z}_2, \un{z}_1) = - \frac{g^2 \, p_1^+}{8 \, s} \, t^b \, \psi_{\beta} (z_2^-,\un{z}_2) \, \left[ \delta^{b'b''} - \frac{i \, p_1^+ \, \mathscr{D}^{b'b''-}_{z_2}}{2 s} \right] \, U_{\un{z}_2}^{b''a''} [z_2^-,z_1^-] \, \delta^2 (\un{z}_2 - \un{z}_1)  \\ 
 & \times \left[ \delta^{a''a'} - \frac{i \, p_1^+ \, \mathscr{D}^{a''a'-}_{z_1}}{2 s} \right]   \Bigg\{  \delta_{\chi, \chi'} \Bigg[ \gamma^+ \, \delta^{a'a} \, \delta^{bb'} - \frac{2 m p_1^+}{s} \, \delta^{a'a} \, \delta^{bb'} \notag \\ 
 &  \hspace*{1.5cm}  - \frac{p_1^+}{s} \Bigg( (\gamma^1 - i \gamma^5 \gamma^2) \, [i \underline{S} \vdot \cev{\underline{\mathscr{D}}}_{z_2}^{bb'} + \gamma^5 \underline{S} \cross \cev{\underline{\mathscr{D}}}_{z_2}^{bb'}  ] \, \delta^{a'a}  - (\gamma^1 + i \gamma^5 \gamma^2) \,  [i \underline{S} \vdot \underline{\mathscr{D}}_{z_1}^{a'a} \textcolor{blue}{-} \, \gamma^5 \underline{S} \cross \underline{\mathscr{D}}_{z_1}^{a'a}   ]  \, \delta^{bb'} \Bigg) \Bigg] \notag \\ 
 & \textcolor{blue}{-} \, \delta_{\chi, -\chi'} \Bigg[ \gamma^+ \, \gamma^5 \, \delta^{a'a} \, \delta^{bb'}  - \frac{2 m p_1^+}{s} \, i \, \gamma^1 \, \gamma^2 \, \delta^{a'a} \, \delta^{bb'} \notag \\ 
 & \hspace*{1.5cm} - \frac{p_1^+}{s} \Bigg( (i \gamma^2 - \gamma^5 \gamma^1) \, [i \underline{S} \vdot \cev{\underline{\mathscr{D}}}_{z_2}^{bb'} + \gamma^5 \underline{S} \cross \cev{\underline{\mathscr{D}}}_{z_2}^{bb'}  ] \, \delta^{a'a} + (i \gamma^2 + \gamma^5 \gamma^1) \,  [i \underline{S} \vdot \underline{\mathscr{D}}_{z_1}^{a'a} \textcolor{blue}{-} \, \gamma^5 \underline{S} \cross \underline{\mathscr{D}}_{z_1}^{a'a}   ]  \, \delta^{bb'} \Bigg) \Bigg]  \notag \\ 
 &  - \chi \delta_{\chi, \chi'} \frac{p_1^+}{s} \, \delta^{a'a} \, \delta^{bb'} \Bigg[ \left[ i \gamma^5 \underline{S} \vdot \cev{\underline{D}}_{z_2} - \underline{S} \cross \cev{\underline{D}}_{z_2} \right] (1 - i \gamma^5 \gamma^1 \gamma^2) +  \left[ i \gamma^5 \underline{S} \vdot \underline{D}_{z_1}  - \underline{S} \cross \underline{D}_{z_1} \right] (1 + i \gamma^5 \gamma^1 \gamma^2) \Bigg] \notag \\ 
 & + \chi \delta_{\chi, - \chi'} \frac{p_1^+}{s} \delta^{a'a} \delta^{bb'} \Bigg[ \left[ i \underline{S} \vdot \cev{\underline{D}}_{z_2} - \gamma^5 \underline{S} \cross \cev{\underline{D}}_{z_2} \right] (1 - i \gamma^5 \gamma^1 \gamma^2) +  \left[ i \underline{S} \vdot \underline{D}_{z_1}  - \gamma^5 \underline{S} \cross \underline{D}_{z_1} \right] (1 + i \gamma^5 \gamma^1 \gamma^2) \Bigg] \Bigg{\}}_{\alpha \beta}  \notag \\ 
 &  \times \bar{\psi}_\alpha (z_1^-,\un{z}_1) \, t^a + {\cal O} \left( \frac{1}{s^3} \right). \notag
\end{align}
Once again let us stress that the transverse spin-dependent $\chi \delta_{\chi, \chi'}$ term in \eq{Oq3} is identical to that in Eq.~(22) of \cite{Kovchegov:2018zeq}, if one takes into account the difference between the light cone coordinate definitions in that reference and here.

Inserting the operator $\mathcal{O}^{\textrm{pol q}\overline{\textrm{q}}}_{\chi', \chi}$ once or twice into the quark $S$-matrix we obtain
\begin{align}
\label{polwilgen4}
& V^{\textrm{pol q}\overline{\textrm{q}}}_{\un{x}, \un{y}; \chi', \chi} = \int\limits_{-\infty}^{\infty} \dd{z}_1^- d^2 z_1 \int\limits_{z_1^-}^\infty d z_2^- d^2 z_2 \ V_{\un{x}} [ \infty, z_2^-] \, \delta^2 (\un{x} - \un{z}_2) \, \mathcal{O}^{\textrm{pol q}\overline{\textrm{q}}}_{\chi', \chi} (z_2^-, z_1^-;  \un{z}_2, \un{z}_1)  \, V_{\un{y}} [ z_1^-, -\infty] \, \delta^2 (\un{y} - \un{z}_1) \\ & + \int\limits_{-\infty}^{\infty} \dd{z}_1^- d^2 z_1 \int\limits_{z_1^-}^{\infty} \dd{z}_2^- d^2 z_2 \int\limits_{z_2^-}^\infty \dd{z}_3^- d^2 z_3 \int\limits_{z_3^-}^{\infty} \dd{z}_4^- d^2 z_4 \sum_{\chi'' = \pm 1} \ V_{\un{x}} [ \infty, z_4^-] \, \delta^2 (\un{x} - \un{z}_4) \, \mathcal{O}^{\textrm{pol q}\overline{\textrm{q}}}_{\chi', \chi''} (z_4^-, z_3^-;  \un{z}_4, \un{z}_3) \notag \\ & \times \, V_{\un{z}_3} [z_3^-,  z_2^-] \, \delta^2 (\un{z}_3 - \un{z}_2) \, \mathcal{O}^{\textrm{pol q}\overline{\textrm{q}}}_{\chi'', \chi} (z_2^-, z_1^-;  \un{z}_2, \un{z}_1)  \,   V_{\un{y}} [ z_1^-, -\infty] \, \delta^2 (\un{y} - \un{z}_1), \notag
\end{align}
up to and including the sub-sub-eikonal order. The corresponding expression in the helicity basis can be constructed with the help of Table~\ref{table:conversion}. 

Since all the derivatives in $\mathcal{O}^{\textrm{pol q}\overline{\textrm{q}}}_{\chi', \chi}$ are acting only on the objects within that operator, and not on the functions multiplying it, we can integrate out most of the transverse coordinates in \eq{polwilgen4}, writing 
\begin{align}
\label{polwilgen5}
& V^{\textrm{pol q}\overline{\textrm{q}}}_{\un{x}, \un{y}; \chi', \chi} =  \int\limits_{-\infty}^{\infty} \dd{z}_1^- \int\limits_{z_1^-}^\infty d z_2^-  \ V_{\un{x}} [ \infty, z_2^-] \,  \mathcal{O}^{\textrm{pol q}\overline{\textrm{q}}}_{\chi', \chi} (z_2^-, z_1^-;  \un{x}, \un{y})  \, V_{\un{y}} [ z_1^-, -\infty]  \\ 
& + \int\limits_{-\infty}^{\infty} \dd{z}_1^- \int\limits_{z_1^-}^{\infty} \dd{z}_2^- \int\limits_{z_2^-}^\infty \dd{z}_3^-  \int\limits_{z_3^-}^{\infty} \dd{z}_4^-  d^2 z \sum_{\chi'' = \pm 1}  V_{\un{x}} [ \infty, z_4^-] \,  \mathcal{O}^{\textrm{pol q}\overline{\textrm{q}}}_{\chi', \chi''} (z_4^-, z_3^-;  \un{x}, \un{z})  \, V_{\un{z}} [z_3^-,  z_2^-] \, \mathcal{O}^{\textrm{pol q}\overline{\textrm{q}}}_{\chi'', \chi} (z_2^-, z_1^-;  \un{z}, \un{y})  \,   V_{\un{y}} [ z_1^-, -\infty] . \notag
\end{align}

%%%%%%%%%%%%%%%%%%%%%%%%%%%%%%%%%%%%%%%%%%%%%%%%%%%%%%%%%%%%%%%%%%%%%%%%%%%%%%%%%%%%%%%

\subsection{Quark and gluon insertion operator}

It is tempting, at this point, to assume that the full sub-sub-eikonal polarized Wilson line is obtained by simply adding equations \eqref{O24} and \eqref{Oq3},
\begin{align}
\mathcal{O}^{\textrm{pol G}}_{\chi', \chi} + \mathcal{O}^{\textrm{pol q}\overline{\textrm{q}}}_{\chi', \chi} ,
\end{align}
and inserting the sum of the two operators once or twice into the Wilson lines, similar to Eqs.~\eqref{polwilgen2} and \eqref{polwilgen5}, keeping only the terms up to and including the sub-sub-eikonal order. 

%%%%%%%%%%%%%%%%%%%%%%%%%%%%%%%%%%%%%%%%%%%%%%%%%%%%%%%%%%%%%%%%%%%%%%%%%%%%%%%%%%%%%%
\begin{figure}[ht]
\centering
\includegraphics[width=0.7\linewidth]{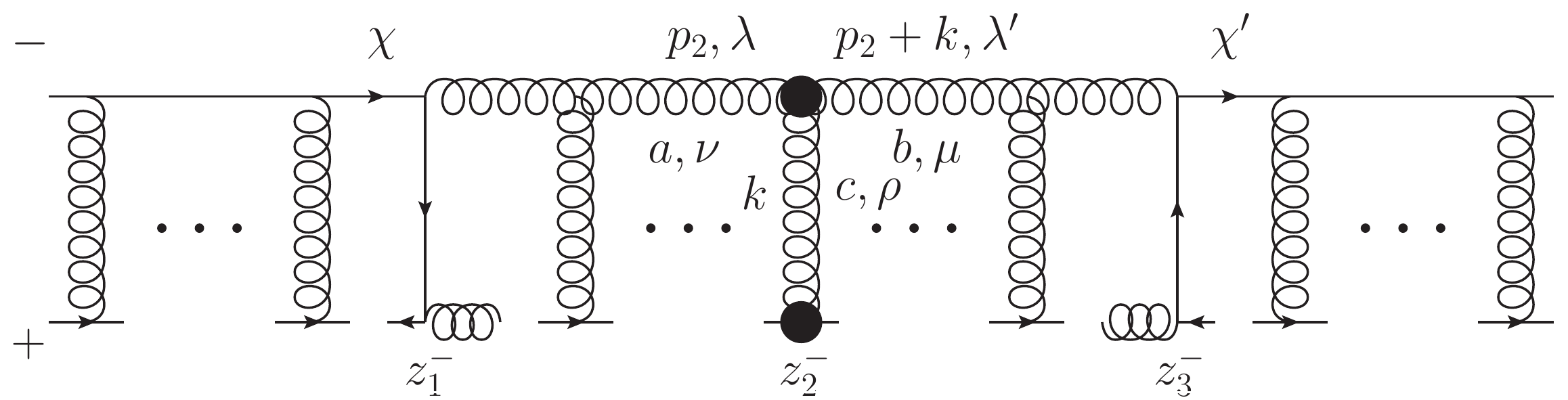}  
\caption{The diagram representing the quark, anti-quark and sub-eikonal gluon field insertions contributing to the quark $S$-matrix at the sub-sub-eikonal order. }
\label{FIG:qGqbar}
\end{figure}
%%%%%%%%%%%%%%%%%%%%%%%%%%%%%%%%%%%%%%%%%%%%%%%%%%%%%%%%%%%%%%%%%%%%%%%%%%%%%%%%%%%%%%

This is almost correct: the two contributions which would be missed by such a strategy are depicted in \fig{FIG:qGqbar} (see also diagram D in Fig.~7 of \cite{Kovchegov:2018zeq}) and \fig{FIG:qqqbarqbar}. The first operator, coming from \fig{FIG:qGqbar}, combines an exchange of the quark and anti-quark, with a sub-eikonal gluon exchange: together this gives a sub-sub-eikonal contribution. The eikonal gluon exchanges between two quark exchanges are already included in \fig{FIG:opoldiagquark} and in the corresponding operator \eqref{Oq3} via the adjoint Wilson line $U$. 

The quark exchanges have been found above in \eq{op2}. The sub-eikonal gluon exchange with the quark projectile can be obtained from \eq{O24} above. However, we need the expression for the gluon projectile. A calculation along the above lines yields the following sub-eikonal gluon insertion contribution, along with the sub-eikonal phase (cf. \cite{Kovchegov:2018znm,Cougoulic:2019aja,Cougoulic:2020tbc}):
\begin{align}\label{Ginsertion}
\frac{i}{2 p_2^-} \left[ 2  g \, \lambda \, \delta_{\lambda, \lambda'} \left( {\cal F}^{12} \right)^{ba}  -  \delta_{\lambda, \lambda'} \, \cev{\underline{\mathscr{D}}}^{bc} \cdot  \underline{\mathscr{D}}^{ca} \right].
\end{align}    
Here ${\cal F}^{12} = \sum_a {\cal F}^{a \, 12}  \, T^a$ is the gluon field strength tensor in the adjoint representation, while the adjoint covariant derivatives $\cev{\underline{\mathscr{D}}}^{ab} = \cev{\un{\pd}} \, \delta^{ab} - g f^{acb} \un{A}^c$  and  $\underline{\mathscr{D}}^{ab} = \un{\pd} \, \delta^{ab} + g f^{acb} \un{A}^c$ act on the operators to be added to the left and to the right of the operator \eqref{Ginsertion}. 

Including the sub-eikonal quark exchange contributions we arrive at the following gauge-covariant contribution of the diagrams like that shown in \fig{FIG:qGqbar} to the quark $S$-matrix at the sub-sub-eikonal level: 
\begin{align}\label{VqGq1}
V^{\textrm{pol qG}\overline{\textrm{q}}}_{\un{x}, \un{y}; \chi', \chi} &= \frac{i g^2 (p_1^+)^2}{16 s^2} \int\limits_{-\infty}^{\infty} \dd{z}_1^- d^2 z_1 \int\limits_{z_1^-}^\infty d z_2^- d^2 z_2 \int\limits_{z_2^-}^\infty d z_3^- d^2 z_3  \ V_{\un{x}} [ \infty, z_3^-] \, \delta^2 (\un{x} - \un{z}_3) \, t^b \, \psi_{\beta} (z_3^-,\un{z}_3) \\
&\times U_{\un{z}_2}^{b'b} [z_3^-,z_2^-] \, \delta^2 (\un{z}_2 - \un{z}_3) \Bigg\{ \delta_{\chi, \chi'} \left[ \gamma^+ \, \cev{\underline{\mathscr{D}}}_{z_2}^{bc} \cdot  \underline{\mathscr{D}}_{z_2}^{ca} + \gamma^+ \gamma^5 \, 2 g \,  \left( {\cal F}^{12} (z_2) \right)^{ba} \right] \notag \\
&- \delta_{\chi, -\chi'} \left[ \gamma^+ \gamma^5 \, \cev{\underline{\mathscr{D}}}_{z_2}^{bc} \cdot  \underline{\mathscr{D}}_{z_2}^{ca} + \gamma^+ \,  2 g \,  \left( {\cal F}^{12} (z_2) \right)^{ba} \right]  \Bigg\}_{\alpha \beta}  \notag \\ 
& \times U_{\un{z}_2}^{aa'} [z_2^-, z_1^-] \, \delta^2 (\un{z}_2 - \un{z}_1) \, \bar{\psi}_\alpha (z_1^-,\un{z}_1) \, t^{a'} \, V_{\un{y}} [ z_1^-, -\infty] \, \delta^2 (\un{y} - \un{z}_1). \notag
\end{align}
Since the (covariant) derivatives in \eq{VqGq1} are only acting on ${\un z}_2$, we can integrate out ${\un z}_1$ and ${\un z}_3$, obtaining
\begin{align}\label{VqGq2}
& V^{\textrm{pol qG}\overline{\textrm{q}}}_{\un{x}, \un{y}; \chi', \chi} = \frac{i g^2 (p_1^+)^2}{16 s^2} \int\limits_{-\infty}^{\infty} \dd{z}_1^- \int\limits_{z_1^-}^\infty d z_2^- \int\limits_{z_2^-}^\infty d z_3^- d^2 z_2  \ V_{\un{x}} [ \infty, z_3^-] \, t^b \, \psi_{\beta} (z_3^-,\un{x}) \, U_{\un{x}}^{b'b} [z_3^-,z_2^-] \, \delta^2 (\un{z}_2 - \un{x}) \\ 
& \times \left\{ \delta_{\chi, \chi'} \left[ \gamma^+ \, \cev{\underline{\mathscr{D}}}_{z_2}^{bc} \cdot  \underline{\mathscr{D}}_{z_2}^{ca} + \gamma^+ \gamma^5 \, 2 g \,  \left( {\cal F}^{12} (z_2) \right)^{ba} \right] - \delta_{\chi, -\chi'} \left[ \gamma^+ \gamma^5 \, \cev{\underline{\mathscr{D}}}_{z_2}^{bc} \cdot  \underline{\mathscr{D}}_{z_2}^{ca} + \gamma^+ \,  2 g \,  \left( {\cal F}^{12} (z_2) \right)^{ba} \right]  \right\}_{\alpha \beta}  \notag \\ 
& \times U_{\un{y}}^{aa'} [z_2^-, z_1^-] \, \delta^2 (\un{z}_2 - \un{y}) \, \bar{\psi}_\alpha (z_1^-,\un{y}) \, t^{a'} \, V_{\un{y}} [ z_1^-, -\infty] \notag \\ 
& \equiv \int\limits_{-\infty}^{\infty} \dd{z}_1^- \int\limits_{z_1^-}^\infty d z_2^- \int\limits_{z_2^-}^\infty d z_3^-  d^2 z_2 V_{\un{x}} [ \infty, z_3^-] \, \delta^2 (\un{z}_2 - \un{x}) \mathcal{O}^{\textrm{pol qG$\overline{\textrm{q}}$}}_{\chi',\chi} (z_1^-,z_2^-,z_3^-; \un{x},\un{z}_2, \un{y}) \delta^2 (\un{z}_2 - \un{y})  V_{\un{y}} [ z_1^-, -\infty] , \notag
\end{align}
where in the last line we have defined the operator $\mathcal{O}^{\textrm{pol qG$\overline{\textrm{q}}$}}_{\chi',\chi}$ for the future use.

%%%%%%%%%%%%%%%%%%%%%%%%%%%%%%%%%%%%%%%%%%%%%%%%%%%%%%%%%%%%%%%%%%%%%%%%%%%%%%%%%%%%%%%

\subsection{Double quark insertion operator}

%%%%%%%%%%%%%%%%%%%%%%%%%%%%%%%%%%%%%%%%%%%%%%%%%%%%%%%%%%%%%%%%%%%%%%%%%%%%%%%%%%%%%%%
\begin{figure}[ht]
\centering
\includegraphics[width=0.7\linewidth]{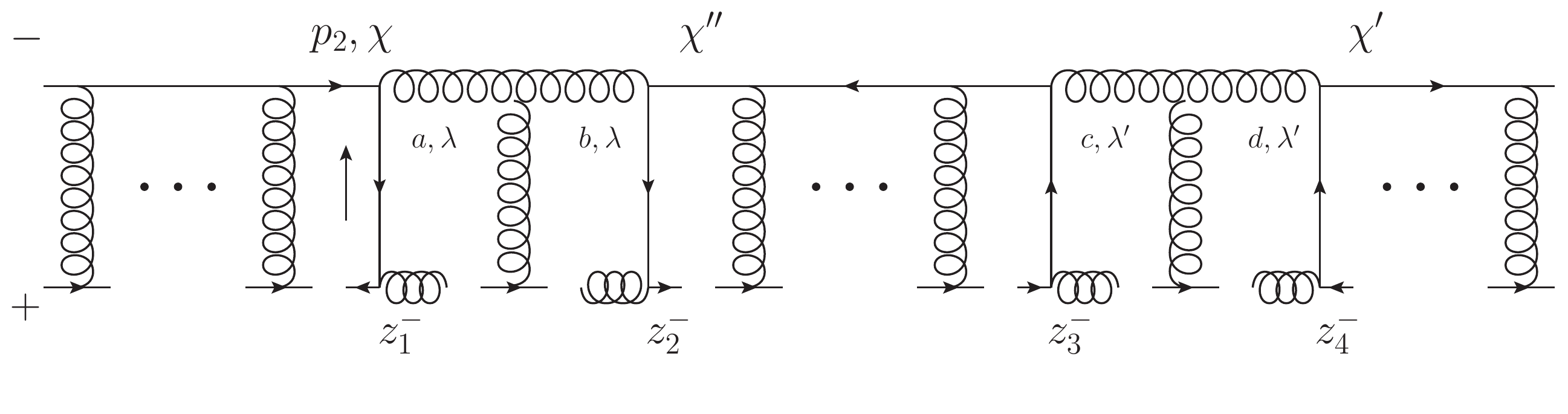}
\caption{The diagram for a double quark exchange following the double anti-quark exchange, with two quark field operators or two anti-quark field operators inserted instead of a quark and anti-quark field operator pair from \fig{FIG:opoldiagquark} inserted twice.}
\label{FIG:qqqbarqbar}
\end{figure}
%%%%%%%%%%%%%%%%%%%%%%%%%%%%%%%%%%%%%%%%%%%%%%%%%%%%%%%%%%%%%%%%%%%%%%%%%%%%%%%%%%%%%%%

The second operator we need to add, shown in \fig{FIG:qqqbarqbar}, is similar to a double insertion of the sub-eikonal part of the quark--anti-quark exchange operator in \eq{Oq3}, as given by the second line of \eq{polwilgen5}, but with the quark and anti-quark fields interchanged in the middle. That is, instead of two quark--anti-quark field operator pairs $(\bar{\psi} \psi)^2$, in \fig{FIG:qqqbarqbar} we have a pair of quark field operators inserted after a pair of anti-quark field operators, $\psi \psi \bar{\psi} \bar{\psi}$, with the appropriate fundamental and adjoint light-cone Wilson lines inserted between the operators. Taking only the sub-eikonal contributions from each pair of (anti-)quark exchanges, a similar calculation to that which yielded \eq{Oq3} now gives
\begin{align}
\label{Vqqqbarqbar}
&V_{\un{x},\un{y};\chi',\chi}^{\textrm{pol \, qq}\bar{\textrm{q}}\bar{\textrm{q}}} = - \frac{g^4 (p_1^+)^2}{64 s^2} \int\limits_{-\infty}^{\infty} \dd{z}_1^- \int\limits_{z_1^-}^{\infty} \dd{z}_2^- \int\limits_{z_2^-}^\infty \dd{z}_3^-  \int\limits_{z_3^-}^{\infty} \dd{z}_4^-   V_{\un{x}} [ \infty, z_4^-] t^d \psi_{\delta} (z_4^-, \un{x}) U^{dc}_{\un{x}} [z_4^-,z_3^-] \bar{\psi}_{\beta} (z_2^-, \un{x}) t^b V^{\dagger}_{\un{x}} [z_3^-,z_2^-] \\
& \times \Big\{ \delta_{\chi,\chi'} \Big[ \left( \gamma^+ \right)_{\alpha \delta}  \left( \gamma^+ \right)_{\beta\gamma} - \left(\gamma^+ \, \gamma^5 \right)_{\alpha\delta} \left(\gamma^+ \, \gamma^5 \right)_{\beta\gamma} \Big] + \delta_{\chi,-\chi'} \Big[ \left( \gamma^+ \right)_{\alpha \delta}  \left(\gamma^+ \, \gamma^5 \right)_{\beta\gamma}  - \left(\gamma^+ \, \gamma^5 \right)_{\alpha\delta} \left( \gamma^+ \right)_{\beta\gamma} \Big]   \Big\} \notag \\
& \times   t^c \psi_{\gamma} (z_3^-, \un{x}) U^{ba}_{\un{x}} [z_2^-,z_1^-] \bar{\psi}_{\alpha} (z_1^-, \un{x}) t^a V_{\un{y}} [z_1^-,-\infty] \delta^2 (\un{x}-\un{y}) \notag \\
& \equiv \int\limits_{-\infty}^{\infty} \dd{z}_1^- \int\limits_{z_1^-}^{\infty} \dd{z}_2^- \int\limits_{z_2^-}^\infty \dd{z}_3^-  \int\limits_{z_3^-}^{\infty} \dd{z}_4^-   V_{\un{x}} [ \infty, z_4^-] \,  \mathcal{O}^{\textrm{pol qq}\bar{\textrm{q}}\bar{\textrm{q}}}_{\chi', \chi} (z_4^-, z_3^-,z_2^-,z_1^-;  \un{x})  \,   V_{\un{y}} [ z_1^-, -\infty] \delta^2 (\un{x}-\un{y}) . \notag 
\end{align}
Again, in the last line of \eq{Vqqqbarqbar} we defined the new operator $\mathcal{O}^{\textrm{pol qq}\bar{\textrm{q}}\bar{\textrm{q}}}_{\chi', \chi}$, for compactness of the future notation.

\subsection{Full sub-sub-eikonal fundamental polarized Wilson line}

We can now assemble all the pieces of the quark $S$-matrix together to construct the full fundamental polarized Wilson line to the sub-sub-eikonal order. We need to add Eqs.~\eqref{polwilgen2}, \eqref{polwilgen5}, \eqref{VqGq2}, and \eqref{Vqqqbarqbar} along with the ``interference terms" for the insertions of the $\mathcal{O}^{\textrm{pol G}}_{\chi', \chi}$ and $\mathcal{O}^{\textrm{pol q}\overline{\textrm{q}}}_{\chi', \chi}$ operators. We arrive at
\begin{align}\label{Full}
& V_{\un{x}, \un{y}; \chi', \chi} =  V_{\un{x}} \, \delta^2 (\un{x} - \un{y}) \, \delta_{\chi,\chi'}  + \int\limits_{-\infty}^{\infty} \dd{z}^- d^2 z \ V_{\un{x}} [ \infty, z^-] \, \delta^2 (\un{x} - \un{z}) \, \mathcal{O}_{\chi', \chi}^{\textrm{pol G}} (z^-, \un{z}) \, V_{\un{y}} [ z^-, -\infty] \, \delta^2 (\un{y} - \un{z}) \\ 
& + \int\limits_{-\infty}^{\infty} \dd{z}_1^- d^2 z_1 \int\limits_{z_1^-}^{\infty} \dd{z}_2^- d^2 z_2 \sum_{\chi'' = \pm 1} \ V_{\un{x}} [ \infty, z_2^-] \, \delta^2 (\un{x} - \un{z}_2) \, \mathcal{O}_{\chi', \chi''}^{\textrm{pol G}} (z_2^-, \un{z}_2) \, V_{\un{z}_1} [z_2^-,  z_1^-] \, \delta^2 (\un{z}_2 - \un{z}_1) \notag \\ 
& \times \, \mathcal{O}_{\chi'', \chi}^{\textrm{pol G}} (z_1^-, \un{z}_1) \,   V_{\un{y}} [ z_1^-, -\infty] \, \delta^2 (\un{y} - \un{z}_1) + \int\limits_{-\infty}^{\infty} \dd{z}_1^- \int\limits_{z_1^-}^\infty d z_2^-  \ V_{\un{x}} [ \infty, z_2^-] \,  \mathcal{O}^{\textrm{pol q}\overline{\textrm{q}}}_{\chi', \chi} (z_2^-, z_1^-;  \un{x}, \un{y})  \, V_{\un{y}} [ z_1^-, -\infty] \notag  \\ 
& + \int\limits_{-\infty}^{\infty} \dd{z}_1^- \int\limits_{z_1^-}^{\infty} \dd{z}_2^- \int\limits_{z_2^-}^\infty \dd{z}_3^-  \int\limits_{z_3^-}^{\infty} \dd{z}_4^-  d^2 z \sum_{\chi'' = \pm 1}  V_{\un{x}} [ \infty, z_4^-] \,  \mathcal{O}^{\textrm{pol q}\overline{\textrm{q}}}_{\chi', \chi''} (z_4^-, z_3^-;  \un{x}, \un{z})  \, V_{\un{z}} [z_3^-,  z_2^-] \, \mathcal{O}^{\textrm{pol q}\overline{\textrm{q}}}_{\chi'', \chi} (z_2^-, z_1^-;  \un{z}, \un{y})  \,   V_{\un{y}} [ z_1^-, -\infty]  \notag \\
& +  \int\limits_{-\infty}^{\infty} \dd{z}_1^- \int\limits_{z_1^-}^{\infty} \dd{z}_2^- \int\limits_{z_2^-}^\infty \dd{z}_3^-  \int\limits_{z_3^-}^{\infty} \dd{z}_4^-   V_{\un{x}} [ \infty, z_4^-] \,  \mathcal{O}^{\textrm{pol qq}\bar{\textrm{q}}\bar{\textrm{q}}}_{\chi', \chi} (z_4^-, z_3^-,z_2^-,z_1^-;  \un{x})  \,   V_{\un{y}} [ z_1^-, -\infty] \delta^2 (\un{x}-\un{y})  \notag \\
&+  \int\limits_{-\infty}^{\infty} \dd{z}_1^- \int\limits_{z_1^-}^\infty d z_2^-  \int\limits_{z_2^-}^\infty d z_3^-  d^2 z \ V_{\un{x}} [ \infty, z_3^-] \, \delta^2 (\un{z} - \un{x}) \mathcal{O}^{\textrm{pol qG$\overline{\textrm{q}}$}}_{\chi',\chi} (z_1^-,z_2^-,z_3^-; \un{x},\un{z}, \un{y}) \delta^2 (\un{z} - \un{y})  V_{\un{y}} [ z_1^-, -\infty] \notag \\ 
& + \int\limits_{-\infty}^{\infty} \dd{z}_1^- \int\limits_{z_1^-}^{\infty} \dd{z}_2^- \int\limits_{z_2^-}^\infty \dd{z}_3^- d^2 z  \sum_{\chi'' = \pm 1} \ V_{\un{x}} [ \infty, z_3^-] \, \delta^2 (\un{x} - \un{z}) \, \mathcal{O}^{\textrm{pol G}}_{\chi', \chi''} (z_3^-; \un{z})  \, V_{\un{z}} [z_3^-,  z_2^-] \, \mathcal{O}^{\textrm{pol q}\overline{\textrm{q}}}_{\chi'', \chi} (z_2^-, z_1^-;  \un{z}, \un{y})  \,   V_{\un{y}} [ z_1^-, -\infty] \notag \\
& + \int\limits_{-\infty}^{\infty} \dd{z}_1^-  \int\limits_{z_1^-}^{\infty} \dd{z}_2^- \int\limits_{z_2^-}^\infty \dd{z}_3^- d^2 z  \sum_{\chi'' = \pm 1} \ V_{\un{x}} [ \infty, z_3^-] \,  \mathcal{O}^{\textrm{pol q}\overline{\textrm{q}}}_{\chi', \chi''} (z_3^-, z_2^-;  \un{x}, \un{z}) \, V_{\un{z}} [z_2^-,  z_1^-] \, \mathcal{O}^{\textrm{pol G}}_{\chi'', \chi} (z_1^-;  \un{z})  \,   V_{\un{y}} [ z_1^-, -\infty] \, \delta^2 (\un{y} - \un{z}) \notag 
\end{align}
with the operators $\mathcal{O}^{\textrm{pol G}}_{\chi', \chi}$ and $\mathcal{O}^{\textrm{pol q}\overline{\textrm{q}}}_{\chi', \chi}$ given by equations \eqref{O24} and \eqref{Oq3}, respectively. Note again, that the corresponding expression in the helicity basis can be constructed with the help of Table~\ref{table:conversion}. 

Equation \eqref{Full} is the main formal result of this work. It gives the quark $S$-matrix for the scattering on an arbitrary background quark and gluon fields up to and including the sub-sub-eikonal order. This result can be used to construct small-$x$ asymptotics of all leading-twist quark TMDs, and, probably, of some higher-twist quark distributions as well. Below we will apply (a part of) it to find the small-$x$ asymptotics of the quark Sivers function.

%%%%%%%%%%%%%%%%%%%%%%%%%%%%%%%%%%%%%%%%%%%%%%%%%%%%%%%%%%%%%%%%%%%%%%%%%%%%%%%%%%%

\section{Quark Sivers Function at the Eikonal Level}
\label{sec:sivop}

%%%%%%%%%%%%%%%%%%%%%%%%%%%%%%%%%%%%%%%%%%%%%%%%%%%%%%%%%%%%%%%%%%%%%%%%%%%%%%%%%%%

\subsection{Quark Sivers Function at Small $x$: a General Expression}
\label{sec:gen_exp}

%%%%%%%%%%%%%%%%%%%%%%%%%%%%%%%%%%%%%%%%%%%%%%%%%%%%%%%%%%%%%%%%%%%%%%%%%%%%%%%%%%%
\begin{figure}[ht]
\captionsetup[subfigure]{justification=centering}
\centering
\begin{subfigure}[b]{.27\linewidth}
\includegraphics[width=\linewidth]{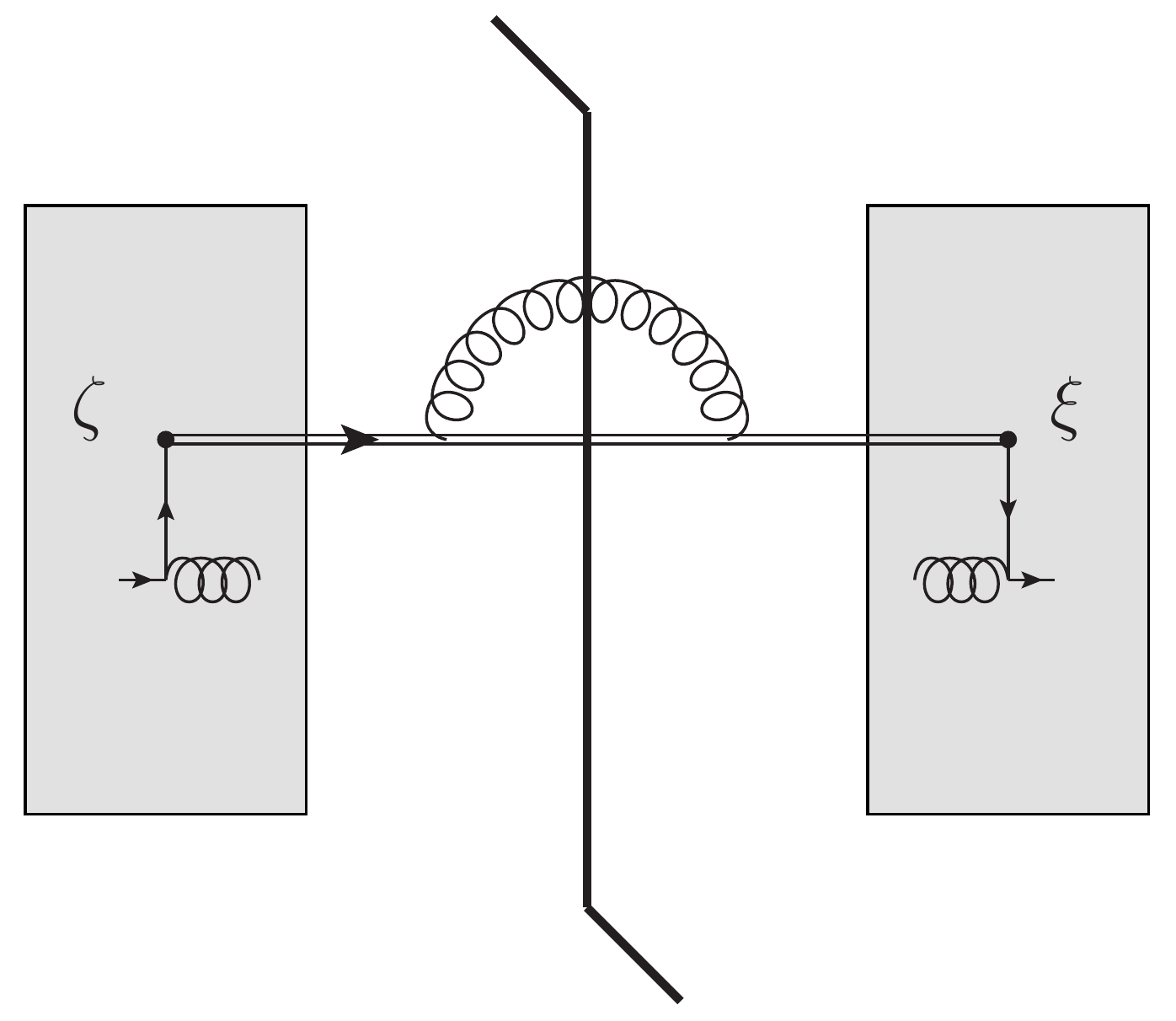}
\caption*{A}\label{FIG:allclassa} 
\end{subfigure}
~
\begin{subfigure}[b]{.31\linewidth}
\includegraphics[width=\linewidth]{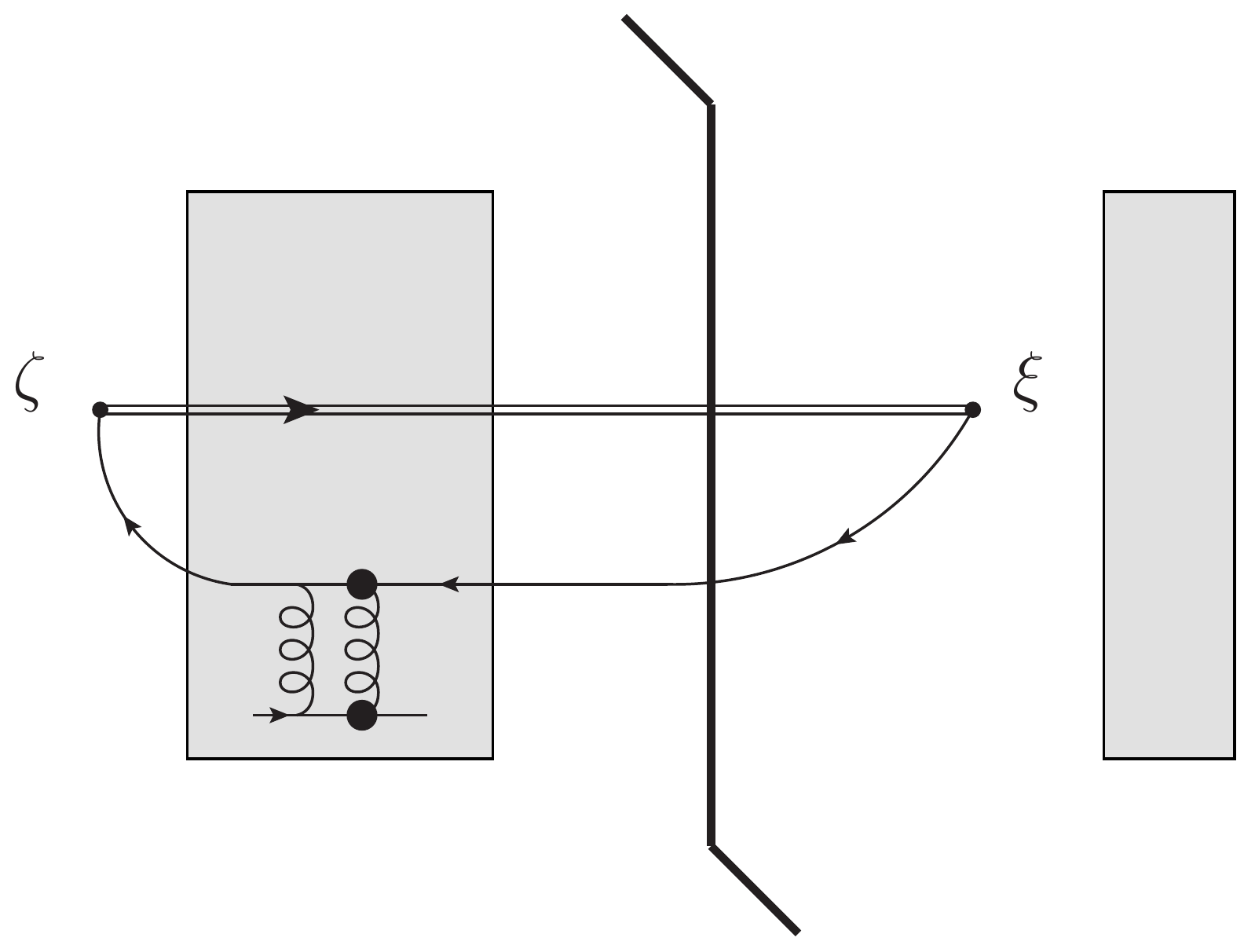}
\caption*{B}
\end{subfigure}
\label{FIG:allclassb} 
~
\begin{subfigure}[b]{.31\linewidth}
\includegraphics[width=\linewidth]{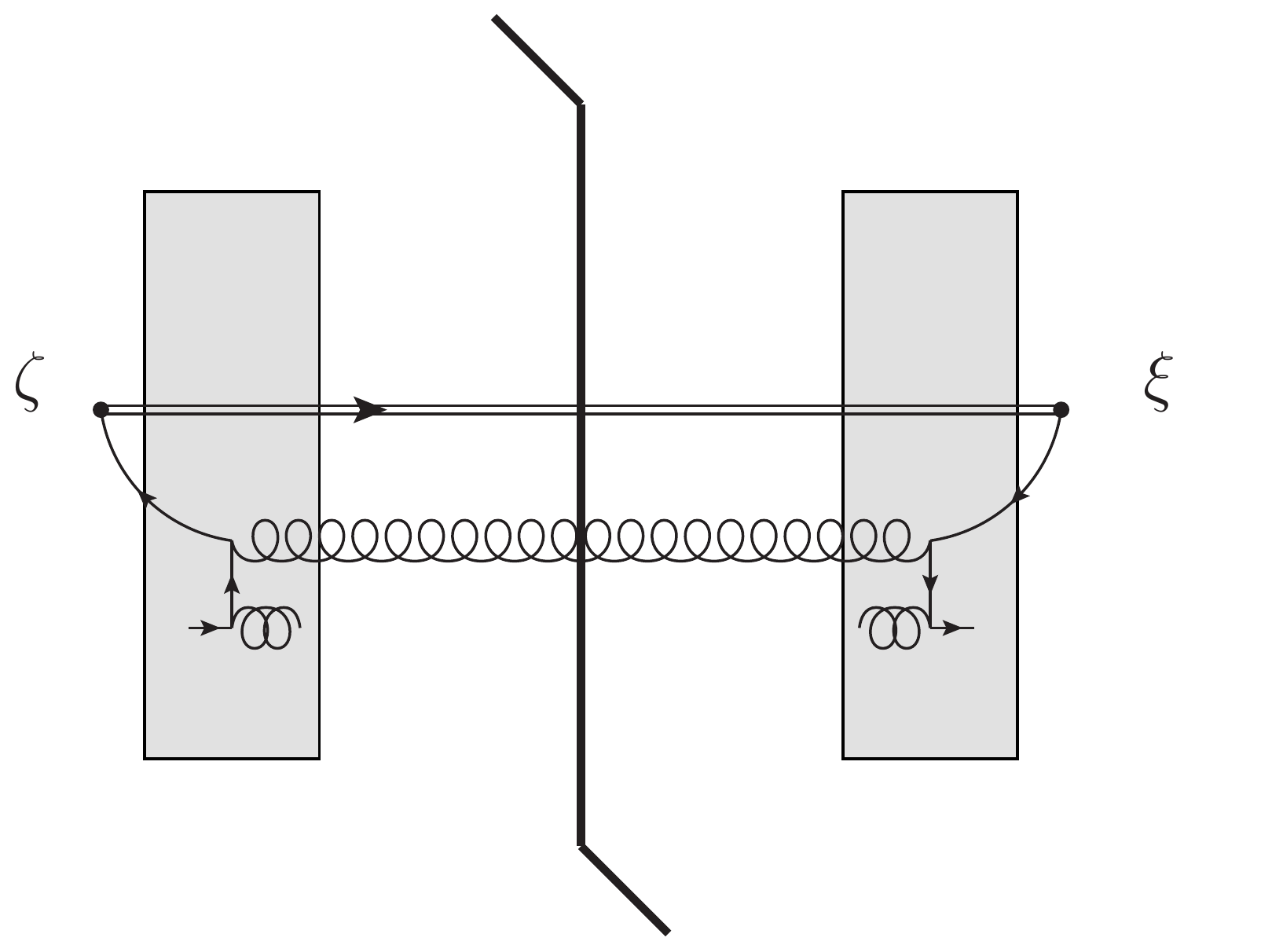}
\caption*{C}\label{FIG:allclassc} 
\end{subfigure}

\begin{subfigure}[b]{.29\linewidth}
\includegraphics[width=\linewidth]{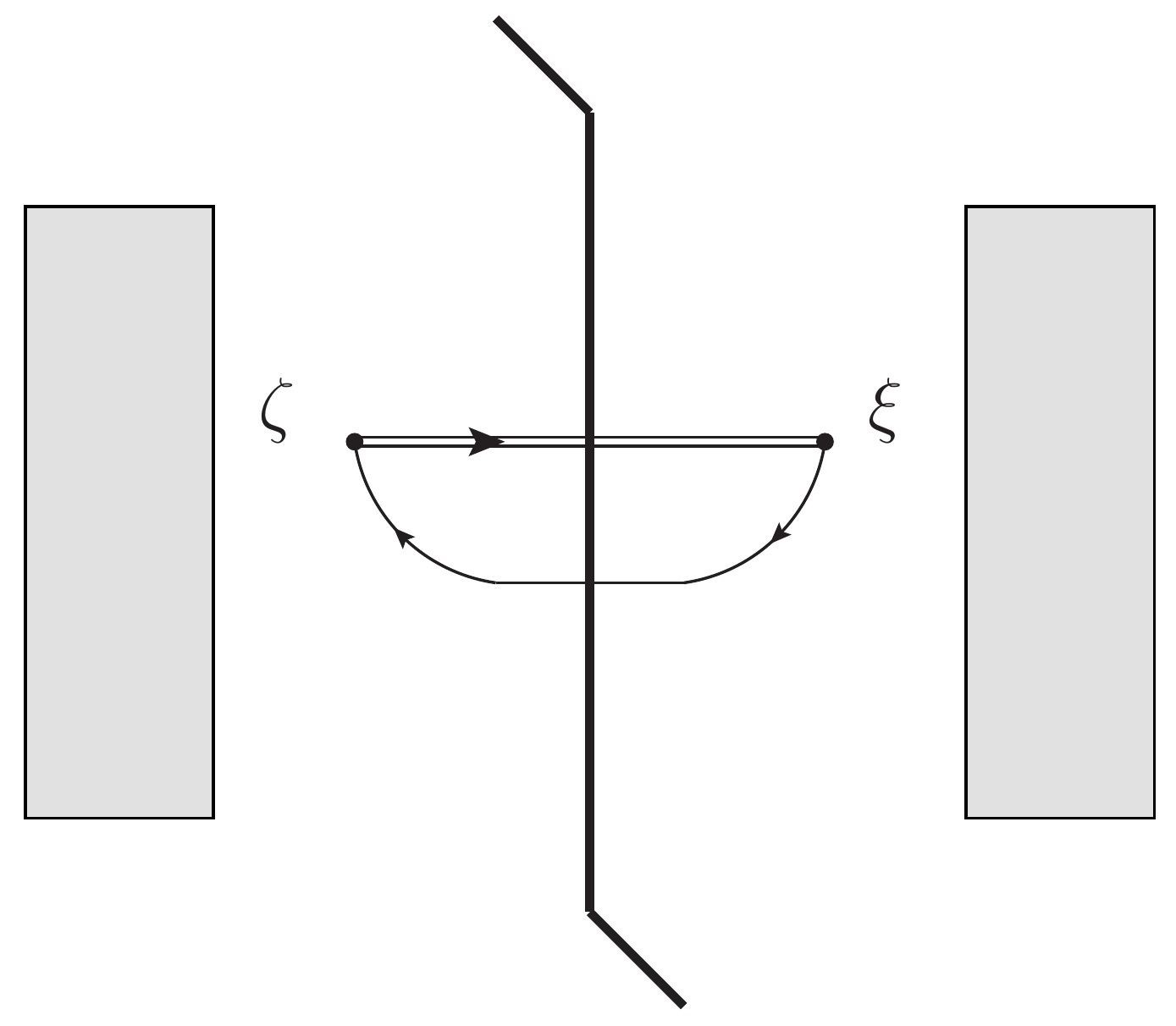}
\caption*{D}\label{FIG:allclassd} 
\end{subfigure}
~
\begin{subfigure}[b]{.33\linewidth}
\includegraphics[width=\linewidth]{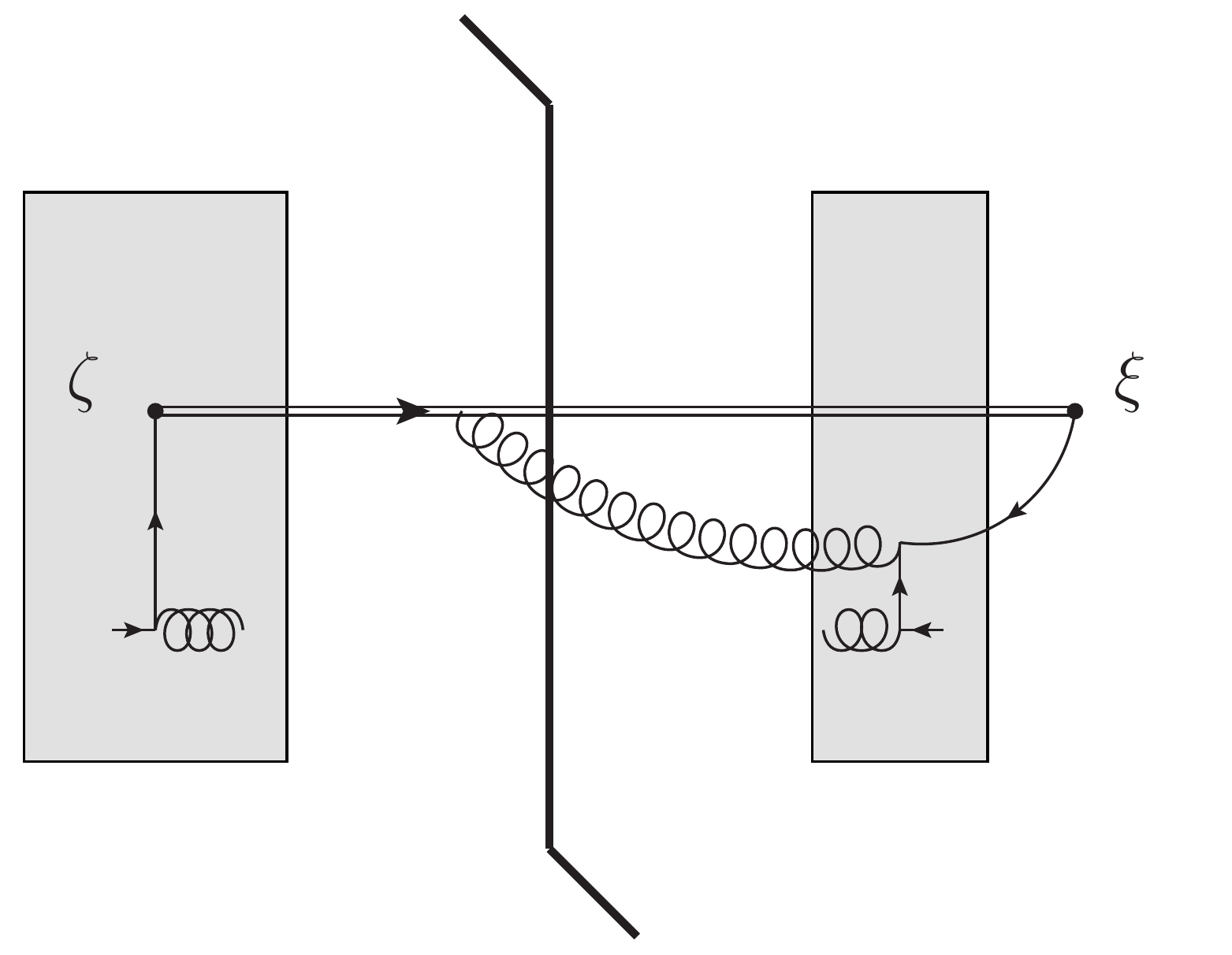}
\caption*{E}\label{FIG:allclasse} 
\end{subfigure}
~
\begin{subfigure}[b]{.30\linewidth}
\includegraphics[width=\linewidth]{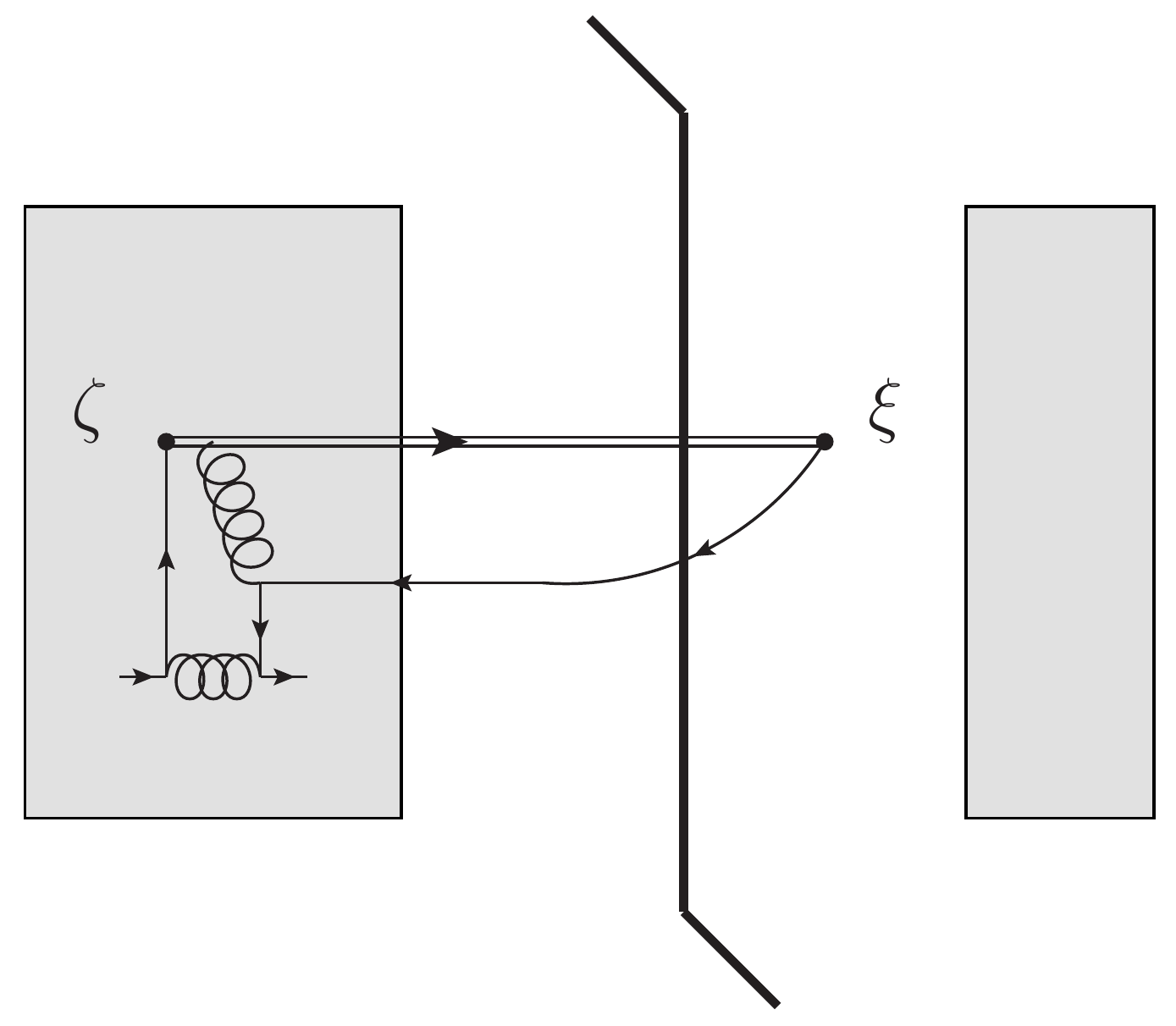}
\caption*{F}\label{FIG:allclassf} 
\end{subfigure}
\caption{Diagram classes of \cite{Kovchegov:2018znm} contributing to a calculation of a quark TMD. The double lines represent the fundamental Wilson lines, the shaded rectangles represent the proton shockwave, the black circles in B represent the sub-eikonal scattering which brings in transverse spin dependence, and the thick vertical line represents the final state cut. }
\label{FIG:allclasses}
\end{figure}
%%%%%%%%%%%%%%%%%%%%%%%%%%%%%%%%%%%%%%%%%%%%%%%%%%%%%%%%%%%%%%%%%%%%%%%%%%%%%%%%%%%

Having constructed the full polarized Wilson line operator, we can now investigate the small-$x$ evolution of the quark Sivers function. The operator definition for the unpolarized quark TMDs for quarks with longitudinal momentum fraction $x$ and transverse momentum $\un{k}$ is (as in \cite{Meissner:2007rx})
\begin{align}\label{S1}
f_1^q (x,k_T^2) - \frac{\un{k} \cross \underline{S}_P}{M_P} f_{1 \: T}^{\perp \: q} (x,k_T^2) = \int \frac{\dd{r^-}\dd[2]{r_{\perp}}}{2 \, (2 \pi)^3} e^{i k \vdot r} \langle P, S | \bar{\psi}(0) \mathcal{U}[0,r] \frac{\gamma^+}{2} \psi(r) | P,S \rangle ,
\end{align}
with $f_1^q$ the unintegrated quark distribution, $f_{1 \, T}^{\perp \,q}$ the quark Sivers function, $M_P$ the proton mass, $\un{S}_P$ the proton spin, $k_T = |\un{k}|$, and $\mathcal{U}[0,r]$ the `staple' gauge link, which we will take here to be the future pointing link used in SIDIS. We work in $A^- = 0$ light cone gauge of the projectile so that the gauge link is just a product of two light-cone Wilson lines, $\mathcal{U}[0,r] = V_{\underline{0}}[0,\infty] V_{\underline{r}} [\infty,r^-]$. 

We employ the standard definition of the quark Sivers TMD in \eq{S1}. Alternatively, we could calculate the corresponding spin asymmetry $A_{UT}^\textrm{Sivers}$  in the SIDIS process at small-$x$ and extract the Sivers TMD by matching that asymmetry onto the TMD factorization expression. In the case of helicity TMDs, the matching on SIDIS was done in \cite{Kovchegov:2015pbl}, leading to the result equivalent to that obtained later in \cite{Kovchegov:2018znm} by taking the standard TMD definition and simplifying it at small $x$. Since the steps in the calculations performed in \cite{Kovchegov:2015pbl} and \cite{Kovchegov:2018znm} are rather general (see also \cite{Kovchegov:2015zha}), here we assume that they apply to all TMDs, including the Sivers function, such that the TMD defined in \eq{S1} would appear in the physical processes, such as SIDIS, even at small $x$.

Using the saturation/CGC formalism, we can rewrite the operator using semi-classical averaging in the proton's wave function as (see \cite{Kovchegov:2018znm} and Appendix~A of \cite{Kovchegov:2019rrz})
\begin{align}\label{S2}
f_1^q (x,k_T^2) - \frac{\un{k} \cross \underline{S}_P}{M_P} f_{1 \: T}^{\perp \: q} (x,k_T^2) = \frac{2 p_1^+}{2 (2 \pi)^3} \sum_X \int \dd{\xi^-} \dd[2]{\xi_{\perp}} \dd{\zeta^-} \dd[2]{\zeta_{\perp}} e^{i k \vdot (\zeta - \xi)}  \Big[\frac{\gamma^+}{2} \Big]_{\alpha \beta} \notag \\
\times  \Big{\langle} \bar{\psi}_{\alpha} (\xi) V_{\underline{\xi}} [\xi^-,\infty] | X \rangle \langle X | V_{\underline{\zeta}} [\infty, \zeta^-] \psi_{\beta} (\zeta) \Big{\rangle} ,
\end{align} 
where we sum over a complete set of partonic states $\ket{X}$, the large angle brackets denote averaging over the proton's wave function \cite{McLerran:1993ni,McLerran:1993ka,McLerran:1994vd,Kovchegov:1996ty,Balitsky:1997mk,Balitsky:1998ya,Kovchegov:2019rrz}, and we have semi-infinite Wilson lines connecting the positions of the quark field operators to infinity. 

We can take the sum over the intermediate states to be a final state cut, and represent the possible contributions diagrammatically in \fig{FIG:allclasses} as in \cite{Kovchegov:2018znm} at the lowest perturbative order (order-$\as$). The contributions are organized by whether $\zeta^-$ and $\xi^-$ are positive, negative, or zero, with $x^- =0$ being the position of the shock wave (assumed to be very thin in the $x^-$ direction). For the left-right asymmetric diagrams B, E, and F, their mirror images need to be added: those are not shown in \fig{FIG:allclasses}. Each diagram class is represented only by one diagram in \fig{FIG:allclasses}. Note that the diagrams in class C also include interactions with the gluons coming from the shock wave and a long-lived anti-quark produced in the final state (see \fig{FIG:diagBCdet} below): such diagrams are also not shown for brevity in \fig{FIG:allclasses}. 

At the eikonal order, only the diagrams B, C and D contribute both to the unpolarized quark TMD and to the Sivers function (see e.g. the unpolarized quark TMD $f_1^q (x,k_T^2)$ calculations in \cite{Mueller:1999wm,Kovchegov:2015zha}). The diagram D does not include any interaction with the target: hence it does not depend on the spin of the target, and cannot contribute to the Sivers function. We are left with the diagrams B and C at the eikonal level, which are shown in \fig{FIG:diagBCdet} in more detail below.

At the sub-eikonal order, the contributing diagrams for helicity TMD was found in \cite{Kovchegov:2018znm} to be the class B diagrams, which is shown in more detail in \fig{FIG:diagbdet} below. At that order the diagrams of classes C, D, and F do not contribute to the TMDs dependent on the target proton polarizations, such as helicity, transversity, and the Sivers function at hand, as argued in \cite{Kovchegov:2018znm}: for class C the sub-eikonal target interactions with the anti-quark line cancel when the quark (or gluon) exchanges are taken across the final state cut, the diagram D has no spin-dependent contributions since it contains no interaction with the target shockwave, and the diagrams in class F are further energy suppressed since the gluon emission and absorption take place inside the shockwave. The usual leading-order diagram calculation for the Sivers function \cite{Brodsky:2002cx,Brodsky:2002rv,Ji:2002aa,Bacchetta:2003rz,Meissner:2007rx,Brodsky:2013oya} comes from diagrams of class A, taking the interference between a re-scattering inside the shockwave and a purely tree level scattering to generate the phase needed to make the transverse spin dependence real. The resulting lowest-order Sivers function is sub-sub-eikonal, $f_{1 \: T}^{\perp \: q} (x,k_T^2) \sim x$ \cite{Meissner:2007rx}, and will not be considered in our calculations below, which are done at the eikonal and sub-eikonal orders. Such a lowest-order diagram can be evolved by adding soft gluon emissions to both class A diagrams and class E diagrams, as depicted in \fig{FIG:allclasses}.  However, in Appendix~A of \cite{Kovchegov:2018znm} it was shown that the contributions of graphs A and E to the double logarithmic approximation (DLA) evolution cancel at the sub-eikonal order. (The double logarithmic approximation is defined as the re-summation of the parameter $\as \, \ln^2 (1/x)$ at small $x$.) We conclude that the sub-sub-eikonal $x$-dependence of the lowest-order quark Sivers function, $f_{1 \: T}^{\perp \: q} (x,k_T^2) \sim x$, is unaffected by the evolution in the DLA, and persists down to small $x$ for the contributions to the Sivers function coming from the diagrams A and E. However, such a sub-sub-eikonal contribution is much smaller than the eikonal and sub-eikonal contributions of interest to us here and will not be considered further. 

To summarize, in the eikonal calculations of the quark Sivers function we only need to consider diagrams B and C, while for the sub-eikonal calculations we only need the diagram B. We begin with the eikonal calculations of the diagrams B and C, depicted in \fig{FIG:diagBCdet} in more detail.

%%%%%%%%%%%%%%%%%%%%%%%%%%%%%%%%%%%%%%%%%%%%%%%%%%%%%%%%%%%%%%%%%%%%%%%%%%%%%%%%%%%%
\begin{figure}[ht]
\centering
\includegraphics[width=0.9\linewidth]{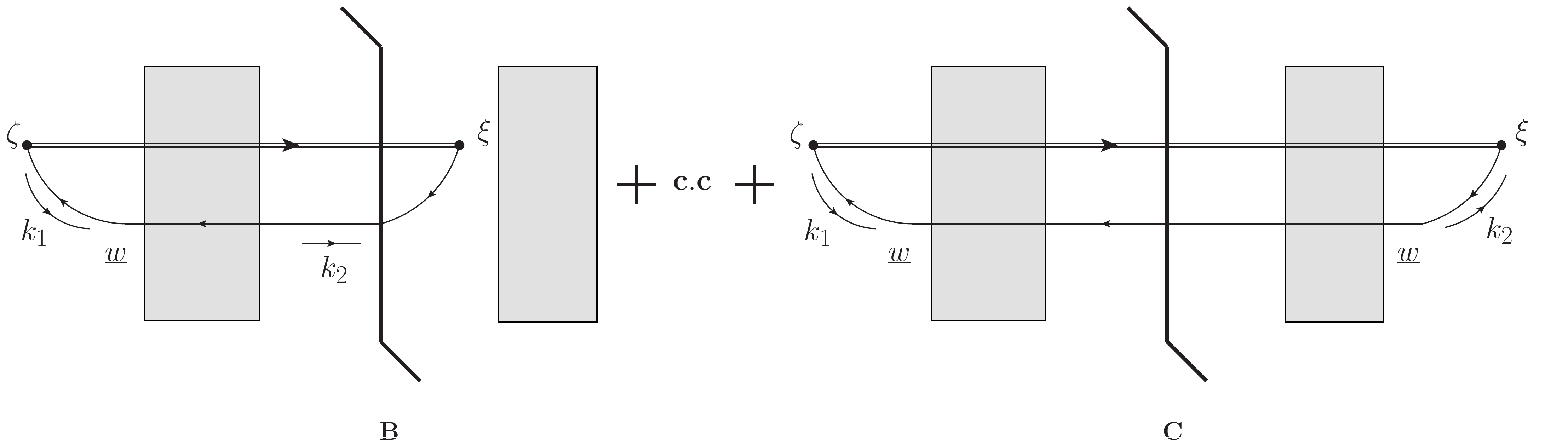}  
\caption{Diagrams of classes B and C with kinematics specified. In the diagram B the antiquark propagates from $\zeta$ with momentum $k_1$, undergoes a transverse spin-dependent interaction with the proton target at the transverse position $\un{w}$, then propagates to $\xi$ with momentum $k_2$. In the diagram C the anti-quark interacts with the shock wave again to the right of the cut, before arriving at the point $\xi$ with momentum $k_2$.}
\label{FIG:diagBCdet}
\end{figure}
%%%%%%%%%%%%%%%%%%%%%%%%%%%%%%%%%%%%%%%%%%%%%%%%%%%%%%%%%%%%%%%%%%%%%%%%%%%%%%%%%%%%

We begin with the class B diagram in \fig{FIG:diagBCdet}, where the light-cone Wilson line crosses the shockwave from $\zeta$ to the final state to the left of the cut, but connects $\xi$ to the final state on the right side of the cut without crossing the shock wave, and is, hence, an identity on the right of the cut. An antiquark with momentum $k_1$ propagates from $\zeta$ through the shockwave, undergoing an interaction at transverse position $\underline{w}$, then propagating with momentum $k_2$ to the final state and connecting on the other side of the cut to $\xi$. Taking the shockwave to be at light cone time $x^- = 0$, the class B diagrams have $\zeta^- < 0$, $\xi^- > 0$ (as in \fig{FIG:diagBCdet}) or $\zeta^- > 0$, $\xi^- < 0$ (in the conjugate diagrams of those like B from \fig{FIG:diagBCdet}). For the B-class diagrams the quark field operator at $\zeta$ is before the shockwave and the anti-quark operator at $\xi$ is after the shockwave on the other side of the cut, allowing us to write the scattering as a dipole composed of a fully infinite Wilson lines connecting the quark field operators at $\xi$ and at $\zeta$. To evaluate the diagram B we will need the quark propagator through the shock wave \cite{Kovchegov:2018znm},
\begin{align}\label{quark_propagator}
\contraction
{}
{\bar\psi^i_\alpha}
{(\eta) \:}
{\psi^j_\beta}
\bar\psi^i_\alpha (\xi) \: \psi^j_\beta (\zeta)
&= \int  d^2 w \, \frac{d^2 k_1 \, d k_1^-}{(2\pi)^3} \, \frac{d^2 k_2}{(2\pi)^2} \, e^{i \frac{{\un k}_1^2}{2 k_1^-} \zeta^- - i \frac{{\un k}_2^2}{2 k_1^-} \xi^- + i \un{k}_1 \cdot (\un{w} - \un{\zeta}) + i \un{k}_2 \cdot (\un{\xi} - \un{w})} \, \theta (k_1^-)
\notag \\ & \hspace{2cm} \times
\left\{ \left[ \frac{\slashed{k_1}}{2 k_1^-} \right]
\left[ \left( \hat{V}_{{\un w}}^\dagger \right)^{ji} \right]
\left[ \frac{ \slashed{k_2}}{2 k_1^-} \right] \right\}_{\beta \alpha} \Bigg|_{k_2^- = k_1^-, k_1^2 =0, k_2^2 =0} ,
\end{align}
where the interaction of the anti-quark with the shock wave is denoted by the Dirac and color matrix $\left( \hat{V}_{{\un w}}^\dagger \right)^{ji}$. Here $\alpha, \beta$ are the Dirac spinor indices, while $i,j$ are the quark color indices. For simplicity we assume that the anti-quark is massless, $m=0$, since the light quark mass is usually negligible in the Sivers function calculations. 

Using the propagator \eqref{quark_propagator} in \eq{S2}, putting $V_{\underline{\xi}} [\xi^-,\infty] =1$, replacing $V_{\underline{\zeta}} [\infty, \zeta^-] \to V_{\underline{\zeta}} [\infty, -\infty] = V_{\underline{\zeta}}$, integrating over $\un \xi$, $\un{k}_2$, $\zeta^-$ and $\xi^-$, and inserting spinor polarization sums \cite{Kovchegov:2018znm}, we see that the diagram B and its complex conjugate give 
\begin{align}
\label{sivwsum}
B + \mbox{c.c.} & \, = \frac{2 p_1^+}{2 (2 \pi)^3} \sum_{\bar{q}} \int\limits_{-\infty}^0 \dd{\zeta^-} \int\limits_0^{\infty} \dd{\xi^-} \int \dd[2]{\zeta_{\perp}}  \dd[2]{\xi_{\perp}} e^{i k \vdot (\zeta - \xi)} \Big[\frac{\gamma^+}{2} \Big]_{\alpha \beta} \Big{\langle} \bar{\psi}_{\alpha} (\xi) V_{\underline{\xi}} [\xi^-,\infty] | \bar{q} \rangle \langle \bar{q} | V_{\underline{\zeta}} [\infty, \zeta^-] \psi_{\beta} (\zeta) \Big{\rangle} + c.c. \notag \\
=& -\frac{2 p_1^+}{2 (2 \pi)^3} \int \dd[2]{\zeta_{\perp}}  \dd[2]{w_{\perp}} \frac{\dd[2]{k_{1 \perp}}\dd{k_1^-}}{(2\pi)^3} e^{i (\underline{k}_1 + \underline{k}) \vdot (\un{w} - \un{\zeta})} \theta (k_1^-) \, \frac{1}{(x p_1^+ k_1^- + \underline{k}_1^2 ) (x p_1^+ k_1^- + \underline{k}^2)}  \notag \\
\times & \sum_{\chi_1 , \chi_2} \bar{v}_{\chi_2} (k_2) \frac{\gamma^+}{2} v_{\chi_1}(k_1) \, \Big{\langle} \tord V_{\underline{\zeta}}^{ij} \, \bar{v}_{\chi_1} (k_1) \left( \hat{V}_{{\un w}}^\dagger \right)^{ji}  v_{\chi_2} (k_2) \Big{\rangle} \Bigg|_{k_2^- = k_1^-, k_1^2 =0, k_2^2 =0, \un{k}_2 = - \un{k}} + \mbox{c.c.},
\end{align}
where c.c. denotes the complex conjugate terms and the intermediate state $\ket{X}$ became an anti-quark state $\ket{\bar{q}}$ as only an anti-quark is produced in the final state. 

Using the $\pm$ reversed transverse Brodsky-Lepage spinors \eq{anti-BLspinors} we evaluate the matrix element for massless anti-quarks as
\begin{align}\label{vG+v}
\bar{v}_{\chi_2} (k_2) \frac{\gamma^+}{2} v_{\chi_1}(k_1) 
\approx \frac{1}{ \sqrt{k_1^- k_2^-}} \big[ \delta_{\chi_1,\chi_2}  \, \underline{k}_1 \vdot \underline{k}_2 - i  \, \delta_{\chi_1, -\chi_2} \, \underline{k}_2 \cross \underline{k}_1 \big] .
\end{align}
For the eikonal calculation at hand we only need the Wilson line contribution to the interaction of the anti-quark with the target. We, therefore, write
\begin{align}\label{vVv}
 \bar{v}_{\chi_1} (k_1)  \left( \hat{V}_{{\un w}}^\dagger \right) \: v_{\chi_2} (k_2) = 2 \sqrt{k_1^- k_2^- } \, \delta_{\chi_1,\chi_2}  \, V_{\underline{w}}^{\dagger} + \ldots
\end{align}
with the ellipsis denoting the sub-eikonal terms and beyond. Note that the Sivers function couples an unpolarized quark to the transverse polarization of the proton: hence we only need the unpolarized quark interaction with the shock wave. This means we will only need the $\delta_{\chi_1,\chi_2}$ term on the right of \eq{vVv}, even at the sub-eikonal order we consider below. 

Employing Eqs.~\eqref{vG+v} and \eqref{vVv} in \eq{sivwsum} and summing over transverse polarizations $\chi_1, \chi_2$ we arrive at
\begin{align}\label{Bcc2}
B + \mbox{c.c.}  =  & \, \frac{4 p_1^+}{(2 \pi)^3} \int \dd[2]{\zeta_{\perp}}  \dd[2]{w_{\perp}} \frac{\dd[2]{k_{1 \perp}}\dd{k_1^-}}{(2\pi)^3} e^{i (\underline{k}_1 + \underline{k}) \vdot (\un{w} - \un{\zeta})} \theta (k_1^-) \, \frac{\un{k} \cdot \un{k}_1}{(x p_1^+ k_1^- + \underline{k}_1^2 ) (x p_1^+ k_1^- + \underline{k}^2)}  \notag \\
 & \times \Big{\langle} \tord \tr \left[ V_{\underline{\zeta}} \, V_{{\un w}}^\dagger \right] + \atord \tr \left[ V_{\underline{\zeta}} \, V_{{\un w}}^\dagger \right] \Big{\rangle} ,
\end{align}
where we have explicitly added in the complex conjugate term. 

Diagram C from \fig{FIG:diagBCdet} can be calculated similarly, except all the interactions with the anti-quark cancel to the left and to the right of the cut, such that one needs to use the free anti-quark propagator, which can be obtained from \eq{quark_propagator} by using $\left( \hat{V}_{{\un w}}^\dagger \right)^{ji}  = \gamma^- \, \delta^{ij}$ in it. In the end one obtains
\begin{align}\label{C}
C =  \frac{4 p_1^+}{(2 \pi)^3} \int \dd[2]{\zeta_{\perp}}  \dd[2]{\xi_{\perp}} \frac{\dd[2]{k_{1 \perp}}\dd{k_1^-}}{(2\pi)^3} e^{i (\underline{k}_1 + \underline{k}) \vdot (\un{\xi} - \un{\zeta})} \theta (k_1^-) \, \frac{\un{k}_1^2}{(x p_1^+ k_1^- + \underline{k}_1^2 )^2}  \, \Big{\langle} \tr \left[ V_{\underline{\zeta}} \, V_{{\un \xi}}^\dagger \right]  \Big{\rangle} .
\end{align}

Combining Eqs.~\eqref{Bcc2} and \eqref{C}, replacing $\un{\xi} \to \un{w}$ in the latter, we arrive at
\begin{align}\label{ff1}
& \left[ f_1^q (x,k_T^2) -  \frac{\un{k} \cross \underline{S}_P}{M_P} f_{1 \: T}^{\perp \: q} (x,k_T^2) \right]_\textrm{eikonal} = \frac{4 p_1^+}{(2 \pi)^3} \int \dd[2]{\zeta_{\perp}}  \dd[2]{w_{\perp}} \frac{\dd[2]{k_{1 \perp}}\dd{k_1^-}}{(2\pi)^3} e^{i (\underline{k}_1 + \underline{k}) \vdot (\un{w} - \un{\zeta})} \theta (k_1^-) \\
& \times \left\{ \frac{\un{k} \cdot \un{k}_1}{(x p_1^+ k_1^- + \underline{k}_1^2 ) (x p_1^+ k_1^- + \underline{k}^2)} \,  \Big{\langle} \tord \tr \left[ V_{\underline{\zeta}} \, V_{{\un w}}^\dagger \right] + \atord \tr \left[ V_{\underline{\zeta}} \, V_{{\un w}}^\dagger \right] \Big{\rangle} + \frac{\un{k}_1^2}{(x p_1^+ k_1^- + \underline{k}_1^2 )^2}  \, \Big{\langle} \tord \tr \left[ V_{\underline{\zeta}} \, V_{{\un w}}^\dagger \right]  \Big{\rangle}  \right\}. \notag
\end{align}
We have also inserted a time-ordering sign into the last correlator, as justified in \cite{Kovchegov:2018znm} (see also \cite{Mueller:2012bn}). 

To extract the Sivers function, which is odd under time reversal, from \eq{ff1} we need the part of its right-hand side which changes sign under $\un{k} \to - \un{k}$. A quick inspection, along with taking $k_1 \rightarrow -k_1$, $\un{\zeta} \leftrightarrow \un{w}$ on the right-hand side of the expression, shows that the part of the correlators in \eq{ff1} contributing to the Sivers function is an odd function under $\un{\zeta} \leftrightarrow \un{w}$. One can write \cite{Hatta:2005as,Kovchegov:2003dm}
\begin{align}
\frac{1}{N_c} \Big{\langle}  \tord \tr [ V_{\underline{\zeta}}  V_{\underline{w}}^{\dagger }] \Big{\rangle} = \mathcal{S}_{\un{\zeta} \un{w}} + i \, \mathcal{O}_{\un{\zeta} \un{w}} ,
\end{align}
with $\mathcal{S}_{\un{\zeta} \un{w}}$ the $\un{\zeta} \leftrightarrow \un{w}$--symmetric, $\cal C$-even Pomeron exchange operator and $\mathcal{O}_{\un{\zeta} \un{w}}$ the $\un{\zeta} \leftrightarrow \un{w}$--antisymmetric, $\cal C$-odd odderon. The combination we have in \eq{ff1} is
\begin{align}
 \Big{\langle}  \tord \tr [ V_{\underline{\zeta}}  V_{\underline{w}}^{\dagger }]  + \atord \tr [V_{\underline{\zeta}}  V_{\underline{w}}^{\dagger } ] \Big{\rangle} = 2 N_c \left( \mathcal{S}_{\un{\zeta} \un{w}} + i \, \mathcal{O}_{\un{\zeta} \un{w}} \right).
\end{align}
The odderon amplitudes being antisymmetric under the $\un{\zeta} \leftrightarrow \un{w}$ exchange, $\mathcal{O}_{\un{\zeta} \un{w}} = - \mathcal{O}_{\un{w} \un{\zeta}}$, are the only ones contributing to the Sivers function. We thus write
\begin{align}\label{siv}
 - \frac{\un{k} \cross \underline{S}_P}{M_P} f_{1 \: T}^{\perp \: q} (x,k_T^2) \Big|_\textrm{eikonal} & = \frac{4 i \, N_c \, p_1^+}{(2 \pi)^3} \int \dd[2]{\zeta_{\perp}}  \dd[2]{w_{\perp}} \frac{\dd[2]{k_{1 \perp}}\dd{k_1^-}}{(2\pi)^3} e^{i (\underline{k}_1 + \underline{k}) \vdot (\un{w} - \un{\zeta})} \theta (k_1^-) \\
& \times \left[ \frac{2 \, \un{k} \cdot \un{k}_1}{(x p_1^+ k_1^- + \underline{k}_1^2 ) (x p_1^+ k_1^- + \underline{k}^2)} + \frac{\un{k}_1^2}{(x p_1^+ k_1^- + \underline{k}_1^2 )^2}  \right] \, \mathcal{O}_{\un{\zeta} \un{w}}. \notag
\end{align}
This is our main result for the eikonal-order quark Sivers function at small $x$.

The odderon is a correlator of eikonal Wilson lines, so the small-$x$ Sivers function we have found is proportional to $1/x$, and should actually grow as we decrease $x$. Moreover, small-$x$ evolution likely leaves this growth unchanged, as the odderon is known to have an intercept equal to zero at the leading \cite{Bartels:1999yt,Kovchegov:2003dm,Hatta:2005as} and next-to-leading \cite{Kovchegov:2012rz} orders in $\alpha_s$ (see also \cite{Janik:1998xj} for an earlier solution with a smaller intercept), at all orders in $\alpha_s$ in the large $N_c$ limit \cite{Caron-Huot:2013fea,Bartels:2013yga}, and at strong coupling in $\mathcal{N}=4$ supersymmetric Yang-Mills theory \cite{Brower:2008cy,Avsar:2009hc,Brower:2014wha}. 

Similar conclusion of the odderon dominance in the small-$x$ quark Sivers function was reached in \cite{Dong:2018wsp} using the expression for the small-$x$ unpolarized quark TMD (unintegrated quark distribution) from \cite{Mueller:1999wm,McLerran:1998nk} (see also \cite{Kovchegov:2015zha,Xiao:2017yya}). While qualitatively our result and that in \cite{Dong:2018wsp} agree on the odderon-driven quark Sivers TMD, a more detailed comparison between the two calculations is left for future work. The importance of $\cal C$-odd correlators in the Sivers asymmetry was also pointed out in \cite{Beppu:2010qn,Dai:2014ala}.

%%%%%%%%%%%%%%%%%%%%%%%%%%%%%%%%%%%%%%%%%%%%%%%%%%%%%%%%%%%%%%%%%%%%%%%%%%%%%%%%%%%

\subsection{Quark Sivers Function at the Eikonal Level: the Spin-Dependent Odderon Contribution}
\label{sec:spinodd}

We have shown that the eikonal small-$x$ evolution for the Sivers function in the operator treatment comes from the spin-dependent odderon. But we have not established that the odderon contribution is nonzero. In \cite{Szymanowski:2016mbq} it was shown that the odderon survives phase space integration if the proton has an asymmetric parton distribution using a diquark model calculation. Here we will show that the odderon is able to generate a nonzero Sivers function in the same diquark model of the proton. The Lagrangian for this model has a spinor field $\psi_P$ for the (point-like) proton, the usual spinor fields for the quarks $\psi_q$, and a complex scalar diquark field $\varphi$ which has mass $M$ roughly equal to the proton mass $M \approx M_P$ and the color quantum numbers of an antiquark. The interaction term between these fields is a Yukawa coupling $\mathscr{L}_{int} = G \, \varphi^{* \, i} \, {\bar \psi}^i_q \, \psi_P$+c.c., with effective coupling constant $G$ for the splitting of the proton into a quark and a diquark. 

%%%%%%%%%%%%%%%%%%%%%%%%%%%%%%%%%%%%%%%%%%%%%%%%%%%%%%%%%%%%%%%%%%%%%%%%%%%%%%%%
\begin{figure}[h]
\begin{center}
\includegraphics[width= 0.5 \textwidth]{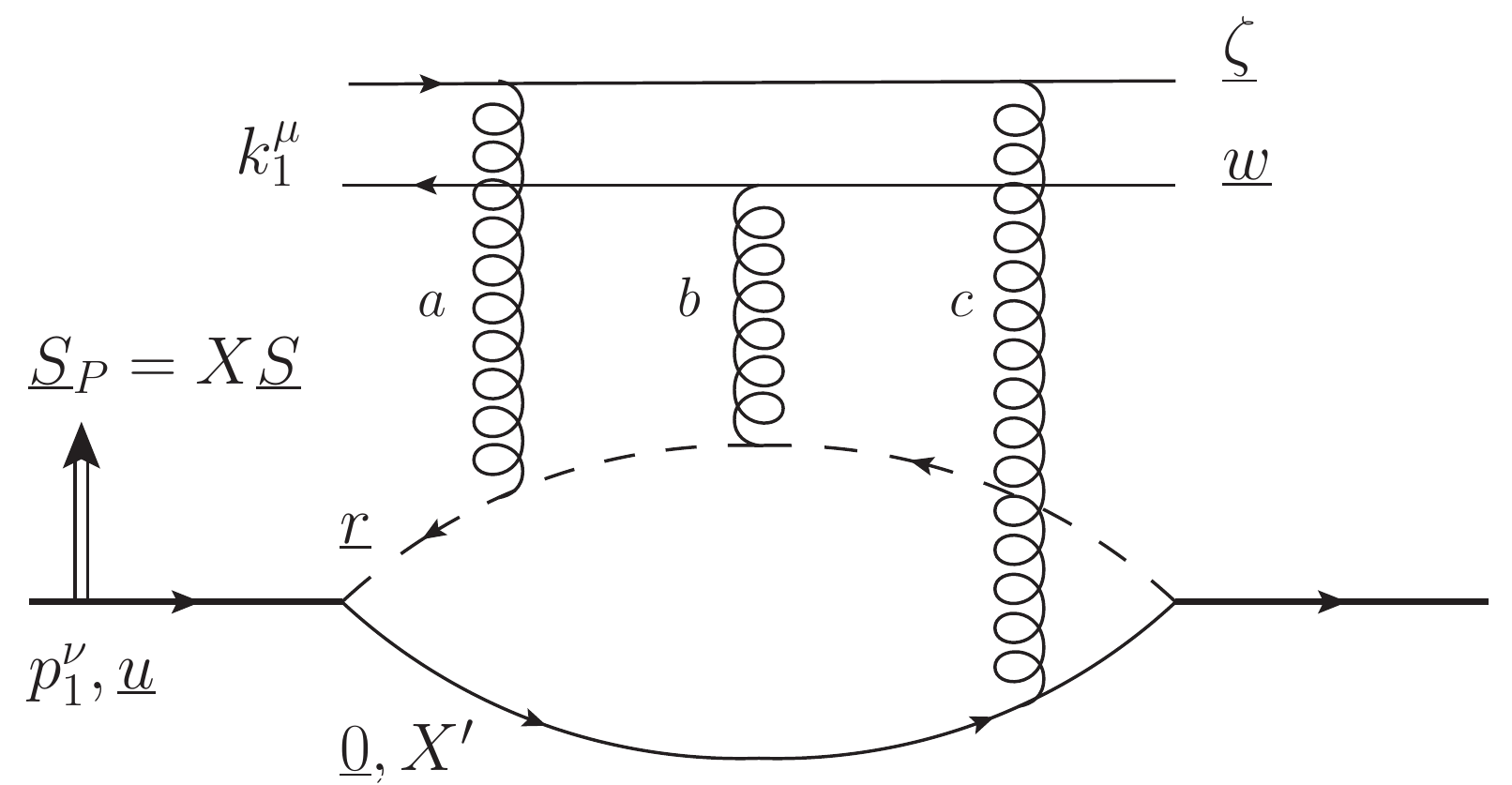} 
\caption{Diagram for the odderon exchange amplitude with the quark-diquark model of the proton, where all possible connections of the three gluons to the quark (solid line at the very bottom) and diquark (dashed line) and to the ${\un \zeta}, {\un w}$ dipole at the top should be summed over and the gluons are in the symmetric $d^{abc}$ color configuration.}
\label{FIG:odddiag}
\end{center}
\end{figure}
%%%%%%%%%%%%%%%%%%%%%%%%%%%%%%%%%%%%%%%%%%%%%%%%%%%%%%%%%%%%%%%%%%%%%%%%%%%%%%%%

We begin by constructing the odderon amplitude $\mathcal{O}_{\un{\zeta} \un{w}}$ to be used in \eq{siv}. At the lowest order it is illustrated in \fig{FIG:odddiag}, with the proton target splitting into a quark (solid line) and diquark (dashed line) in order to interact with the odderon (the 3-gluon exchange at the lowest order). The odderon comes from summing over all possible connections of the three gluons in \fig{FIG:odddiag}, where the polarized `target' proton (thicker solid line) splits into a dipole consisting of a quark and diquark, which exchanges three gluons in the symmetric color configuration $d^{abc} = 2 \textrm{Tr} [ t^a, \{t^b,t^c\}]$ with the `projectile' quark-antiquark dipole. We take the `target' quark to be at transverse position $\un{0}$ and the diquark at $\un{r}$, as shown in \fig{FIG:odddiag}. We want the coupling of the odderon to the transverse spin of the proton, so we need to insert the odderon exchange between the light-cone wave functions of the proton splitting into the quark--diquark pair, and the same wave function, but complex conjugate, describing the merger of the pair back into the proton. The light-cone wave function in the transverse spin basis for a proton at transverse position $\un u$ and with transverse polarization $X$, with the quark carrying the fraction $\gamma$ of the proton $p_1^+$ momentum and the transverse polarization $X'$ (as labeled in \fig{FIG:odddiag}), is \cite{Kovchegov:2020kxg}
\begin{align}\label{coord_wf}
\psi_{X, X'} (\un{r}, \un{0}, \un{u}, \gamma)  = & \, \frac{G \tilde{m}_{\gamma} \sqrt{\gamma} (1 - \gamma )  }{2 \pi}  \delta^{(2)} \left( {\un r} - {\un u} - \gamma \, \un{r} \right) \\ & \times \left[\delta_{X, X'}  K_0 (\tilde{m}_{\gamma} r_{\perp})  - \frac{ i X  {r}^i }{ r_{\perp}}  K_1 (\tilde{m}_{\gamma} r_{\perp}) (i \delta_{X, X'} \delta^{i 2} - \delta_{X, - X'} \delta^{i 1} ) \right] , \notag
\end{align}
where $\tilde{m}_{\gamma}^2 =  (1-\gamma) m + \gamma M^2 - \gamma (1 - \gamma ) M_P^2 \approx \gamma^2 M_P^2$ in the limit of massless light quarks (cf., e.g., \cite{Meissner:2007rx}). The odderon exchange amplitude in the dipole-dipole scattering depicted in \fig{FIG:odddiag} is \cite{Kovchegov:2003dm,Kovchegov:2012ga}
\begin{align}\label{Ohat}
\hat{\mathcal{O}}_{\un{\zeta},\un{w}} = c_0 \, \alpha_s^3 \, \ln^3 \Bigg( \frac{|\un{\zeta} - \un{0} | \,  |\underline{w} - \underline{r}|}{|\un{w} - \un{0} | \, |\underline{\zeta} - \underline{r} | } \Bigg) 
\end{align}
with \cite{Kovchegov:2012ga,Hatta:2005as,Kovner:2005qj,Jeon:2005cf}
\begin{align}\label{c0}
c_0 = - \frac{(N_c^2 - 4)(N_c^2-1)}{4N_c^3}, 
\end{align}

One obtains the spin-dependent odderon amplitude $\mathcal{O}_{\zeta w}$ by convoluting a square of the wave function \eqref{coord_wf} with the three-gluon exchange amplitude \eqref{Ohat}. Since the amplitude \eqref{Ohat} is odd under the $\un{r} \leftrightarrow \un{0}$ interchange, only the $\sim {\un r}$ part of the wave function squared contributes: this is exactly the part of the wave function dependent on the proton polarization $X$ \cite{Kovchegov:2020kxg}. Putting the wave function squared and the amplitude \eqref{Ohat} together we have
\begin{align}\label{Odd1}
\mathcal{O}_{\un{\zeta} \un{w}} &= \int\limits_0^1 \frac{\dd{\gamma}}{4 \pi \, \gamma \, (1-\gamma)} \int \dd[2]{u}_{\perp} \dd[2]{v}_{\perp} \dd[2]{r}_{\perp} \, \sum_{X'} \, \psi_{X, X'} (\un{r}, \un{0}, \un{u}, \gamma) \, \hat{\mathcal{O}}_{\un{\zeta},\un{w}} \, \psi_{X, X'}^* (\un{r}, \un{0}, \un{v}, \gamma) \notag \\
&= \int\limits_0^1 \dd{\gamma} \frac{G^2 \, \tilde{m}^2_{\gamma} \, (1-\gamma)}{(2\pi)^3} \int \dd[2]{r}_{\perp} \frac{X \, \un{S} \cross \un{r}}{r_{\perp}} K_0 (\tilde{m}_{\gamma} r_{\perp}) K_1 (\tilde{m}_{\gamma} r_{\perp}) \, c_0 \, \alpha_s^3 \, \ln^3 \Bigg( \frac{\zeta_{\perp} |\underline{w} - \underline{r}|}{w_{\perp} |\underline{\zeta} - \underline{r} | } \Bigg) .
\end{align}

Substituting the odderon amplitude \eqref{Odd1} into \eq{siv} we get
\begin{align}
 -& \frac{\un{k} \cross \underline{S}_P}{M_P} f_{1 \: T}^{\perp \: q} (x,k_T^2) \Big|_\textrm{eikonal} = \int\limits_0^1 \dd{\gamma} \frac{4 i N_c p_1^+ G^2  (1-\gamma) \tilde{m}_\gamma^2 c_0 \alpha_s^3}{(2 \pi)^6}  \int \dd[2]{r}_{\perp} \frac{\underline{S}_P \cross \underline{r}}{r_{\perp}} K_0 (\tilde{m}_{\gamma} r_{\perp} ) K_1 ( \tilde{m}_{\gamma} r_{\perp} )  \int \dd[2]{\zeta}_{\perp} \dd[2]{w}_{\perp} \notag \\
&\times \int \frac{\dd[2]{k}_{1 \perp}\dd{k}_1^-}{(2\pi)^3} \theta (k_1^-)  e^{i (\underline{k}_1 + \underline{k} ) \vdot ( \underline{w} - \underline{\zeta})} \, \ln^3 \Bigg( \frac{\zeta_{\perp} | \underline{w} - \underline{r} |}{w_{\perp} | \underline{\zeta} - \underline{r} |} \Bigg) \left[ \frac{2 \, \un{k} \cdot \un{k}_1}{(x p_1^+ k_1^- + \underline{k}_1^2 ) (x p_1^+ k_1^- + \underline{k}^2)} + \frac{\un{k}_1^2}{(x p_1^+ k_1^- + \underline{k}_1^2 )^2}  \right] ,
\end{align}
where the factor of $\underline{S}_P$ comes from the spin quantization axis $\underline{S}$ multiplied by the proton polarization $X$. For simplicity we assume that the $\underline{r}$-integral is dominated by the perturbatively short distances, $\tilde{m}_{\gamma} r_{\perp} \ll 1$, such that we can expand the modified Bessel functions to the lowest non-trivial order and obtain
\begin{align}\label{OddSiv1}
 - \frac{\un{k} \cross \underline{S}_P}{M_P} f_{1 \: T}^{\perp \: q}  (x,k_T^2)  \Big|_\textrm{eikonal} &  =  \frac{1}{x} \int\limits_0^1 \dd{\gamma} \frac{4 i N_c G^2  (1-\gamma) \tilde{m}_\gamma c_0 \alpha_s^3}{(2 \pi)^7}  \int \dd[2]{r}_{\perp} \frac{\underline{S}_P \cross \underline{r}}{\underline{r}^2} \, \ln \Big(\frac{1}{r_{\perp}\tilde{m}_{\gamma}} \Big)  \int \dd[2]{(\zeta-w)}_{\perp}  \\
&\times \int \frac{\dd[2]{k}_{1 \perp}}{(2\pi)^2}  e^{i (\underline{k}_1 + \underline{k} ) \vdot ( \underline{w} - \underline{\zeta})}   \left[ \frac{2 \, \un{k} \cdot \un{k}_1}{\underline{k}^2 - \underline{k}_1^2} \, \ln \frac{\un{k}^2}{\un{k}_1^2} + 1 \right] \int d^2 \left( \frac{\zeta + w}{2} \right)  \ln^3 \Bigg( \frac{\zeta_{\perp} | \underline{w} - \underline{r} |}{w_{\perp} | \underline{\zeta} - \underline{r} |} \Bigg)    . \notag 
\end{align}
In arriving at \eq{OddSiv1} we have also integrated over $k_1^-$. 

The integral over the impact parameter $(\zeta + w)/2$ can be rewritten as a momentum-space integral,
\begin{align}\label{ln3}
& \int d^2 \left( \frac{\zeta + w}{2} \right)  \ln^3 \Bigg( \frac{\zeta_{\perp} | \underline{w} - \underline{r} |}{w_{\perp} | \underline{\zeta} - \underline{r} |} \Bigg) = - \frac{1}{2 \pi} \, \int \frac{\dd[2]{l}_{1 \perp} \dd[2]{l}_{2 \perp} }{\underline{l}_1^2 \, \underline{l}_2^2  \, (\underline{l}_1 + \underline{l}_2)^2} \notag \\
&\times  \Bigg[ e^{i \underline{l}_1 \vdot \underline{x}} - e^{-i \underline{l}_1 \vdot \underline{x}} + e^{i \underline{l}_2 \vdot \underline{x}} - e^{-i \underline{l}_2 \vdot \underline{x}} - e^{i (\underline{l}_1 + \underline{l}_2) \vdot \underline{x}} + e^{-i (\underline{l}_1 + \underline{l}_2) \vdot \underline{x}} \Bigg] \notag \\
&\times \Bigg[ e^{i \underline{l}_1 \vdot \underline{r}} - e^{-i \underline{l}_1 \vdot \underline{r}} + e^{i \underline{l}_2 \vdot \underline{r}} - e^{-i \underline{l}_2 \vdot \underline{r}} - e^{i (\underline{l}_1 + \underline{l}_2) \vdot \underline{r}} + e^{-i (\underline{l}_1 + \underline{l}_2) \vdot \underline{r}} \Bigg],
\end{align}
where we have defined $\un{x} \equiv \un{\zeta} - {\un w}$. Using \eq{ln3} in \eq{OddSiv1} yields
\begin{align}\label{OddSiv2}
 - \frac{\un{k} \cross \underline{S}_P}{M_P} f_{1 \: T}^{\perp \: q}  (x,k_T^2)  \Big|_\textrm{eikonal} &  = - \frac{1}{x} \int\limits_0^1 \dd{\gamma} \frac{4 i N_c  G^2  (1-\gamma) \tilde{m}_\gamma c_0 \alpha_s^3}{(2 \pi)^{8}}  \int \dd[2]{r}_{\perp} \frac{\underline{S}_P \cross \underline{r}}{\underline{r}^2} \, \ln \Big(\frac{1}{r_{\perp}\tilde{m}_{\gamma}} \Big)  \int \dd[2]{x}_{\perp}  \\
&\times \int \frac{\dd[2]{k}_{1 \perp}}{(2\pi)^2}  e^{i (\underline{k}_1 + \underline{k} ) \vdot \un{x}}   \left[ \frac{2 \, \un{k} \cdot \un{k}_1}{\underline{k}^2 - \underline{k}_1^2} \, \ln \frac{\un{k}^2}{\un{k}_1^2} + 1 \right] \, \int \frac{\dd[2]{l}_{1 \perp} \dd[2]{l}_{2 \perp} }{\underline{l}_1^2 \, \underline{l}_2^2  \, (\underline{l}_1 + \underline{l}_2)^2} \notag \\
&\times  \Bigg[ e^{i \underline{l}_1 \vdot \underline{x}} - e^{-i \underline{l}_1 \vdot \underline{x}} + e^{i \underline{l}_2 \vdot \underline{x}} - e^{-i \underline{l}_2 \vdot \underline{x}} - e^{i (\underline{l}_1 + \underline{l}_2) \vdot \underline{x}} + e^{-i (\underline{l}_1 + \underline{l}_2) \vdot \underline{x}} \Bigg] \notag \\
&\times \Bigg[ e^{i \underline{l}_1 \vdot \underline{r}} - e^{-i \underline{l}_1 \vdot \underline{r}} + e^{i \underline{l}_2 \vdot \underline{r}} - e^{-i \underline{l}_2 \vdot \underline{r}} - e^{i (\underline{l}_1 + \underline{l}_2) \vdot \underline{r}} + e^{-i (\underline{l}_1 + \underline{l}_2) \vdot \underline{r}} \Bigg] . \notag 
\end{align}
Next we argue that $\ln (1/|\underline{r}| \tilde{m}_{\gamma})$ is a slowly-varying function compared to powers and exponentials of $r_\perp$ present in the integrand and approximate this logarithm by one, $\ln (1/|\underline{r}| \tilde{m}_{\gamma}) \approx 1$. The integral over $\un{r}$ then becomes straightforward: performing it and the integral over $\un{x}$ we find
\begin{align}\label{OddSiv3}
 - \frac{\un{k} \cross \underline{S}_P}{M_P} f_{1 \: T}^{\perp \: q}  (x,k_T^2)  \Big|_\textrm{eikonal} &  =  \frac{1}{x} \int\limits_0^1 \dd{\gamma} \frac{8 N_c G^2  (1-\gamma) \tilde{m}_\gamma c_0 \alpha_s^3}{(2 \pi)^{5}}  \int \frac{\dd[2]{k}_{1 \perp}}{(2\pi)^2}    \left[ \frac{2 \, \un{k} \cdot \un{k}_1}{\underline{k}^2 - \underline{k}_1^2} \, \ln \frac{\un{k}^2}{\un{k}_1^2} + 1 \right]  \\
&\times \, \int \frac{\dd[2]{l}_{1 \perp} \dd[2]{l}_{2 \perp} }{\underline{l}_1^2 \, \underline{l}_2^2  \, (\underline{l}_1 + \underline{l}_2)^2} \ \un{S}_P \cross \Bigg[ \frac{\un{l}_1}{\un{l}_1^2} + \frac{\un{l}_2}{\un{l}_2^2} - \frac{\un{l}_1 + \un{l}_2}{(\un{l}_1 + \un{l}_2)^2} \Bigg]  \Bigg[ \delta^2 ( \underline{l}_1 + \underline{k}_1 + \underline{k} ) -\delta^2 ( \underline{l}_1 - \underline{k}_1 - \underline{k} ) \notag \\
&+ \delta^2 ( \underline{l}_2 + \underline{k}_1 + \underline{k} ) - \delta^2 ( \underline{l}_2 - \underline{k}_1 - \underline{k} ) - \delta^2 ( \underline{l}_1 + \underline{l}_2 + \underline{k}_1 + \underline{k} ) + \delta^2 ( \underline{l}_1 + \underline{l}_2 - \underline{k}_1 - \underline{k} ) \Bigg] . \notag 
\end{align}
Next we integrate over $\un{l}_1$ and $\un{l}_2$ while regulating the singularities by the infrared (IR) cutoff $\Lambda$. We get
\begin{align}\label{OddSiv4}
 - \frac{\un{k} \cross \underline{S}_P}{M_P} f_{1 \: T}^{\perp \: q}  (x,k_T^2)  \Big|_\textrm{eikonal} &  =  \frac{1}{x} \int\limits_0^1 \dd{\gamma} \frac{48 N_c G^2  (1-\gamma) \tilde{m}_\gamma c_0 \alpha_s^3}{(2 \pi)^{4}}  \\
&\times \,  \int \frac{\dd[2]{k}_{1 \perp}}{(2\pi)^2}    \left[ \frac{2 \, \un{k} \cdot \un{k}_1}{\underline{k}^2 - \underline{k}_1^2} \, \ln \frac{\un{k}^2}{\un{k}_1^2} + 1 \right]  \,  \frac{\un{S}_P \times (\un{k} + \un{k}_1)}{|\un{k} + \un{k}_1 |^6} \, \ln \frac{(\un{k} + \un{k}_1)^2}{\Lambda^2}  . \notag 
\end{align}

To integrate over $\un{k}_1$ analytically we, again, have to neglect a logarithm, by putting $\ln [(\un{k} + \un{k}_1)^2 / \Lambda^2 ] \approx 1$. The remaining $\un{k}_1$-integral is dominated by the singularity at $\un{k}_1 = - \un{k}$, which we regulate by the IR cutoff $\Lambda$. Keeping only the most singular part in the expansion in $1/\Lambda$ we obtain
\begin{align}\label{OddSiv5}
 - \frac{\un{k} \cross \underline{S}_P}{M_P} f_{1 \: T}^{\perp \: q}  (x,k_T^2)  \Big|_\textrm{eikonal}   =  \frac{1}{x} \int\limits_0^1 \dd{\gamma} \frac{3 N_c G^2  (1-\gamma) \tilde{m}_\gamma c_0 \alpha_s^3}{(2 \pi)^{5}} \, \frac{\un{S}_P \cross \un{k}}{\un{k}^2 \Lambda^2} .
\end{align}
The IR cutoff $\Lambda$ can, in principle, depend on $\gamma$ since the natural mass scale for an IR cutoff would be $\tilde{m}_{\gamma}$. However, at sufficiently small $\gamma$ one must replace this cutoff with the QCD confinement scale $\Lambda_{QCD}$, since, formally, in the quark--diquark model the transverse distances are bound by $r \lesssim 1/m_\gamma$ and may get much larger than the confinement scale as $\gamma \to 0$. For simplicity we take the IR cutoff to be some momentum scale independent of $\gamma$, as this will only change an overall numerical factor. Taking $\tilde{m}_{\gamma} = \gamma M_P$, and integrating over $\gamma$ we have the Sivers function due to the odderon exchange in the quark--diquark model
\begin{align}\label{oddsiv}
f_{1 \: T}^{\perp \: q}  (x,k_T^2)  \Big|_\textrm{eikonal}   =  \frac{1}{x}  \frac{N_c G^2 \, c_0 \alpha_s^3}{2 (2 \pi)^{5}} \, \frac{M_P^2}{\un{k}^2 \Lambda^2} .
\end{align}

We have obtained a nonzero, eikonal contribution to the Sivers function corresponding to the same spin-dependent odderon as studied in \cite{Szymanowski:2016mbq}. Both the eikonality and the dependence on $\un{k}$ and $M_P$ in \eq{oddsiv} are different from the Born-level results calculated in the same diquark model (see Eq.~(A8) in \cite{Meissner:2007rx}). In particular, the Sivers function in \eq{oddsiv} has a very interesting behavior in the $\Lambda_{QCD},M_P \rightarrow 0$ limit. Since the IR cutoff $\Lambda$ must be proportional to $\Lambda_{QCD}$, we conclude that $\Lambda \sim \Lambda_{QCD} \sim M_P$. Therefore, $M_P^2/\Lambda^2$ ratio will remain constant in the $\Lambda_{QCD},M_P \rightarrow 0$ limit and the Sivers function \eq{oddsiv} will not vanish in the limit of zero proton mass. This may be a feature unique to the quark--diquark model of the proton, but it calls for further analysis.

%%%%%%%%%%%%%%%%%%%%%%%%%%%%%%%%%%%%%%%%%%%%%%%%%%%%%%%%%%%%%%%%%%%%%%%%%%%%%%%%%%%

\subsection{Sivers Function at the Sub-Eikonal Level: a New Evolution}
\label{sec:sub-eik}

In this Section we will construct a sub-eikonal correction to \eq{oddsiv}. Part of the motivation for this is that the effects of the odderon have been historically hard to find in the data. The recent announcement of the odderon discovery \cite{TOTEM:2020zzr} was many years in the making and required very careful measurements and extrapolation in energy. It is, therefore, important to understand the background for the odderon contribution \eqref{oddsiv} in order to discover the latter in the future measurements of the Sivers function. In addition, the existing data on the hadronic single transverse spin asymmetry (STSA) $A_N$ measured in $p^\uparrow + p$ collisions \cite{Adams:1991rw,Adams:1991cs,Abelev:2008af,Adler:2005in} exhibits no evidence for the odderon (with the possible exception of the AnDY Collaboration data \cite{Bland:2013pkt}): assuming that the $x$-dependence of the Sivers function is related to that in $A_N$, that is, that $A_N \sim x  \, f_{1 \: T}^{\perp}$, we see that the odderon contribution \eqref{oddsiv} would predict an $x$-independent $A_N$. However, most of the data \cite{Adams:1991rw,Adams:1991cs,Abelev:2008af,Adler:2005in} show $A_N$ which falls off with decreasing $x$ of the transversely polarized proton. This behavior of the data can be roughly approximated as $A_N \sim x$, which translates into the Sivers function $f_{1 \: T}^{\perp}$ being independent of $x$. Since $f_{1 \: T}^{\perp} \sim x^0$ is the sub-eikonal scaling, it is clear that a sub-eikonal study of the Sivers function is in order.

\subsubsection{Sub-Eikonal Sivers Function}

\begin{figure}[ht]
\centering
\includegraphics[width=0.5\linewidth]{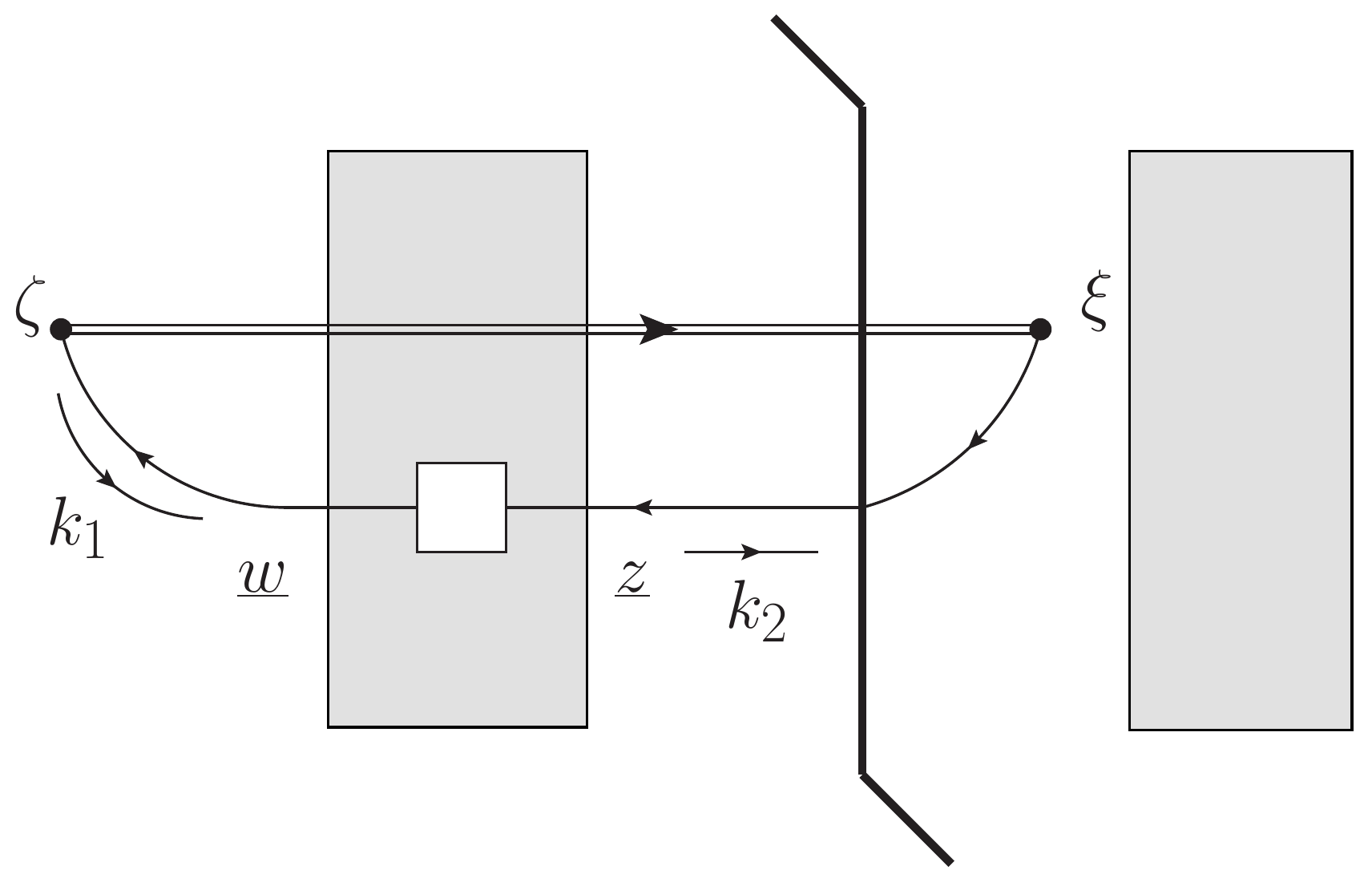}  
\caption{Diagram of class B with kinematics specified. The antiquark propagates from $\zeta$ with momentum $k_1$, undergoes a transverse spin dependent interaction with the proton at transverse positions $\un{w}$ and $\un{z}$ before and after scattering, respectively, then propagates to $\xi$ with momentum $k_2$. The sub-eikonal interaction with the proton shock wave is denoted by the white box.}
\label{FIG:diagbdet}
\end{figure}

The analysis of Sec.~\ref{sec:gen_exp} also applies at the sub-eikonal level \cite{Kovchegov:2018znm}. At the sub-eikonal level only the diagram B from \fig{FIG:diagbdet} contributes \cite{Kovchegov:2018znm}. One essential difference we have here as compared to, say, the helicity calculation in \cite{Kovchegov:2018znm}, is that the sub-eikonal and sub-sub-eikonal terms in the quark $S$-matrix operator \eqref{Full} are non-local in the transverse plane. Thus, the calculation in Sec.~\ref{sec:gen_exp} has to be augmented by introducing two different coordinates of the anti-quark to the left and to the right of the shock wave, $\un w$ and $\un z$, as shown in \fig{FIG:diagbdet}. This is taken into account by replacing 
\begin{align}
V_{{\un w}}^\dagger   \to \int d^2 z_\perp \, V_{{\un z}, {\un w}}^\dagger
\end{align} 
along with $e^{i \un{k} \cdot (\un{w} - \un{\zeta})} \to e^{i \un{k} \cdot (\un{z} - \un{\zeta})}$ in \eq{Bcc2}. We arrive at the Sivers function given by the diagram B and its complex conjugate,
\begin{align}
\label{Sivers_sub_eik}
 - \frac{\un{k} \cross \underline{S}_P}{M_P}  & \, f_{1 \: T}^{\perp \: q} (x,k_T^2) \Big|_\textrm{sub-eikonal} 
\subset  \frac{4 p_1^+}{(2 \pi)^3} \int \dd[2]{\zeta_{\perp}}  \dd[2]{w_{\perp}} \dd[2]{z_{\perp}}  \frac{\dd[2]{k_{1 \perp}}\dd{k_1^-}}{(2\pi)^3} \theta (k_1^-) \, e^{i \underline{k}_1 \vdot (\un{w} - \un{\zeta}) + i \underline{k} \vdot (\un{z} - \un{\zeta})} \, \Big{\langle} \tord \tr \left[  V_{\underline{\zeta}} V_{\un{z}, \underline{w}}^{\textrm{pol} \: \dagger } \right] \Big{\rangle}  \notag \\
\times & \frac{{\un k}_1 \cdot \un{k} }{\left[ x p_1^+ k_1^- + \underline{k}_1^2 \right] \left[ x p_1^+ k_1^- + \underline{k}^2 \right] } + \mbox{c.c.} \approx \frac{4 i}{(2 \pi)^5} \int \dd[2]{\zeta_{\perp}}  \dd[2]{w_{\perp}} \dd[2]{z_{\perp}} \int\limits_\frac{\Lambda^2}{p_1^+}^{p_2^-} \frac{\dd{k_1^-}}{k_1^-} \, e^{i \underline{k} \vdot (\un{z} - \un{\zeta})} \, \frac{\un{w} - \un{\zeta}}{|\un{w} - \un{\zeta}|^2} \cdot \frac{\un{k}}{k_T^2} \notag \\ & \times \, \llangle \tord \tr \left[  V_{\underline{\zeta}} V_{\un{z}, \underline{w}}^{\textrm{pol} \: \dagger } \right] \rrangle + \mbox{c.c.} ,
\end{align}
where, in the last step, we have extracted the DLA contribution only, and defined
\begin{align}
\llangle \ldots \rrangle = z s \, \langle \ldots \rangle = p_1^+ k_1^-  \, \langle \ldots \rangle
\end{align}
with $z = k_1^-/p_2^-$.

%%%%%%%%%%%%%%%%%%%%%%%%%%%%%%%%%%%%%%%%%%%%%%%%%%%%%%%%%%%%%%%%%%%%%%%%%%%%%%%%%%%%
%%%%%%%%%%%%%%%%%%%%%%%%%%%%%%%%%%%%%%%%%%%%%%%%%%%%%%%%%%%%%%%%%%%%%%%%%%%%%%%%%%%%

At the sub-eikonal order, the polarized Wilson line in \eq{Sivers_sub_eik} contains only the $\sim \delta_{\chi, \chi'}$ contribution, since the Sivers function couples an unpolarized quark to the proton spin.\footnote{From this point on our analysis of the sub-eikonal contribution is incomplete and ultimately leads to incorrect asymptotics for the Sivers function in \eq{Sivers_sub_eik8}. One may expect that only operators with the $\sim \delta_{\chi, \chi'}$ prefactor should enter into the dipole amplitudes contributing to the Sivers function as the latter is a distribution of unpolarized quarks. However it was found in \cite{Cougoulic:2022gbk} that the sub-eikonal operator with the $\sim \delta_{\chi, -\chi'}$ prefactor containing $F^{12}$, which naively describes quark helicity coupling to the background chromomagnetic field, mixes with the naively unpolarized $\sim \delta_{\chi, \chi'}$ operator ($\sim \cev{D}_z^i  \vec{D}^i_z$ in the gluon sector) under evolution. Therefore, it is incorrect to neglect the contribution of the $\sim \delta_{\chi, -\chi'}$ operator to the Sivers function and its small-$x$ evolution. In \cite{Kovchegov:2022kyy} we restore this term, as well as terms proportional to the sub-eikonal $A^+$ field which we have neglected here in simplifying the covariant derivatives. These terms also contribute to evolution, and yield a new set of evolution equations superseding Eqs.~\eqref{Fi_evol5} below \cite{Kovchegov:2022kyy}. We numerically solve the new evolution equations in \cite{Kovchegov:2022kyy} to obtain the correct small-$x$ asymptotics for the sub-eikonal contribution to the flavor non-singlet Sivers function. Further details for the observation made in this footnote can be found in \cite{Kovchegov:2022kyy}.} We thus write, neglecting the quark mass ($m=0$) and employing Eqs.~\eqref{O24} and \eqref{Oq3} in \eq{Full}, truncated to the sub-eikonal order, keeping the $\delta_{\chi, \chi'}$ term only, and neglecting the eikonal contribution already accounted for above,
\begin{align}\label{V_subeik_1}
& V_{\un{x}, \underline{y}}^{\textrm{pol}} = -  \frac{i \, p_1^+}{2 \, s}   \int\limits_{-\infty}^{\infty} \dd{z}^- d^2 z \ V_{\un{x}} [ \infty, z^-] \, \delta^2 (\un{x} - \un{z}) \, \cev{D}_z^i  \vec{D}^i_z \, V_{\un{y}} [ z^-, -\infty] \, \delta^2 (\un{y} - \un{z}) \\ 
& - \frac{g^2 \, p_1^+}{4 \, s} \, \delta^2 (\un{x} - \un{y})  \, \int\limits_{-\infty}^{\infty} \dd{z}_1^- \int\limits_{z_1^-}^\infty d z_2^-  \ V_{\un{x}} [ \infty, z_2^-] \,  t^b \, \psi_{\beta} (z_2^-,\un{x}) \, U_{\un{x}}^{ba} [z_2^-,z_1^-] \, \left[ \frac{\gamma^+}{2} \right]_{\alpha \beta} \, \bar{\psi}_\alpha (z_1^-,\un{x}) \, t^a \, V_{\un{x}} [ z_1^-, -\infty] . \notag 
\end{align}
In the $A^-= 0$ gauge we are working in, the transverse components of the gluon field, $\un{A}$, are sub-eikonal. Using this to simplify the gluon part of \eq{V_subeik_1}, while keeping only sub-eikonal terms dependent on the target transverse polarization, we get 
\begin{align}\label{V_subeik_2}
& V_{\un{x}, \underline{y}}^{\textrm{pol}} = - \frac{g \, p_1^+}{2 \, s}   \int\limits_{-\infty}^{\infty} \dd{z}^- d^2 z \ V_{\un{x}} [ \infty, z^-] \, \delta^2 (\un{x} - \un{z}) \left[ \underline{A} (z^-, \un{z})  \cdot \vec{\un{\nabla}}_z - \cev{\un{\nabla}}_z \cdot \underline{A} (z^-, \un{z}) \right] V_{\un{y}} [ z^-, -\infty] \, \delta^2 (\un{y} - \un{z}) \\ 
& - \frac{g^2 \, p_1^+}{4 \, s} \, \delta^2 (\un{x} - \un{y})  \, \int\limits_{-\infty}^{\infty} \dd{z}_1^- \int\limits_{z_1^-}^\infty d z_2^-  \ V_{\un{x}} [ \infty, z_2^-] \,  t^b \, \psi_{\beta} (z_2^-,\un{x}) \, U_{\un{x}}^{ba} [z_2^-,z_1^-] \,  \left[ \frac{\gamma^+}{2} \right]_{\alpha \beta} \, \bar{\psi}_\alpha (z_1^-,\un{x}) \, t^a \, V_{\un{x}} [ z_1^-, -\infty] . \notag 
\end{align}
Further, integrating over $\un{z}$ yields
\begin{align}\label{V_subeik_3}
& V_{\un{x}, \underline{y}}^{\textrm{pol}} = - \frac{g \, p_1^+}{2 \, s} \, \left[ \un{\nabla}_x \delta^2 (\un{x} - \un{y}) \right] \cdot \int\limits_{-\infty}^{\infty} \dd{z}^- \, V_{\un{x}} [ \infty, z^-]  \left[ \underline{A} (z^-, \un{x})  + \underline{A} (z^-, \un{y}) \right] V_{\un{y}} [ z^-, -\infty]  \\ 
& - \frac{g^2 \, p_1^+}{4 \, s} \, \delta^2 (\un{x} - \un{y})  \, \int\limits_{-\infty}^{\infty} \dd{z}_1^- \int\limits_{z_1^-}^\infty d z_2^-  \ V_{\un{x}} [ \infty, z_2^-] \,  t^b \, \psi_{\beta} (z_2^-,\un{x}) \, U_{\un{x}}^{ba} [z_2^-,z_1^-] \,  \left[ \frac{\gamma^+}{2} \right]_{\alpha \beta} \, \bar{\psi}_\alpha (z_1^-,\un{x}) \, t^a \, V_{\un{x}} [ z_1^-, -\infty] . \notag 
\end{align}

We see that while the quark sector operator in \eq{V_subeik_3} is local in the transverse plane, the gluon sector operator is non-local due to the derivative of the $\delta$-function. Substituting the gluon part of \eq{V_subeik_3} into the first line of \eq{Sivers_sub_eik} we arrive at, after multiple integrations by parts,
\begin{align}\label{glue_simp}
& - \frac{4}{(2 \pi)^3} \frac{g \, p_1^+}{2} \, \int \dd[2]{\zeta_{\perp}}  \dd[2]{w_{\perp}} \dd[2]{z_{\perp}}  \frac{\dd[2]{k_{1 \perp}}\dd{k_1^-}}{(2\pi)^3 \, k_1^-} \theta (k_1^-) \, e^{i \underline{k}_1 \vdot (\un{w} - \un{\zeta}) + i \underline{k} \vdot (\un{z} - \un{\zeta})} \, \frac{{\un k}_1 \cdot \un{k} }{\underline{k}_1^2 \, \underline{k}^2} \, \int\limits_{-\infty}^{\infty} \dd{z}^- \\ 
& \times \left[ \un{\nabla}_z \delta^2 (\un{z} - \un{w}) \right] \cdot  \Big\langle \tord \tr \left[  V_{\underline{\zeta}} \, V_{\un{w}} [ - \infty, z^-]  \left[ \underline{A} (z^-, \un{z})  + \underline{A} (z^-, \un{w}) \right] V_{\un{z}} [ z^-, \infty] \right] \Big\rangle  + \mbox{c.c.}  \notag \\ 
& = - \frac{4}{(2 \pi)^3} \frac{g \, p_1^+}{2 } \, \int \dd[2]{\zeta_{\perp}}  \dd[2]{w_{\perp}}  \frac{\dd[2]{k_{1 \perp}}\dd{k_1^-}}{(2\pi)^3 \, k_1^-} \theta (k_1^-) \, e^{i (\un{k} + \underline{k}_1) \vdot (\un{w} - \un{\zeta}) } \, \frac{{\un k}_1 \cdot \un{k} }{\underline{k}_1^2 \, \underline{k}^2} \, \int\limits_{-\infty}^{\infty} \dd{z}^- \notag \\ 
& \Big\langle \tr \Big[  V_{\underline{\zeta}} \, \Big\{ i (\un{k}_1 - \un{k}) \cdot V_{\un{w}} [ - \infty, z^-]  \, \underline{A} (z^-, \un{w}) \, V_{\un{w}} [ z^-, \infty]  \notag \\ 
&  + \left( \un{\nabla}_w V_{\un{w}} [ - \infty, z^-] \right) \cdot  \underline{A} (z^-, \un{w}) \, V_{\un{w}} [ z^-, \infty]  - V_{\un{w}} [ - \infty, z^-]  \, \underline{A} (z^-, \un{w})  \cdot \left( \un{\nabla}_w V_{\un{w}} [ z^-, \infty] \right)  \Big\} \Big] \Big\rangle  + \mbox{c.c.}  \notag \\
& = - \frac{4}{(2 \pi)^3} \frac{g \, p_1^+}{2 } \, \int \dd[2]{\zeta_{\perp}}  \dd[2]{w_{\perp}}  \frac{\dd[2]{k_{1 \perp}}\dd{k_1^-}}{(2\pi)^3 \, k_1^-} \theta (k_1^-) \, e^{i (\un{k} + \underline{k}_1) \vdot (\un{w} - \un{\zeta}) } \, \frac{{\un k}_1 \cdot \un{k} }{\underline{k}_1^2 \, \underline{k}^2} \, \int\limits_{-\infty}^{\infty} \dd{z}^- \notag \\ 
& \Big\langle \tr \Big[  V_{\underline{\zeta}} \, \Big\{ \!\! \left( (\un{\nabla}_w + i \un{k}_1) \, V_{\un{w}} [ - \infty, z^-] \right) \cdot  \underline{A} (z^-, \un{w}) \, V_{\un{w}} [ z^-, \infty]  - V_{\un{w}} [ - \infty, z^-]  \, \underline{A} (z^-, \un{w})  \cdot \left( (\un{\nabla}_w + i \un{k}) \, V_{\un{w}} [ z^-, \infty] \right) \!\! \Big\} \Big] \Big\rangle  + \mbox{c.c.}  . \notag
\end{align}
It appears that the non-locality of the gluon operator in \eq{V_subeik_3} translates into the factors of $\un{k}_1$ and $\un k$ in \eq{glue_simp}. 

We conclude that the sub-eikonal contribution to the quark Sivers function is
\begin{align}
\label{Sivers_sub_eik2}
- \frac{\un{k} \cross \underline{S}_P}{M_P} f_{1 \: T}^{\perp \: q} (x,k_T^2)  \Big|_\textrm{sub-eikonal} 
\subset  \frac{4}{(2 \pi)^3} \int \dd[2]{\zeta_{\perp}}  \dd[2]{w_{\perp}} &  \frac{\dd[2]{k_{1 \perp}}\dd{k_1^-}}{(2\pi)^3 \, k_1^-} \theta (k_1^-) \, e^{i (\un{k} + \underline{k}_1) \vdot (\un{w} - \un{\zeta}) } \, \frac{{\un k}_1 \cdot \un{k} }{\underline{k}_1^2 \, \underline{k}^2} \\ & \times \, \llangle \tord \tr \left[  V_{\underline{\zeta}} V_{\un{w}; \un{k}, \un{k}_1}^{\textrm{pol} \: \dagger } \right] + \atord \tr \left[  V_{\un{\zeta}; \un{k}, \un{k}_1}^{\textrm{pol}} V_{\underline{w}}^\dagger  \right] \rrangle \notag
\end{align}
with
\begin{align}\label{V_subeik_4}
& V_{\un{w}; \un{k}, \un{k}_1}^{\textrm{pol}} = - \frac{g \, p_1^+}{2 \, s} \, \int\limits_{-\infty}^{\infty} \dd{z}^- \, \Big\{ V_{\un{w}} [ \infty, z^-] \, \underline{A} (z^-, \un{w}) \cdot \left( (\un{\nabla}_w - i \un{k}_1) \, V_{\un{w}} [ z^-, -\infty] \right) \\ 
& \hspace*{3.5cm} -  \left( (\un{\nabla}_w - i \un{k}) V_{\un{w}} [ \infty, z^-]  \right) \cdot \underline{A} (z^-, \un{w}) \, V_{\un{w}} [ z^-, -\infty] \Big\}  \notag \\ 
& - \frac{g^2 \, p_1^+}{4 \, s} \, \int\limits_{-\infty}^{\infty} \dd{z}_1^- \int\limits_{z_1^-}^\infty d z_2^-  \ V_{\un{w}} [ \infty, z_2^-] \,  t^b \, \psi_{\beta} (z_2^-,\un{x}) \, U_{\un{w}}^{ba} [z_2^-,z_1^-] \,  \left[ \frac{\gamma^+}{2} \right]_{\alpha \beta} \, \bar{\psi}_\alpha (z_1^-,\un{w}) \, t^a \, V_{\un{w}} [ z_1^-, -\infty] . \notag 
\end{align}

The part of the expression on the right-hand side of \eq{Sivers_sub_eik2} that contributes to the Sivers function should change sign after $\un{k} \to - \un{k}$. Hence, to make \eqref{Sivers_sub_eik2} an equality, we need to anti-symmetrize its right-hand side under $\un{k} \to - \un{k}$. Simultaneously changing $\un{k}_1 \to - \un{k}_1$ and $\un{w} \leftrightarrow \un{\zeta}$ we arrive at
\begin{align}
\label{Sivers_sub_eik3}
- \frac{\un{k} \cross \underline{S}_P}{M_P} & \, f_{1 \: T}^{\perp \: q} (x,k_T^2) \Big|_\textrm{sub-eikonal} 
=  \frac{2}{(2 \pi)^3} \int \dd[2]{\zeta_{\perp}}  \dd[2]{w_{\perp}}  \frac{\dd[2]{k_{1 \perp}}\dd{k_1^-}}{(2\pi)^3 \, k_1^-} \theta (k_1^-) \, e^{i (\un{k} + \underline{k}_1) \vdot (\un{w} - \un{\zeta}) } \, \frac{{\un k}_1 \cdot \un{k} }{\underline{k}_1^2 \, \underline{k}^2} \\ & \times \, \llangle \tord \tr \left[  V_{\underline{\zeta}} V_{\un{w}; \un{k}, \un{k}_1}^{\textrm{pol} \: \dagger } \right] - \tord \tr \left[  V_{\underline{w}} V_{\un{\zeta}; -\un{k}, - \un{k}_1}^{\textrm{pol} \: \dagger } \right] + \atord \tr \left[  V_{\un{\zeta}; \un{k}, \un{k}_1}^{\textrm{pol}} V_{\underline{w}}^\dagger  \right] -  \atord \tr \left[  V_{\un{w}; -\un{k}, -\un{k}_1}^{\textrm{pol}} V_{\underline{\zeta}}^\dagger  \right] \rrangle . \notag
\end{align}

Further, define
\begin{subequations}\label{Vi2}
\begin{align}
& V_{\un{w}}^{i \, \textrm{pol}} \equiv \frac{i g \, p_1^+}{2 \, s} \, \int\limits_{-\infty}^{\infty} \dd{z}^- \, V_{\un{w}} [ \infty, z^-] \, A^i (z^-, \un{w}) \, V_{\un{w}} [ z^-, -\infty] , \label{Vi} \\
& V_{\un{w}}^{[2] \, \textrm{pol}} = - \frac{g \, p_1^+}{2 \, s} \, \int\limits_{-\infty}^{\infty} \dd{z}^- \, \Big\{ V_{\un{w}} [ \infty, z^-] \, \underline{A} (z^-, \un{w}) \cdot \left( \un{\nabla}_w \, V_{\un{w}} [ z^-, -\infty] \right)  -  \left( \un{\nabla}_w V_{\un{w}} [ \infty, z^-]  \right) \cdot \underline{A} (z^-, \un{w}) \, V_{\un{w}} [ z^-, -\infty] \Big\} \notag   \\ 
& \hspace*{1.4cm}  - \frac{g^2 \, p_1^+}{4 \, s} \, \int\limits_{-\infty}^{\infty} \dd{z}_1^- \int\limits_{z_1^-}^\infty d z_2^-  \ V_{\un{w}} [ \infty, z_2^-] \,  t^b \, \psi_{\beta} (z_2^-,\un{w}) \, U_{\un{w}}^{ba} [z_2^-,z_1^-] \,  \left[ \frac{\gamma^+}{2} \right]_{\alpha \beta} \, \bar{\psi}_\alpha (z_1^-,\un{w}) \, t^a \, V_{\un{w}} [ z_1^-, -\infty] , \label{V2}
\end{align}
\end{subequations}
such that 
\begin{align}\label{VVV}
 V_{\un{w}; \un{k}, \un{k}_1}^{\textrm{pol}} =  ({k}_1 - {k})^i  \, V_{\un{w}}^{i \, \textrm{pol}} + V_{\un{w}}^{[2] \, \textrm{pol}} .
\end{align}

Equation \eqref{Sivers_sub_eik3} becomes
\begin{align}
\label{Sivers_sub_eik4}
- \frac{\un{k} \cross \underline{S}_P}{M_P} & \, f_{1 \: T}^{\perp \: q} (x,k_T^2)  \Big|_\textrm{sub-eikonal}  
=  \frac{2}{(2 \pi)^3} \int \dd[2]{\zeta_{\perp}}  \dd[2]{w_{\perp}}  \frac{\dd[2]{k_{1 \perp}}\dd{k_1^-}}{(2\pi)^3 \, k_1^-} \theta (k_1^-) \, e^{i (\un{k} + \underline{k}_1) \vdot (\un{w} - \un{\zeta}) } \, \frac{{\un k}_1 \cdot \un{k} }{\underline{k}_1^2 \, \underline{k}^2} \\ 
& \times \, \Bigg\{ ({k}_1 - {k})^i  \,  \llangle \tord \tr \left[  V_{\underline{\zeta}} V_{\un{w}}^{i \, \textrm{pol} \: \dagger } \right] + \tord \tr \left[  V_{\underline{w}} V_{\un{\zeta}}^{i \, \textrm{pol} \: \dagger } \right] + \atord \tr \left[  V_{\un{\zeta}}^{i \, \textrm{pol}} V_{\underline{w}}^\dagger  \right] +  \atord \tr \left[  V_{\un{w}}^{i \, \textrm{pol}} V_{\underline{\zeta}}^\dagger  \right] \rrangle \notag \\
& +  \llangle \tord \tr \left[  V_{\underline{\zeta}} V_{\un{w}}^{[2] \, \textrm{pol} \: \dagger } \right] - \tord \tr \left[  V_{\underline{w}} V_{\un{\zeta}}^{[2] \, \textrm{pol} \: \dagger } \right] + \atord \tr \left[  V_{\un{\zeta}}^{[2] \, \textrm{pol}} V_{\underline{w}}^\dagger  \right] -  \atord \tr \left[  V_{\un{w}}^{[2] \, \textrm{pol}} V_{\underline{\zeta}}^\dagger  \right] \rrangle \Bigg\}. \notag
\end{align}

Next we define two polarized dipole amplitudes
\begin{subequations}\label{Fdef}
\begin{align}
& F^i_{\un{w}, \un{\zeta}} (z) \equiv \frac{1}{2 N_c} \, \mbox{Re} \, \llangle \tord \tr \left[  V_{\underline{\zeta}} V_{\un{w}}^{i \, \textrm{pol} \: \dagger } \right] + \tord \tr \left[  V_{\underline{w}} V_{\un{\zeta}}^{i \, \textrm{pol} \: \dagger } \right] \rrangle (z) ,  \label{Fidef} \\
& F^{[2]}_{\un{w}, \un{\zeta}} (z) \equiv \frac{1}{2 N_c} \, \mbox{Im} \, \llangle \tord \tr \left[  V_{\underline{\zeta}} V_{\un{w}}^{[2] \, \textrm{pol} \: \dagger } \right] - \tord \tr \left[  V_{\underline{w}} V_{\un{\zeta}}^{[2] \, \textrm{pol} \: \dagger } \right] \rrangle (z).  \label{F2def} 
\end{align}
\end{subequations}
Apart from the transverse positions of the Wilson lines, the amplitudes depend on the longitudinal momentum fraction $z$, which can be roughly thought of as the smallest of the momentum fractions of the quark and anti-quark lines (see \cite{Kovchegov:2015pbl,Kovchegov:2021lvz} for a more precise definition of the argument $z$). Note that 
\begin{align}\label{Fsymm}
F^i_{\un{w}, \un{\zeta}} (z) = F^i_{\un{\zeta}, \un{w}} (z), \ \ \ F^{[2]}_{\un{w}, \un{\zeta}} (z) = - F^{[2]}_{\un{\zeta}, \un{w}} (z).
\end{align}
Employing the definitions \eqref{Fdef} along with \eq{Fsymm} we rewrite \eq{Sivers_sub_eik4} as
\begin{align}
\label{Sivers_sub_eik5}
- \frac{\un{k} \cross \underline{S}_P}{M_P} & \, f_{1 \: T}^{\perp \: q} (x,k_T^2) \Big|_\textrm{sub-eikonal}  \\ & 
=  \frac{8 N_c}{(2 \pi)^3} \int \dd[2]{\zeta_{\perp}}  \dd[2]{w_{\perp}} \frac{\dd[2]{k_{1 \perp}}}{(2\pi)^3 } \, e^{i (\un{k} + \underline{k}_1) \vdot (\un{w} - \un{\zeta}) } \, \frac{{\un k}_1 \cdot \un{k} }{\underline{k}_1^2 \, \underline{k}^2} \int\limits_\frac{\Lambda^2}{s}^1 \frac{dz}{z}   \, \left[ ({k}_1 - {k})^i  \, F^i_{\un{w}, \un{\zeta}} (z)  + i \, F^{[2]}_{\un{w}, \un{\zeta}} (z) \right]. \notag
\end{align}
The DLA small-$x$ evolution of a sub-eikonal operator only couples it to sub-eikonal operators at the next step, otherwise the evolution would not generate longitudinal logarithms of energy. Hence, the small-$x$ evolution of the operator(s) in Eqs.~\eqref{V_subeik_4} and \eqref{Vi2} (or, equivalently, the evolution for the polarized dipole amplitudes in Eqs.~\eqref{Fdef}) will only couple to the same operators.

%%%%%%%%%%%%%%%%%%%%%%%%%%%%%%%%%%%%%%%%%%%%%%%%%%%%%%%%%%%%%

\subsubsection{Initial Conditions}

For a single quark target, the lowest-order gluon field dependent on the transverse polarization of the quark at $\un{b}=0, \, b^- =0$ and an arbitrary $b^+$ position in $A^- =0$ gauge is
\begin{subequations}\label{AplusAperp}
\begin{align}
A^{a +} (x) = & \, i \frac{g \, t^a}{4 \pi} \, \frac{M_P}{p_1^+} \, X \delta_{X,X'} \, \left[ 1 + \frac{i}{2 p_1^+} \, \pd^+  \right] \pd^+ \delta(x^-) \, e^{-i (x^+ - b^+) \frac{\nabla_\perp^2}{2 p_1^+}} \, \un{S} \times \un{x} \left[ 2 \ln \left( \frac{1}{x_{\perp} \Lambda } \right) - 1 \right], \\
A^{a i} (x) = & \, i \frac{g \, t^a}{2 \pi} \, \frac{M_P}{p_1^+} \, \left[ \delta(x^-) + \frac{i}{2 p_1^+} \, \pd^+ \delta(x^-) \right]  \, X \delta_{X,X'} \, \epsilon^{kj} S^k \, e^{-i (x^+ - b^+) \frac{\nabla_\perp^2}{2 p_1^+}} \left[ \frac{x^i x^j}{x_\perp^2} - \delta^{ij} \, \ln \left( \frac{1}{x_{\perp} \Lambda } \right) \right] \notag \\
& - \frac{g \, t^a}{2 \pi} \, \frac{M_P}{(p_1^+)^2} \, X \delta_{X,X'} \, \epsilon^{ij} S^j \, \ln \left( \frac{1}{x_{\perp} \Lambda } \right) \,  \pd^+ \delta(x^-) , \label{Aperp}
\end{align} 
\end{subequations}
where $X$ and $X'$ are the quark's polarizations before and after the gluon field emission, respectively, and $M_P$ is the ``quark" mass. This field is indeed sub-eikonal ($\sim 1/p_1^+$) at the leading order and may contribute to the operators in Eqs.~\eqref{Vi2}. Note that it does not contribute to the $F^{12}$ sub-eikonal term in \eq{O24}. Hence, the transverse spin dependence at the sub-eikonal level only contributes to the gluon operator in \eq{V_subeik_1}, and, consequently, the operators in Eqs.~\eqref{V_subeik_4} and \eqref{Vi2}. 

Note that the field \eqref{AplusAperp} depends on $x^+$ and on the rapidly varying $b^+$ (with $b^+$ even not specified for a quasi-classical target \cite{McLerran:1993ni,McLerran:1993ka,McLerran:1994vd}). However, this dependence only appears at the sub-sub-eikonal order, and, even at that order, it does not affect the field strength tensor $F^{-i}$ needed for the transversity operator in \eq{O24}. The $x^+$-dependence may appear in any gluon field if we take into account the energy-suppressed phase, as in \eq{AplusAperp}. However, at the leading orders in eikonality in each channel, in the operators entering \eq{O24}, the $x^+$-dependence does not appear.  

Another puzzling feature of the field in Eqs.~\eqref{AplusAperp} is that the sub-eikonal ($\sim 1/p_1^+$) terms in it appear to come in with a factor of $i$, that is, they are imaginary. This feature, while requiring further interpretation in the future, can be attributed to the conventional wisdom that the transverse spin dependence, $X \delta_{X,X'}$, usually comes in with a factor of $i$ associated with it. This is the well-known $i$ which needs a complex phase to give a real contribution to the Sivers function \cite{Qiu:1991pp,Brodsky:2002cx,Brodsky:2002rv,Brodsky:2013oya}. Here it comes in through the transverse spin-dependent part of the gluon field. 

To construct the initial conditions for our evolution, we will work with the single quark target. Substituting the field from \eq{Aperp} along with the polarization-independent eikonal field 
\begin{align}\label{Aeik}
A^{a +} (x) = - \frac{g \, t^a}{\pi} \, \delta(x^-) \, \ln \left( \frac{1}{x_{\perp} \Lambda } \right)
\end{align}
into \eq{Fidef} we immediately see that the two-gluon exchange contribution vanishes, and the first non-trivial polarized dipole amplitudes arise at the three-gluon level. We obtain, after integrating over the impact parameters $\un{b}$ between the dipole and the target quark,
\begin{align}\label{Fi0}
\int d^2 b_\perp \, F^{i \, (0)}_{\un{w}, \un{\zeta}} = - \as^3 \, c_0 \, N_c \, (2 \pi)^2 \, M_P \, \epsilon^{ij} S_P^j \, (\un{w} - \un{\zeta})^2 \ln \left( \frac{1}{|\un{w} - \un{\zeta}| \Lambda} \right) ,
\end{align}
with $c_0$ given by \eq{c0} above. Note that the three gluons have to be in the $d^{abc}$ color state, similar to the odderon. One may, therefore, think of \eq{Fi0} as of the lowest-order contribution to the sub-eikonal spin-dependent odderon. 

Similar calculation of the initial condition for $F^{[2]}_{\un{w}, \un{\zeta}}$ from \eq{F2def} due to the fields in Eqs.~\eqref{Aperp} and \eqref{Aeik} of a single quark target readily gives zero in the gluon sector to any order in the gluon exchanges. There is also a quark sector, as one can see from \eq{V2}. However, the quark part of the operator in \eq{V2} contains only the $\gamma^+$ Dirac matrix and, like other operators in Eqs.~\eqref{Vi2}, is local in the transverse plane: therefore, this operator is similar to that in the unpolarized quark parton distribution function (PDF), and cannot couple to the transverse spin of the target proton. Therefore, it appears that one can discard the quark part of the operator \eqref{V2}, both in the initial conditions and in evolution.

This can also be seen by the order-by-order evaluation of the quark part of $F^{[2]}_{\un{w}, \un{\zeta}}$. At the lowest order, the quark operator in \eq{V2} comes in with a two-quark exchange in the $t$-channel: this contribution to $F^{[2]}_{\un{w}, \un{\zeta}}$ in \eq{F2def} is zero, since, after averaging over the quark impact parameters, each trace in \eq{F2def} will become transverse coordinate independent and the difference of the two traces will be zero. Adding eikonal gluon exchanges will make each trace in \eq{F2def} a function of $\un{w} - \un{\zeta}$. However, since the quark operator in \eq{V2} is local in the transverse plane and comes in with $\gamma^+$ only, it will not generate any dependence on the transverse spin, such that each trace in \eq{F2def} will be a function of $|\un{w} - \un{\zeta}|$, and their difference will again cancel. 

We thus conclude that the initial condition for the second polarized dipole amplitude is zero, 
\begin{align}\label{F20}
F^{[2] (0)}_{\un{w}, \un{\zeta}} = 0. 
\end{align}
Below we will show that the DLA evolution of $F^{[2]}_{\un{w}, \un{\zeta}}$ is closed in the gluon sector, it does not mix with $F^i_{\un{w}, \un{\zeta}}$: therefore, the zero initial conditions \eqref{F20} imply that $F^{[2]}_{\un{w}, \un{\zeta}} = 0$ even after evolution.

%%%%%%%%%%%%%%%%%%%%%%%%%%%%%%%%%%%%%%%%%%%%%%%%%%%%%%%%%%%%%%%%%%%%%%%%%%%%%%%%%%%

\subsubsection{Small-$x$ Evolution: General Expression}

The small-$x$ evolution of the polarized dipoles in Eqs.~\eqref{Fdef} will ultimately be calculated in the large-$N_c$ limit. We will, therefore, assume that the evolution is gluon-driven and neglect the quark field insertion operator in the second line of \eq{V2}. (The same approximation was done for the large-$N_c$ limit of the small-$x$ helicity evolution in \cite{Kovchegov:2015pbl}: see also a discussion of this approximation in \cite{Kovchegov:2020hgb}.) Above we have argued that the initial condition for the entire $F^{[2]}_{\un{w}, \un{\zeta}}$ is zero: the only way it may become non-zero is by mixing with $F^i_{\un{w}, \un{\zeta}}$ through evolution. The mixing of the quark part of \eqref{V2} in $F^{[2]}_{\un{w}, \un{\zeta}}$ with $F^i_{\un{w}, \un{\zeta}}$, even if non-zero, has to include an (eikonal) interaction of the unpolarized (anti-)quark in the dipole with the target, otherwise the contribution to the dipole amplitude $F^{[2]}_{\un{w}, \un{\zeta}}$ from \eq{F2def} will be zero, since the latter has to be an odd function under $\un{w} \leftrightarrow \un{\zeta}$ interchange, and, therefore, has to be a function of both transverse positions $\un{w}$ and  $\un{\zeta}$. An interaction of the unpolarized (anti-)quark in the original parent dipole with the target would make the quark operator evolution sub-leading in $N_c$ and can be discarded. Therefore, we will neglect the quark operator from the second line of \eq{V2} in our evolution analysis below.

To construct the small-$x$ evolution of the operators in Eqs.~\eqref{Vi2} we follow \cite{Balitsky:1995ub} and rewrite the gluon field as a sum of the background field $B^\mu$ and the quantum field $a^\mu$,
\begin{align}\label{decomp}
A^\mu = B^\mu + a^\mu,
\end{align}
and integrate out the quantum fields $a^\mu$. To do so, we will need the propagator 
\begin{align}\label{plus_perp_prop}
\int\limits_{-\infty}^0 dx_1^- 
\int\limits_0^\infty dx_2^- \,
\contraction[2ex]
{}{a_{\bot}^{i \, a}  \:}
{(x_1^- , \un{x}_1)  \:}
{a^{+ \, b}}
a_{\bot}^{i \, a} (x_1^- , \un{x}_1)  \:
a^{+ \, b} (x_2^- , \un{x}_0) = \frac{i}{2 \pi^3} \, \int d k^- \, \theta (k^-) \, \int d^2 x_2 \, d^2 x_{2'} \, \ln \left( \frac{1}{x_{21} \Lambda } \right) \, \frac{x_{2'0}^i}{x_{2'0}^2} \, U^{\textrm{pol} \, ba}_{2',2}
\end{align} 
with the sub-eikonal polarized gluon Wilson line operator (cf. Eqs.~\eqref{Ginsertion} and \eqref{V_subeik_3})
\begin{align}\label{Upol}
& U^{\textrm{pol} \, ba}_{\un{x}, \un{y}} = - \frac{i p_1^+}{2 s} \int\limits_{-\infty}^{\infty} \dd{z}^- d^2 z \ U^{bb'}_{\un{x}} [ \infty, z^-] \, \delta^2 (\un{x} - \un{z}) \, \cev{\underline{\mathscr{D}}}^{b'c}_z \cdot  \underline{\mathscr{D}}^{ca'}_z \, U^{a'a}_{\un{y}} [ z^-, -\infty] \, \delta^2 (\un{y} - \un{z}) \\ 
& = - \frac{i p_1^+\, g}{2 s} f^{b'da'} \int\limits_{-\infty}^{\infty} \dd{z}^- d^2 z \ U^{bb'}_{\un{x}} [ \infty, z^-] \, \delta^2 (\un{x} - \un{z}) \, \left[ \underline{A}^d (z^-, \un{z})  \cdot \vec{\un{\nabla}}_z - \cev{\un{\nabla}}_z \cdot \underline{A}^d (z^-, \un{z}) \right] \, U^{a'a}_{\un{y}} [ z^-, -\infty] \, \delta^2 (\un{y} - \un{z}) \notag \\
& = - \frac{i p_1^+\, g}{2 s} f^{b'da'} \, \left[ \un{\nabla}_x \delta^2 (\un{x} - \un{y}) \right] \cdot  \int\limits_{-\infty}^{\infty} \dd{z}^- \ U^{bb'}_{\un{x}} [ \infty, z^-]  \, \left[ \underline{A}^d (z^-, \un{x}) + \underline{A}^d (z^-, \un{y}) \right] \, U^{a'a}_{\un{y}} [ z^-, -\infty] . \notag
\end{align}
The propagator is obtained by the technique outlined in \cite{Kovchegov:2017lsr}, employing the $\delta_{\lambda, \lambda'}$ part of the the operator in \eq{Ginsertion} to describe the interaction with the shock wave. While, strictly speaking, the gluon field on the right-hand side of \eq{Upol} should be $B^\mu$, we denote it $A^\mu$ since at the next step of evolution it will again be separated into the background and quantum fields.

Substituting \eq{Upol} with $s = p_1^+ k^-$ into \eq{plus_perp_prop} and integrating over $\un{x}_{2'}$ yields
\begin{align}\label{plus_perp_prop2}
& \int\limits_{-\infty}^0 dx_1^- 
\int\limits_0^\infty dx_2^- \,
\contraction[2ex]
{}{a_{\bot}^{i \, a}  \:}
{(x_1^- , \un{x}_1)  \:}
{a^{+ \, b}}
a_{\bot}^{i \, a} (x_1^- , \un{x}_1)  \:
a^{+ \, b} (x_2^- , \un{x}_0) = \frac{g}{4 \pi^3} \,  f^{b'da'} \, \int \frac{d k^-}{k^-} \, \int d^2 x_2 \, d^2 x_{2'} \, \ln \left( \frac{1}{x_{21} \Lambda } \right) \, \frac{x_{2'0}^i}{x_{2'0}^2}  \\ 
& \times \,   \int\limits_{-\infty}^{\infty} \dd{z}^- \ U^{bb'}_{\un{x}_{2'}} [ \infty, z^-]  \, \left[ - \left[ \un{\nabla}_{2} \delta^2 (\un{x}_{22'}) \right] \cdot \underline{A}^d (z^-, \un{x}_{2'}) + \left[ \un{\nabla}_{2'} \delta^2 (\un{x}_{22'}) \right] \cdot \underline{A}^d (z^-, \un{x}_{2}) \right] \, U^{a'a}_{\un{x}_{2}} [ z^-, -\infty] \notag \\ 
& = \frac{g}{4 \pi^3}  \, f^{b'da'} \, \int\limits_0^{p_2^-} \frac{d k^-}{k^-} \, \int d^2 x_2 \, \int\limits_{-\infty}^{\infty} \dd{z}^- \ \Bigg\{ \ln \left( \frac{1}{x_{21} \Lambda } \right) \, \frac{x_{20}^i}{x_{20}^2} \, \Big[ U^{bb'}_{\un{x}_2} [ \infty, z^-] \, \underline{A}^d (z^-, \un{x}_2) \cdot  \left( \un{\nabla}_2 U^{a'a}_{\un{x}_{2}} [ z^-, -\infty]  \right)   \notag \\
& - \left( \un{\nabla}_2 U^{bb'}_{\un{x}_2} [ \infty, z^-]   \right) \cdot  \underline{A}^d (z^-, \un{x}_2) \, U^{a'a}_{\un{x}_{2}} [ z^-, -\infty]  \Big] - U^{bb'}_{\un{x}_2} [ \infty, z^-] \, {A}^{j \, d} (z^-, \un{x}_2) \, U^{a'a}_{\un{x}_{2}} [ z^-, -\infty]  \notag  \\
& \times  \left[ \ln \left( \frac{1}{x_{21} \Lambda } \right) \, \frac{\delta^{ij} \, x_{20}^2 - 2 \, x_{20}^i x_{20}^j}{x_{20}^4} + \frac{x_{20}^i}{x_{20}^2} \, \frac{x_{21}^j}{x_{21}^2} \right] \Bigg\}.  \notag
\end{align} 
One will also need the standard eikonal propagator $\contraction{}{a^+ \:}{}{a^+} a^+ \:a^+$, which is the same as for the unpolarized evolution,
\begin{align}\label{plus_plus_prop}
& \int\limits_{-\infty}^0 dx_1^- 
\int\limits_0^\infty dx_2^- \,
\contraction[2ex]
{}{a^{+ \, a}  \:}
{(x_1^- , \un{x}_1)  \:}
{a^{+ \, b}}
a^{+ \, a} (x_1^- , \un{x}_1)  \:
a^{+ \, b} (x_2^- , \un{x}_0) = - \frac{1}{\pi^3} \, \int \frac{d k^-}{k^-} \, \int d^2 x_2 \, \frac{\un{x}_{21}}{x_{21}^2} \cdot  \frac{\un{x}_{20}}{x_{20}^2} \, U^{ba}_{2}.
\end{align}

%%%%%%%%%%%%%%%%%%%%%%%%%%%%%%%%%%%%%%%%%%%%%%%%%%%%%%%%%%%%%%%%%%%%%%%%%%%%%%%%
\begin{figure}[ht]
\begin{center}
\includegraphics[width= \textwidth]{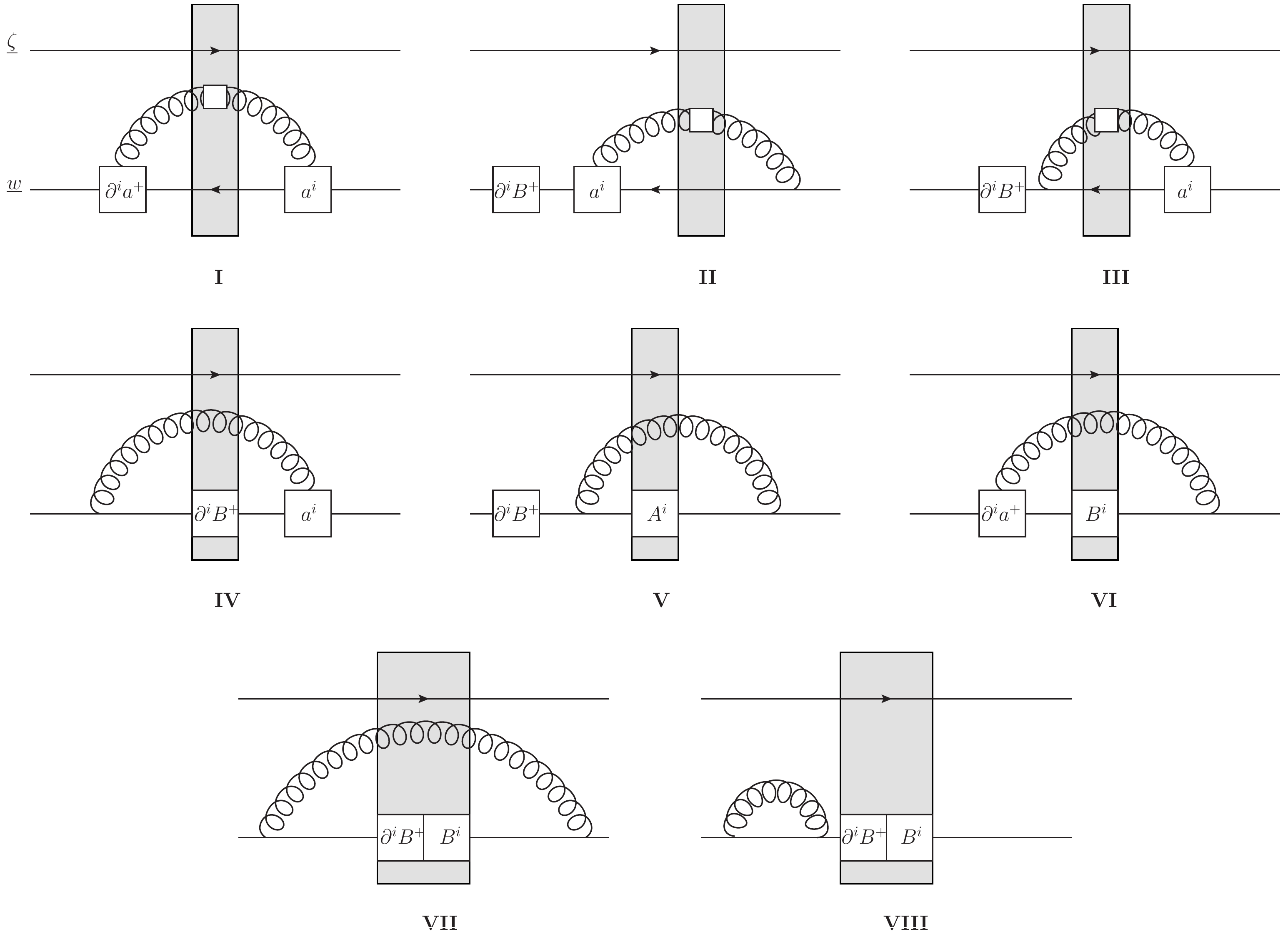} 
\caption{Diagrams illustrating the main types of contractions in the small-$x$ evolution of the polarized dipole amplitude $F^{[2]}_{\un{w}, \un{\zeta}}$ from \eqref{F2def}. Solid straight lines represent the fundamental Wilson lines, the boxes with $a^i$, $\partial^i a^+$, $B^i$, and $\partial^i B^+$ represent the operator insertions in \eq{V2glue}, while the box on the gluon line in the shock wave represents an insertion of the entire operator from \eq{plus_perp_prop2}.}
\label{FIG:V2contractions}
\end{center}
\end{figure}
%%%%%%%%%%%%%%%%%%%%%%%%%%%%%%%%%%%%%%%%%%%%%%%%%%%%%%%%%%%%%%%%%%%%%%%%%%%%%%%%

We begin with the gluon part of the $V_{\un{w}}^{[2] \, \textrm{pol}}$ from \eq{V2}, which we rewrite as
\begin{align}\label{V2glue}
V_{\un{w}}^{[2] \, \textrm{pol}} =  - \frac{g \, p_1^+}{2 \, s} \, \frac{ig}{2} \, \int\limits_{-\infty}^{\infty} \dd{z}^- \int\limits_{z^-}^{\infty} \dd{w}^- \, \Big\{ & \, V_{\un{w}} [ \infty, w^-] \, A^i (w^-, \un{w}) \, V_{\un{w}} [ w^-, z^-] \left[ {\nabla}_w^i \, A^+ (z^-, \un{w}) \right] V_{\un{w}} [ z^-, -\infty]  \\ & -  V_{\un{w}} [ \infty, w^-]  \left[ {\nabla}^i_w  \, A^+ (w^-, \un{w}) \right] V_{\un{w}} [w^-, z^-] A^i (z^-, \un{w}) \, V_{\un{w}} [ z^-, -\infty] \Big\} . \notag 
\end{align}
Substituting \eq{V2glue} into \eq{F2def} and employing the decomposition \eqref{decomp} we obtain a number of contractions. They are diagrammatically represented in \fig{FIG:V2contractions}, where only representative graphs from most diagram classes are shown and only for the first term in \eq{V2glue}, for brevity. 

Since the initial condition for $F^{[2]}_{\un{w}, \un{\zeta}}$ is zero, per \eq{F20}, to get a non-zero $F^{[2]}_{\un{w}, \un{\zeta}}$ we need to find evolution steps mixing it with $F^{i}_{\un{w}, \un{\zeta}}$. To this end we note that in the regime opposite to that in \eq{plus_perp_prop2}, that is, for $x_1^- > 0 > x_2^-$, one has
\begin{align}\label{plus_perp_prop3}
& \int\limits_0^\infty  dx_1^- 
\int\limits_{-\infty}^0 dx_2^- \,
\contraction[2ex]
{}{a_{\bot}^{i \, b}  \:}
{(x_1^- , \un{x}_1)  \:}
{a^{+ \, a}}
a_{\bot}^{i \, b} (x_1^- , \un{x}_1)  \:
a^{+ \, a} (x_2^- , \un{x}_0) = \\ 
& = \frac{g}{4 \pi^3}  \, f^{b'da'} \, \int \frac{d k^-}{k^-} \, \int d^2 x_2 \, \int\limits_{-\infty}^{\infty} \dd{z}^- \ \Bigg\{ - \ln \left( \frac{1}{x_{21} \Lambda } \right) \, \frac{x_{20}^i}{x_{20}^2} \, \Big[ U^{bb'}_{\un{x}_2} [ \infty, z^-] \, \underline{A}^d (z^-, \un{x}_2) \cdot  \left( \un{\nabla}_2 U^{a'a}_{\un{x}_{2}} [ z^-, -\infty]  \right)   \notag \\
& - \left( \un{\nabla}_2 U^{bb'}_{\un{x}_2} [ \infty, z^-]   \right) \cdot  \underline{A}^d (z^-, \un{x}_2) \, U^{a'a}_{\un{x}_{2}} [ z^-, -\infty]  \Big] - U^{bb'}_{\un{x}_2} [ \infty, z^-] \, {A}^{j \, d} (z^-, \un{x}_2) \, U^{a'a}_{\un{x}_{2}} [ z^-, -\infty]  \notag  \\
& \times  \left[ \ln \left( \frac{1}{x_{21} \Lambda } \right) \, \frac{\delta^{ij} \, x_{20}^2 - 2 \, x_{20}^i x_{20}^j}{x_{20}^4} + \frac{x_{20}^i}{x_{20}^2} \, \frac{x_{21}^j}{x_{21}^2} \right] \Bigg\}.  \notag
\end{align} 
The first term in the curly brackets of \eq{plus_perp_prop3} has a different sign compared to that in \eq{plus_perp_prop2}. The second terms are the same in both equations. Since we are interested in evolution mixing $F^{[2]}_{\un{w}, \un{\zeta}}$ with $F^{i}_{\un{w}, \un{\zeta}}$, below, when discussing the evolution of $F^{[2]}_{\un{w}, \un{\zeta}}$, we will only talk about the second term in Eqs.~\eqref{plus_perp_prop2} and \eqref{plus_perp_prop3}, which is the same in both expressions.

To give a longitudinal logarithmic integral $dk^-/k^-$, the $\perp+$ gluon propagator, like those in Eqs.~\eqref{plus_perp_prop2} and \eqref{plus_perp_prop3}, has to cross the shock wave \cite{Kovchegov:2015pbl}: hence, only the diagrams with the gluon crossing the shock wave are shown in \fig{FIG:V2contractions} for the $\perp+$ propagator. The quantum field $a^\mu$ can only be involved in contractions (see, e.g., diagram I in \fig{FIG:V2contractions}), as it is integrated out. The $B^\mu$-field outside the shock wave is put to zero. This way the diagrams II, III and V, along with other similar diagrams where we have a factor of the gluon field ``outside" the shock wave and of the gluon propagator, are all zero. One can further argue that the diagrams I, IV, and VI cancel between the two terms in \eq{V2glue}. Once again we are talking about the term in the $\perp+$ gluon propagator which is the same in Eqs.~\eqref{plus_perp_prop2} and \eqref{plus_perp_prop3}: the other terms do not cancel, but do not mix $F^{[2]}_{\un{w}, \un{\zeta}}$ with $F^{i}_{\un{w}, \un{\zeta}}$. This leaves us with the diagrams VII and VIII, where the gluon emission is of the eikonal type, similar to the unpolarized small-$x$ evolution \cite{Mueller:1994rr,Mueller:1994jq,Mueller:1995gb,Balitsky:1995ub,Balitsky:1998ya,Kovchegov:1999yj,Kovchegov:1999ua}. These diagrams represent all possible eikonal gluon emissions and absorptions. Eikonal gluons do not have to cross the shock wave to generate longitudinal logarithms: hence, diagram VIII is allowed, along with other virtual diagrams. The eikonal evolution leaves the operator \eqref{V2glue} intact in the shock wave, as denoted by the two adjacent boxes inside the shock wave shown in the diagrams VII and VIII: again, no mixing with $F^{i}_{\un{w}, \un{\zeta}}$ is generated. Summarizing the analysis we have just made, we see that the DLA evolution for the dipole amplitude $F^{[2]}_{\un{w}, \un{\zeta}}$ in the large-$N_c$ limit couples only to the same amplitudes $F^{[2]}_{\un{w}, \un{\zeta}}$ and does not mix with $F^{i}_{\un{w}, \un{\zeta}}$. That is, the evolution is of the following type,
\begin{align}\label{F2evol}
F^{[2]}_{\un{w}, \un{\zeta}} = F^{[2] (0)}_{\un{w}, \un{\zeta}} + K \otimes F^{[2]}_{\un{w}, \un{\zeta}},
\end{align}
with $K$ some integral kernel. Since, by \eq{F20}, the initial condition (the inhomogeneous term in \eq{F2evol}) for this evolution is zero, $F^{[2] (0)}_{\un{w}, \un{\zeta}}=0$, this means that 
\begin{align}\label{F2=0}
F^{[2]}_{\un{w}, \un{\zeta}} =0
\end{align}
with the DLA and large-$N_c$ accuracy. We thus discard the dipole amplitude $F^{[2]}_{\un{w}, \un{\zeta}}$ and proceed by constructing the evolution for $F^i_{\un{w}, \un{\zeta}}$. 

%%%%%%%%%%%%%%%%%%%%%%%%%%%%%%%%%%%%%%%%%%%%%%%%%%%%%%%%%%%%%%%%%%%%%%%%%%%%%%%%
\begin{figure}[ht]
\begin{center}
\includegraphics[width= 0.9 \textwidth]{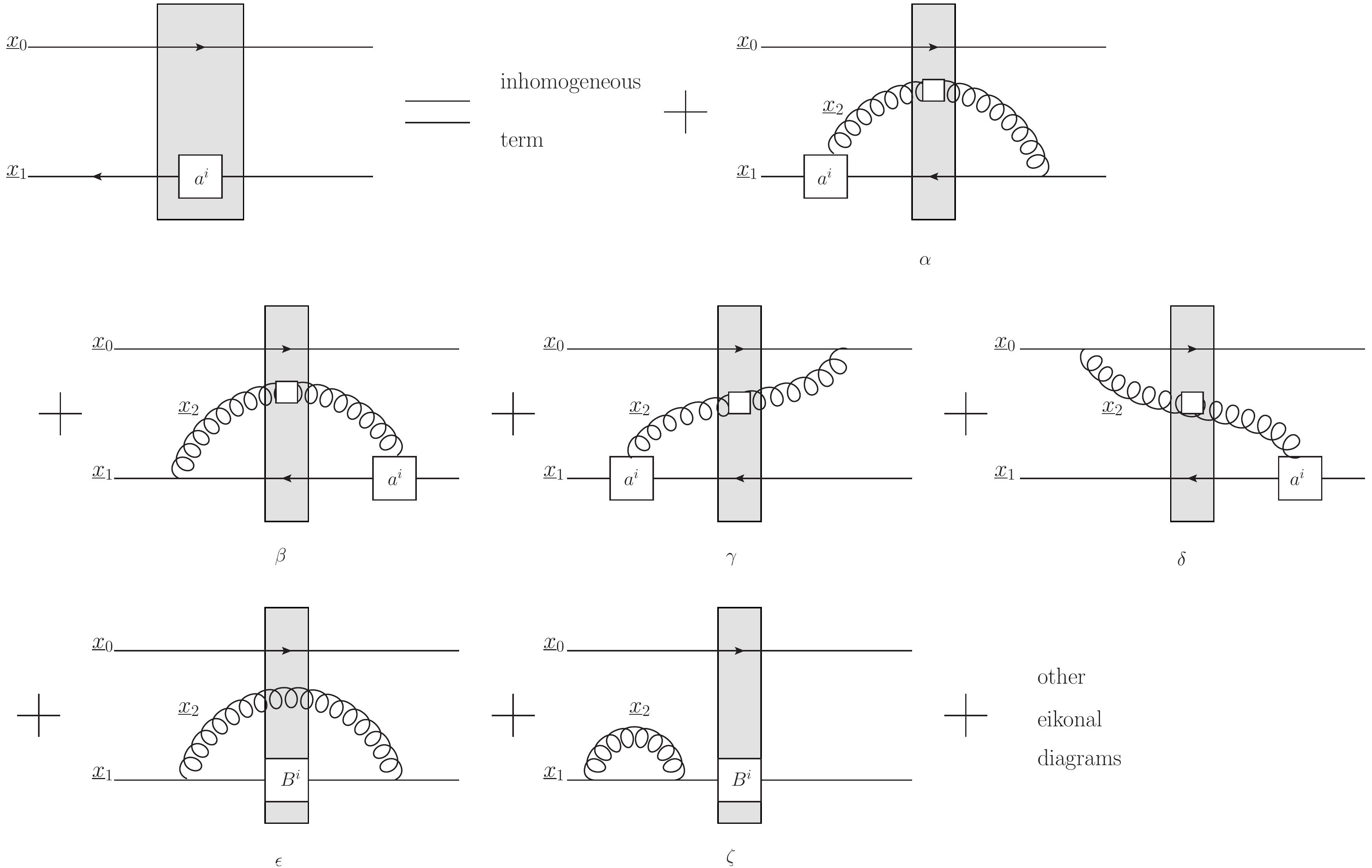} 
\caption{Diagrams illustrating the main types of contractions in the small-$x$ evolution of the polarized dipole amplitude $F^{i}_{\un{w}, \un{\zeta}}$ from \eqref{Fidef}. Solid straight lines represent the fundamental Wilson lines, the boxes with $a^i$ and $B^i$ represent the operator insertions in \eq{Vi}, while the box on the gluon line in the shock wave represents an insertion of the entire operator \eqref{Vi}.}
\label{FIG:Vi_contractions}
\end{center}
\end{figure}
%%%%%%%%%%%%%%%%%%%%%%%%%%%%%%%%%%%%%%%%%%%%%%%%%%%%%%%%%%%%%%%%%%%%%%%%%%%%%%%%

The diagrams contributing to the evolution of the polarized dipole amplitude $F^{i}_{\un{w}, \un{\zeta}}$ from \eqref{Fidef} are shown in \fig{FIG:Vi_contractions}. Again, we do not show all the eikonal diagrams explicitly. For brevity, we also depict the diagrams contributing only to one of the traces in \eq{Fidef}. We begin by calculating the sub-eikonal diagrams $\alpha, \beta, \gamma$, and $\delta$ in \fig{FIG:Vi_contractions}.

By analogy to \eq{Vi}, we define an adjoint polarized Wilson line of the same type by
\begin{align}
U_{\un{w}}^{i \, \textrm{pol}} \equiv \frac{i g \, p_1^+}{2 \, s} \, \int\limits_{-\infty}^{\infty} \dd{z}^- \, U_{\un{w}} [ \infty, z^-] \, {\cal A}^i (z^-, \un{w}) \, U_{\un{w}} [ z^-, -\infty] . \label{Ui}
\end{align}
With the help of this definition, repeating the steps outlines in more detail in \cite{Kovchegov:2017lsr,Kovchegov:2018znm}, we arrive at the following contributions of the diagrams $\alpha$ and $\beta$
\begin{align}\label{alpha_beta}
\alpha + \beta = - \frac{\as \, N_c}{2 \pi^2} \int\limits_{\frac{\Lambda^2}{s}}^z \frac{d z'}{z'} \, \int d^2 x_2 \, & \,  \left[ \ln \left( \frac{1}{x_{21} \Lambda } \right) \, \frac{\delta^{ij} \, x_{21}^2 - 2 \, x_{21}^i x_{21}^j}{x_{21}^4} + \frac{x_{21}^i}{x_{21}^2} \, \frac{x_{21}^j}{x_{21}^2} \right] \\ 
& \times \mbox{Re} \, \llangle \frac{1}{N_c^2} \, \tord \tr \left[ V_{\un{0}} t^a V_{\un{1}}^\dagger t^b \right] \, \left( U_{\un{2}}^{j \, \textrm{pol}} \right)^{ba} \rrangle (z') + (1 \leftrightarrow 0). \notag
\end{align}
Similarly, diagrams $\gamma$ and $\delta$ give
\begin{align}\label{gamma_delta}
\gamma + \delta =  \frac{\as \, N_c}{2 \pi^2} \int\limits_{\frac{\Lambda^2}{s}}^z \frac{d z'}{z'} \, \int d^2 x_2 \, & \, \left[ \ln \left( \frac{1}{x_{21} \Lambda } \right) \, \frac{\delta^{ij} \, x_{20}^2 - 2 \, x_{20}^i x_{20}^j}{x_{20}^4} + \frac{x_{20}^i}{x_{20}^2} \, \frac{x_{21}^j}{x_{21}^2} \right] \\ 
& \times \mbox{Re} \, \llangle \frac{1}{N_c^2} \,  \tord \tr \left[ V_{\un{0}} t^a V_{\un{1}}^\dagger t^b \right] \, \left( U_{\un{2}}^{j \, \textrm{pol}} \right)^{ba} \rrangle (z') - (1 \leftrightarrow 0). \notag
\end{align}
In obtaining Eqs.~\eqref{alpha_beta} and \eqref{gamma_delta} we have defined $z' = k^-/p_2^-$ with the upper cutoff on the $z'$ integral given by $z$, the minus momentum fraction at the previous step of the evolution. The lower limit of the $z'$ integral, $\Lambda^2/s$, involves an IR cutoff $\Lambda$ for the transverse momenta. We have also been using an abbreviated notation for the light-cone Wilson lines $V_{\un{1}} = V_{\un{x}_1}, U_{\un{2}} = U_{\un{x}_2}$, etc. Transverse vectors are defined by ${\un x}_{ij} = \un{x}_i - \un{x}_j$ with $x_{ij} = |\un{x}_{ij}|$. Most importantly, in deriving Eqs.~\eqref{alpha_beta} and \eqref{gamma_delta} we have neglected the first term in the propagator \eqref{plus_perp_prop2} as being the gluon analogue of the (gluon part of the) operator \eqref{V2}, which we have established to be zero, per \eq{F2=0}.

The eikonal diagrams contribution is well-known. The diagrams $\epsilon$, $\zeta$, etc., yield \cite{Mueller:1994rr,Mueller:1994jq,Mueller:1995gb,Balitsky:1995ub,Balitsky:1998ya,Kovchegov:1999yj,Kovchegov:1999ua,Kovchegov:2017lsr,Kovchegov:2018znm}
\begin{align}\label{eik_diag}
\epsilon + \zeta + \ldots = &  \, \frac{\as \, N_c}{2 \pi^2} \int\limits_{\frac{\Lambda^2}{s}}^z \frac{d z'}{z'} \, \int d^2 x_2 \, \frac{x_{10}^2}{x_{21}^2 \, x_{20}^2} \\ 
& \times \, \mbox{Re} \, \llangle \frac{1}{N_c^2} \,  \tord \tr \left[ V_{\un{0}} t^a V_{\un{1}}^{i \, \textrm{pol} \, \dagger} t^b \right] \, \left( U_{\un{2}} \right)^{ba} - \frac{C_F}{N_c^2} \,  \tord \tr \left[ V_{\un{0}} \,  V_{\un{1}}^{i \, \textrm{pol} \, \dagger} \right] \rrangle (z') + (1 \leftrightarrow 0). \notag 
\end{align}

Combining equations \eqref{alpha_beta}, \eqref{gamma_delta} and \eqref{eik_diag} we arrive at the evolution equation for the dipole amplitude $F^i_{10}$ in the operator form,
\begin{align}\label{Fi_evol}
F^i_{10} (z) = & \, F^{i \, (0)}_{10} (z) + \frac{\as \, N_c}{2 \pi^2} \int\limits_{\frac{\Lambda^2}{s}}^z \frac{d z'}{z'}  \int d^2 x_2 \, \frac{x_{10}^2}{x_{21}^2 \, x_{20}^2}  \, \mbox{Re} \, \llangle \frac{1}{N_c^2} \, \tr \left[ V_{\un{0}} t^a V_{\un{1}}^{i \, \textrm{pol} \, \dagger} t^b \right] \, \left( U_{\un{2}} \right)^{ba} - \frac{C_F}{N_c^2} \, \tr \left[ V_{\un{0}} \,  V_{\un{1}}^{i \, \textrm{pol} \, \dagger} \right] \rrangle (z') \notag  \\ 
& - \frac{\as \, N_c}{2 \pi^2} \int\limits_{\frac{\Lambda^2}{s}}^z \frac{d z'}{z'} \, \int d^2 x_2 \,  \left[ \ln \left( \frac{1}{x_{21} \Lambda } \right) \, \left( \frac{\delta^{ij} \, x_{21}^2 - 2 \, x_{21}^i x_{21}^j}{x_{21}^4} - \frac{\delta^{ij} \, x_{20}^2 - 2 \, x_{20}^i x_{20}^j}{x_{20}^4} \right)  + \left( \frac{x_{21}^i}{x_{21}^2}  - \frac{x_{20}^i}{x_{20}^2}  \right) \, \frac{x_{21}^j}{x_{21}^2} \right] \notag \\ 
& \times \mbox{Re} \, \llangle  \frac{1}{N_c^2} \, \tr \left[ V_{\un{0}} t^a V_{\un{1}}^\dagger t^b \right] \, \left( U_{\un{2}}^{j \, \textrm{pol}} \right)^{ba} \rrangle (z') + (1 \leftrightarrow 0), 
\end{align}
with the inhomogeneous term given by \eq{Fi0} (for the impact-parameter integrated version of \eq{Fi_evol}). Here $+ (1 \leftrightarrow 0)$ applies to everything on the right, except for the inhomogeneous term. We have also dropped the time-ordering signs for brevity: they are still implied in all correlation functions. 
  
Combining Eqs.~\eqref{Fi_evol} and \eqref{F2=0} with \eq{Sivers_sub_eik5} we write the sub-eikonal contribution to the Sivers function as  
\begin{align}
\label{Sivers_sub_eik7}
- \frac{\un{k} \cross \underline{S}_P}{M_P}  f_{1 \: T}^{\perp \: q} (x,k_T^2) \Big|_\textrm{sub-eikonal}  
=  \frac{4 N_c}{(2 \pi)^3} \int \dd[2]{\zeta_{\perp}}  \dd[2]{w_{\perp}} \frac{\dd[2]{k_{1 \perp}}}{(2\pi)^3}  \, e^{i (\un{k} + \underline{k}_1) \vdot (\un{w} - \un{\zeta}) } \, \frac{{\un k}_1 \cdot \un{k} }{\underline{k}_1^2 \, \underline{k}^2} \, ({k}_1 - {k})^i \int\limits_\frac{\Lambda^2}{s}^1 \frac{dz}{z}   \, F^i_{\un{w}, \un{\zeta}} (z)  . 
\end{align}

%%%%%%%%%%%%%%%%%%%%%%%%%%%%%%%%%%%%%%%%%%%%%%%%%%%%%%%%%%%%%%%%%%%%%%%%%%%%%%%%%%%

\subsubsection{Small-$x$ Evolution in the Large-$N_c$ Limit}  
  
Similar to the unpolarized evolution \cite{Balitsky:1995ub,Balitsky:1998ya,Kovchegov:1999yj,Kovchegov:1999ua,Jalilian-Marian:1997dw,Jalilian-Marian:1997gr,Weigert:2000gi,Iancu:2001ad,Iancu:2000hn,Ferreiro:2001qy} the evolution equation \eqref{Fi_evol} is not closed: not all the operators on its right-hand side are the same as the dipole operator on its left-hand side (see also \eqref{Fidef}). To obtain a closed evolution equation we will take the large-$N_c$ limit. Employing
 \begin{align}\label{UVV1}
U_{\underline{x}}^{ba} = 2 \tr \left[t^bV_{\underline{x}}t^aV_{\underline{x}}^{\dagger}\right],
\end{align}
along with the Fierz identity, one can readily show that (cf. \cite{Kovchegov:2018znm})
\begin{align}\label{UVV2}
U_{\underline{x}}^{i \, \textrm{pol} \, ba} = 2 \tr \left[ t^b V_{\underline{x}}^{i \, \textrm{pol}} t^a V_{\underline{x}}^{\dagger} \right] + 2 \tr \left[ t^b V_{\underline{x}}  t^a V_{\underline{x}}^{i \, \textrm{pol} \, \dagger} \right].
\end{align}

Using Eqs.~\eqref{UVV1} and \eqref{UVV2} in \eq{Fi_evol} we obtain 
\begin{align}\label{Fi_evol2}
F^i_{10} (z) = & \, F^{i \, (0)}_{10} (z) + \frac{\as \, N_c}{2 \pi^2} \int\limits_{\frac{\Lambda^2}{s}}^z \frac{d z'}{z'}  \int d^2 x_2 \, \Bigg\{  \frac{x_{10}^2}{x_{21}^2 \, x_{20}^2}  \, \mbox{Re} \, \llangle \frac{1}{2 N_c^2} \, \tr \left[ V_{\un{2}}  V_{\un{1}}^{i \, \textrm{pol} \, \dagger}  \right] \, \tr \left[ V_{\un{0}} \, V_{\un{2}}^\dagger \right]  - \frac{1}{2 N_c} \, \tr \left[ V_{\un{0}} \,  V_{\un{1}}^{i \, \textrm{pol} \, \dagger} \right] \rrangle (z') \notag  \\ 
& -  \left[ \ln \left( \frac{1}{x_{21} \Lambda } \right) \, \left( \frac{\delta^{ij} \, x_{21}^2 - 2 \, x_{21}^i x_{21}^j}{x_{21}^4} - \frac{\delta^{ij} \, x_{20}^2 - 2 \, x_{20}^i x_{20}^j}{x_{20}^4} \right)  + \left( \frac{x_{21}^i}{x_{21}^2}  - \frac{x_{20}^i}{x_{20}^2}  \right) \, \frac{x_{21}^j}{x_{21}^2} \right] \notag \\ 
& \times \mbox{Re} \, \llangle  \frac{1}{2 N_c^2} \, \tr \left[ V_{\un{0}} \, V_{\un{2}}^{\dagger} \right] \, \tr \left[ V_{\un{2}}^{j \, \textrm{pol}} \, V_{\un{1}}^\dagger \right]  + \frac{1}{2 N_c^2} \, \tr \left[ V_{\un{2}} \, V_{\un{1}}^\dagger \right] \, \tr \left[ V_{\un{0}} \, V_{\un{2}}^{j \, \textrm{pol} \, \dagger} \right] + \ldots \rrangle (z') \Bigg\} + (1 \leftrightarrow 0), 
\end{align} 
where the ellipsis denote the $N_c$-suppressed terms. Taking the large-$N_c$ limit yields
\begin{align}\label{Fi_evol3}
F^i_{10} (z) & \, = F^{i \, (0)}_{10} (z) + \frac{\as \, N_c}{2 \pi^2} \int\limits_{\frac{\Lambda^2}{s}}^z \frac{d z'}{z'}  \int d^2 x_2 \, \Bigg\{  \frac{x_{10}^2}{x_{21}^2 \, x_{20}^2}  \, \frac{1}{2}  \mbox{Re} \, \Bigg[ \llangle \frac{1}{N_c} \, \tr \left[ V_{\un{2}}  V_{\un{1}}^{i \, \textrm{pol} \, \dagger}  \right] \rrangle \, \llangle \frac{1}{N_c} \, \tr \left[ V_{\un{0}} \, V_{\un{2}}^\dagger \right] \rrangle \\ 
& - \llangle \frac{1}{N_c} \, \tr \left[ V_{\un{0}} \,  V_{\un{1}}^{i \, \textrm{pol} \, \dagger} \right] \rrangle \Bigg] -  \left[ \ln \left( \frac{1}{x_{21} \Lambda } \right) \, \left( \frac{\delta^{ij} \, x_{21}^2 - 2 \, x_{21}^i x_{21}^j}{x_{21}^4} - \frac{\delta^{ij} \, x_{20}^2 - 2 \, x_{20}^i x_{20}^j}{x_{20}^4} \right)  + \left( \frac{x_{21}^i}{x_{21}^2}  - \frac{x_{20}^i}{x_{20}^2}  \right) \, \frac{x_{21}^j}{x_{21}^2} \right] \notag \\ 
& \times \frac{1}{2} \mbox{Re} \, \Bigg[ \llangle  \frac{1}{N_c} \, \tr \left[ V_{\un{0}} \, V_{\un{2}}^{\dagger} \right] \rrangle \, \llangle \frac{1}{N_c} \,  \tr \left[ V_{\un{2}}^{j \, \textrm{pol}} \, V_{\un{1}}^\dagger \right] \rrangle + \llangle \frac{1}{N_c} \, \tr \left[ V_{\un{2}} \, V_{\un{1}}^\dagger \right] \rrangle \, \llangle \frac{1}{N_c} \, \tr \left[ V_{\un{0}} \, V_{\un{2}}^{j \, \textrm{pol} \, \dagger} \right] \rrangle \Bigg] \Bigg\} + (1 \leftrightarrow 0),  \notag 
\end{align} 
where all the correlators on the right are functions of $z'$, with this dependence not shown explicitly.

We are interested in the solution of this equation outside the saturation region: therefore, we linearize it by replacing $\tr \left[ V_{\un{0}} \, V_{\un{2}}^{\dagger} \right]$ and $\tr \left[ V_{\un{2}} \, V_{\un{1}}^\dagger \right]$ by $N_c$. This gives the linearized evolution
\begin{align}\label{Fi_evol4}
F^i_{10} (z) & \, = F^{i \, (0)}_{10} (z) + \frac{\as \, N_c}{2 \pi^2} \int\limits_{\frac{\Lambda^2}{s}}^z \frac{d z'}{z'}  \int d^2 x_2 \, \Bigg\{  \frac{x_{10}^2}{x_{21}^2 \, x_{20}^2}  \, \frac{1}{2} \mbox{Re} \, \Bigg[ \llangle \frac{1}{N_c} \, \tr \left[ V_{\un{2}}  V_{\un{1}}^{i \, \textrm{pol} \, \dagger}  \right] \rrangle  - \llangle \frac{1}{N_c} \, \tr \left[ V_{\un{0}} \,  V_{\un{1}}^{i \, \textrm{pol} \, \dagger} \right] \rrangle \Bigg]  \\ 
& - \left[ \ln \left( \frac{1}{x_{21} \Lambda } \right) \, \left( \frac{\delta^{ij} \, x_{21}^2 - 2 \, x_{21}^i x_{21}^j}{x_{21}^4} - \frac{\delta^{ij} \, x_{20}^2 - 2 \, x_{20}^i x_{20}^j}{x_{20}^4} \right)  + \left( \frac{x_{21}^i}{x_{21}^2}  - \frac{x_{20}^i}{x_{20}^2}  \right) \, \frac{x_{21}^j}{x_{21}^2} \right] \notag \\ 
& \times \frac{1}{2} \mbox{Re} \, \Bigg[ \llangle \frac{1}{N_c} \,  \tr \left[ V_{\un{2}}^{j \, \textrm{pol}} \, V_{\un{1}}^\dagger \right] \rrangle + \llangle \frac{1}{N_c} \, \tr \left[ V_{\un{0}} \, V_{\un{2}}^{j \, \textrm{pol} \, \dagger} \right] \rrangle \Bigg] \Bigg\} + (1 \leftrightarrow 0).  \notag 
\end{align} 

We are interested in the double-logarithmic evolution, for which the transverse integrals in the kernel reduce to logarithms \cite{Kovchegov:2015pbl}. Whether the transverse integrals are logarithmic in the IR or in the ultraviolet (UV) depends, for most of the terms, on the transverse distance dependence of the correlators on the right-hand side of \eq{Fi_evol4}. Inspired by the inhomogeneous term \eqref{Fi0}, we will assume that
\begin{align}\label{ansatz}
\int d^2 b_\perp \,  \llangle \frac{1}{N_c} \, \tr \left[ V_{\un{0}} \,  V_{\un{1}}^{i \, \textrm{pol} \, \dagger} \right] \rrangle (z) = \epsilon^{ij} \, S_P^j \, x_{10}^2 \, F (x_{10}^2, z) , 
\end{align}
where the function $F (x_{10}^2, z)$ may include logarithms of $x_{10}^2$ or perturbatively small powers of $x_{10}^2$ (e.g. $x_{10}^{\textrm{const} \, \sqrt{\as}}$), along with the $z$-dependence, but no order-one powers of $x_{10}^2$. Note that \eq{ansatz} along with \eq{Fidef} imply that 
\begin{align}\label{FiF}
\int d^2 b_\perp \,  F^i_{10} (z) = \epsilon^{ij} \, S_P^j \, x_{10}^2 \, F (x_{10}^2, z) . 
\end{align}

For the correlators scaling as shown in \eq{ansatz} we can analyze different terms in the kernel of \eq{Fi_evol4}, extracting the logarithmic contribution. 

\begin{itemize}

\item The eikonal kernel ${x_{10}^2}/(x_{21}^2 \, x_{20}^2)$ is logarithmic in the UV when $\un{x}_2 \to \un{x}_1$ and in the IR when $x_{21} \approx x_{20} \gg x_{10}$. It is not logarithmic when $\un{x}_2 \to \un{x}_0$, since the two terms in the square brackets multiplying this kernel in \eq{Fi_evol4} cancel. We thus approximate, with the DLA accuracy and after integrating over the impact parameters,
\begin{align}\label{simpl10}
& \int d^2 x_2 \, \frac{x_{10}^2}{x_{21}^2 \, x_{20}^2}  \, \int d^2 b_\perp \frac{1}{2} \mbox{Re} \, \Bigg[ \llangle \frac{1}{N_c} \, \tr \left[ V_{\un{2}}  V_{\un{1}}^{i \, \textrm{pol} \, \dagger}  \right] \rrangle  - \llangle \frac{1}{N_c} \, \tr \left[ V_{\un{0}} \,  V_{\un{1}}^{i \, \textrm{pol} \, \dagger} \right] \rrangle \Bigg] \\ 
& \approx - \pi \int\limits_{\frac{1}{z' s}}^{x_{10}^2} \frac{d x_{21}^2}{x_{21}^2} \, \int d^2 b_\perp \frac{1}{2} \mbox{Re} \,  \llangle \frac{1}{N_c} \, \tr \left[ V_{\un{0}} \,  V_{\un{1}}^{i \, \textrm{pol} \, \dagger} \right] \rrangle +  \pi \int\limits_{x_{10}^2}^{\frac{z}{z'} x_{10}^2} d x_{21}^2 \, \frac{x_{10}^2}{x_{21}^4}  \, \int d^2 b_\perp \frac{1}{2} \mbox{Re} \, \llangle \frac{1}{N_c} \, \tr \left[ V_{\un{2}}  V_{\un{1}}^{i \, \textrm{pol} \, \dagger}  \right] \rrangle  ,  \notag 
\end{align}
where the last term on the right is logarithmic because we assume that the correlator multiplying is scales $\sim x_{21}^2$, per \eq{ansatz}. The IR (upper) limit of the integral in the second term comes from the light-cone ($x^-$-) lifetime ordering condition \cite{Kovchegov:2015pbl,Cougoulic:2019aja}, 
\begin{align}\label{lifetime}
z \, x_{10}^2 \gg z' \, x_{21}^2 , 
\end{align}
which is essential for the DLA. Similarly, the UV limit of the $x_{21}^2$-integral in the first term on the right of \eq{simpl10} comes from requiring that the emitted gluon's lifetime is longer than the extent of the shock wave, $z' \, x_{21}^2 > 1/s$. One may also think of $1/(z' s)$ as the shortest transverse distance in the problem. 

There is one important caveat left. Consider the first trace in the last line of \eq{simpl10}. It describes the amplitude in the dipole $10$. The subsequent emissions in that dipole will have their lifetimes capped by $z' \, x_{21}^2$, with $x_{21}$ being the size of the ``neighbor" dipole, not the one we continue the evolution in. We, therefore, define the  ``neighbor" dipole amplitude $\Gamma (x_{10}^2, x_{21}^2, z)$ \cite{Kovchegov:2015pbl,Kovchegov:2016zex,Cougoulic:2019aja} by
\begin{align}
\epsilon^{ij} \, S^j \, x_{10}^2 \, \Gamma (x_{10}^2, x_{21}^2, z) \equiv \int d^2 b_\perp \mbox{Re} \,  \llangle \frac{1}{N_c} \, \tr \left[ V_{\un{0}} \,  V_{\un{1}}^{i \, \textrm{pol} \, \dagger} \right] \rrangle (z; z \, x_{21}^2 ),
\end{align}
where the lifetime dependence is shown explicitly in the argument. Equation \eqref{simpl10} becomes
\begin{align}\label{simpl1}
& \int d^2 x_2 \, \frac{x_{10}^2}{x_{21}^2 \, x_{20}^2}  \, \int d^2 b_\perp \frac{1}{2} \mbox{Re} \, \Bigg[ \llangle \frac{1}{N_c} \, \tr \left[ V_{\un{2}}  V_{\un{1}}^{i \, \textrm{pol} \, \dagger}  \right] \rrangle  - \llangle \frac{1}{N_c} \, \tr \left[ V_{\un{0}} \,  V_{\un{1}}^{i \, \textrm{pol} \, \dagger} \right] \rrangle \Bigg] \\ 
& \approx \epsilon^{ij} \, S^j \, x_{10}^2 \,  \left[ - \frac{\pi}{2} \int\limits_{\frac{1}{z' s}}^{x_{10}^2} \frac{d x_{21}^2}{x_{21}^2} \, \Gamma (x_{10}^2, x_{21}^2, z') +  \frac{\pi}{2} \int\limits_{x_{10}^2}^{\frac{z}{z'} x_{10}^2} \, \frac{d x_{21}^2 }{x_{21}^2}  \,  F (x_{21}^2, z')  \right] .   \notag 
\end{align}

\item Next consider the first expression in the second term in the kernel of \eq{Fi_evol4}, that is, 
\begin{align}\label{simpl20}
& - \int d^2 x_2 \, \ln \left( \frac{1}{x_{21} \Lambda } \right) \, \left( \frac{\delta^{ij} \, x_{21}^2 - 2 \, x_{21}^i x_{21}^j}{x_{21}^4} - \frac{\delta^{ij} \, x_{20}^2 - 2 \, x_{20}^i x_{20}^j}{x_{20}^4} \right) \\ & \times \, \frac{1}{2} \mbox{Re} \, \Bigg[ \llangle \frac{1}{N_c} \,  \tr \left[ V_{\un{2}}^{j \, \textrm{pol}} \, V_{\un{1}}^\dagger \right] \rrangle + \llangle \frac{1}{N_c} \, \tr \left[ V_{\un{0}} \, V_{\un{2}}^{j \, \textrm{pol} \, \dagger} \right] \rrangle \Bigg] . \notag
\end{align}
This kernel has no UV divergences at either $\un{x}_2 \to \un{x}_1$ or $\un{x}_2 \to \un{x}_0$, since the potentially divergent terms vanish after angular averaging. (Strictly-speaking angular averaging would lead to delta-functions $\delta^2 (\un{x}_{21})$ and $\delta^2 (\un{x}_{20})$ in the kernel: however, zero-size daughter dipoles generated this way would have zero lifetimes, or, more precisely, the delta-functions would imply that $z' \, x_{21}^2 = 1/s$ or $z' \, x_{20}^2 = 1/s$, not allowing for any further DLA evolution due to impossibility of imposing lifetime ordering like \eqref{lifetime} beyond zero lifetime: such contribution may need to be added to the inhomogeneous term, but is not included in the evolution.) Note that employing \eq{ansatz} one can show that $\ln (1/\Lambda)$ vanishes in \eq{simpl20}: hence the logarithm present in the integrand does not necessarily make the result of the integration logarithmic. Lastly, to determine the IR asymptotics, $x_{21} \approx x_{20} \gg x_{10}$, one has to split \eq{simpl20}: in each term we expand in $x_{10}/x_{21}$ or $x_{10}/x_{20}$ and average over the angles of ${\un x}_{21}$ or ${\un x}_{20}$, depending on whether the impact-parameter integrated correlator depends on $x_{21}^2$ or $x_{20}^2$, respectively. We thus get (after the impact parameter integration)
\begin{subequations}
\begin{align}
- \int d^2 x_2 \, \ln \left( \frac{1}{x_{21} \Lambda } \right)  & \, \left( \frac{\delta^{ij} \, x_{21}^2 - 2 \, x_{21}^i x_{21}^j}{x_{21}^4} - \frac{\delta^{ij} \, x_{20}^2 - 2 \, x_{20}^i x_{20}^j}{x_{20}^4} \right)  \, \int d^2 b_\perp \frac{1}{2} \mbox{Re} \, \llangle \frac{1}{N_c} \,  \tr \left[ V_{\un{2}}^{j \, \textrm{pol}} \, V_{\un{1}}^\dagger \right] \rrangle \\ & = \pi \int\limits_{x_{10}^2}^{\frac{z}{z'} x_{10}^2} d x_{21}^2 \, \ln \left( \frac{1}{x_{21} \Lambda } \right) \,  \frac{x_{10}^2}{x_{21}^2} \, \epsilon^{ij} \, S^j \, F (x_{21}^2, z') ,  \notag \\
- \int d^2 x_2 \, \ln \left( \frac{1}{x_{21} \Lambda } \right)  & \, \left( \frac{\delta^{ij} \, x_{21}^2 - 2 \, x_{21}^i x_{21}^j}{x_{21}^4} - \frac{\delta^{ij} \, x_{20}^2 - 2 \, x_{20}^i x_{20}^j}{x_{20}^4} \right)  \, \int d^2 b_\perp \frac{1}{2} \mbox{Re} \, \llangle \frac{1}{N_c} \, \tr \left[ V_{\un{0}} \, V_{\un{2}}^{j \, \textrm{pol} \, \dagger} \right] \rrangle \\ & = - \pi \int\limits_{x_{10}^2}^{\frac{z}{z'} x_{10}^2} d x_{20}^2 \, \left[ \ln \left( \frac{1}{x_{20} \Lambda } \right) + 1 \right]  \,  \frac{x_{10}^2}{x_{20}^2} \, \epsilon^{ij} \, S^j \, F (x_{20}^2, z') , \notag
\end{align}
\end{subequations}
such that \eq{simpl20} in DLA approximates to 
\begin{align}\label{simpl2}
& - \int d^2 x_2 \, \ln \left( \frac{1}{x_{21} \Lambda } \right) \, \left( \frac{\delta^{ij} \, x_{21}^2 - 2 \, x_{21}^i x_{21}^j}{x_{21}^4} - \frac{\delta^{ij} \, x_{20}^2 - 2 \, x_{20}^i x_{20}^j}{x_{20}^4} \right) \, \int d^2 b_\perp \\ & \times \, \frac{1}{2} \mbox{Re} \, \Bigg[ \llangle \frac{1}{N_c} \,  \tr \left[ V_{\un{2}}^{j \, \textrm{pol}} \, V_{\un{1}}^\dagger \right] \rrangle + \llangle \frac{1}{N_c} \, \tr \left[ V_{\un{0}} \, V_{\un{2}}^{j \, \textrm{pol} \, \dagger} \right] \rrangle \Bigg] \approx - \epsilon^{ij} \, S^j \, x_{10}^2 \, \pi \int\limits_{x_{10}^2}^{\frac{z}{z'} x_{10}^2}  \, \frac{d x_{20}^2}{x_{20}^2} \,  F (x_{20}^2, z'). \notag
\end{align}

\item Finally, let us consider the last term in the kernel of \eq{Fi_evol4},
\begin{align}\label{simpl30}
- \int d^2 x_2 \,  \left( \frac{x_{21}^i}{x_{21}^2}  - \frac{x_{20}^i}{x_{20}^2}  \right) \, \frac{x_{21}^j}{x_{21}^2} \, \frac{1}{2} \mbox{Re} \, \Bigg[ \llangle \frac{1}{N_c} \,  \tr \left[ V_{\un{2}}^{j \, \textrm{pol}} \, V_{\un{1}}^\dagger \right] \rrangle + \llangle \frac{1}{N_c} \, \tr \left[ V_{\un{0}} \, V_{\un{2}}^{j \, \textrm{pol} \, \dagger} \right] \rrangle \Bigg] . 
\end{align}
This term contains a UV divergence at $\un{x}_2 \to \un{x}_1$, for the second term in the square brackets. There is no UV divergence at $\un{x}_2 \to \un{x}_0$. The contribution coming from the IR region, $x_{21} \approx x_{20} \gg x_{10}$, can be evaluated using the above technique. In the end we obtain
\begin{align}\label{simpl3}
& - \int d^2 x_2 \,  \left( \frac{x_{21}^i}{x_{21}^2}  - \frac{x_{20}^i}{x_{20}^2}  \right) \, \frac{x_{21}^j}{x_{21}^2} \, \int d^2 b_\perp \, \frac{1}{2} \mbox{Re} \, \Bigg[ \llangle \frac{1}{N_c} \,  \tr \left[ V_{\un{2}}^{j \, \textrm{pol}} \, V_{\un{1}}^\dagger \right] \rrangle + \llangle \frac{1}{N_c} \, \tr \left[ V_{\un{0}} \, V_{\un{2}}^{j \, \textrm{pol} \, \dagger} \right] \rrangle \Bigg] \\
& \approx  - \pi \int\limits_{\frac{1}{z' s}}^{x_{10}^2} \frac{d x_{21}^2}{x_{21}^2} \, \int d^2 b_\perp \frac{1}{4} \mbox{Re} \,  \llangle \frac{1}{N_c} \, \tr \left[ V_{\un{0}} \,  V_{\un{1}}^{i \, \textrm{pol} \, \dagger} \right] \rrangle +  \pi \int\limits_{x_{10}^2}^{\frac{z}{z'} x_{10}^2} d x_{21}^2 \, \frac{x_{10}^2}{x_{21}^4}  \, \int d^2 b_\perp \frac{1}{4} \mbox{Re} \, \llangle \frac{1}{N_c} \, \tr \left[ V_{\un{2}}  V_{\un{1}}^{i \, \textrm{pol} \, \dagger}  \right] \rrangle \notag \\ 
& = - \epsilon^{ij} \, S^j \, x_{10}^2 \,  \left[ \frac{\pi}{4} \int\limits_{\frac{1}{z' s}}^{x_{10}^2} \frac{d x_{21}^2}{x_{21}^2} \, \Gamma (x_{10}^2, x_{21}^2, z') -  \frac{\pi}{4} \int\limits_{x_{10}^2}^{\frac{z}{z'} x_{10}^2} \, \frac{d x_{21}^2 }{x_{21}^2}  \,  F (x_{21}^2, z')  \right] .   \notag 
\end{align}

\end{itemize}

Substituting Eqs.~\eqref{simpl1}, \eqref{simpl2}, and \eqref{simpl3}, into \eq{Fi_evol4} integrated over all impact parameters we arrive at
\begin{subequations}\label{Fi_evol5}
\begin{align}
& F (x^2_{10} , z)  = F^{(0)} (x^2_{10} , z) - \frac{\as \, N_c}{4 \pi} \int\limits_{\frac{\Lambda^2}{s}}^z \frac{d z'}{z'}  \left\{ \int\limits_{x_{10}^2}^{\frac{z}{z'} x_{10}^2} \, \frac{d x_{21}^2 }{x_{21}^2}  \,  F (x_{21}^2, z') + 3 \, \int\limits_{\frac{1}{z' s}}^{x_{10}^2} \frac{d x_{21}^2}{x_{21}^2} \, \Gamma (x_{10}^2, x_{21}^2, z') \right\} , \\
& \Gamma (x_{10}^2, x_{21}^2, z') = F^{(0)} (x^2_{10} , z') - \frac{\as \, N_c}{4 \pi} \int\limits_{\frac{\Lambda^2}{s}}^{z'} \frac{d z''}{z''}  \left\{ \int\limits_{x_{10}^2}^{\frac{z'}{z''} x_{21}^2} \, \frac{d x_{32}^2 }{x_{32}^2}  \,  F (x_{32}^2, z'')  \right. \\ 
& \left. \hspace*{8cm} + \, 3 \, \int\limits_{\frac{1}{z'' s}}^{\min \{x_{10}^2, \frac{z'}{z''} x_{21}^2 \}} \frac{d x_{32}^2}{x_{32}^2} \, \Gamma (x_{10}^2, x_{32}^2, z'') \right\}. \notag
\end{align} 
\end{subequations}
The second equation, for the ``neighbor" dipole amplitude $\Gamma$, is derived by analogy to the first one \cite{Kovchegov:2015pbl,Kovchegov:2016zex}. Each integral is non-zero only when the upper integration limit is larger than the lower one. 

The initial condition (the inhomogeneous term) of Eqs.~\eqref{Fi_evol5} can be read off from Eqs.~\eqref{Fi0} and \eqref{ansatz},
\begin{align}\label{F0init}
F^{(0)} (x^2_{10} , z) = - \as^3 \, c_0 \, (2 \pi)^2 \, M_P \,  \ln \left( \frac{1}{x_{10} \Lambda} \right) .
\end{align}

%%%%%%%%%%%%%%%%%%%%%%%%%%%%%%%%%%%%%%%%%%%%%%%%%%%%%%%%%%%%%%%%%%%%%%%%%%%%%%%%%%%

\subsubsection{Small-$x$ Sub-Eikonal Asymptotics of the Sivers function}  
  
Solution of Eqs.~\eqref{Fi_evol5}, while possible both analytically and numerically, appears to be somewhat involved. Instead we will argue that the high-energy asymptotics should not depend on the initial conditions. We therefore, replace equations \eqref{ansatz} and \eqref{FiF} by
\begin{align}\label{ansatz2}
\int d^2 b_\perp \,  F^i_{10} (z) = \int d^2 b_\perp \,  \llangle \frac{1}{N_c} \, \tr \left[ V_{\un{0}} \,  V_{\un{1}}^{i \, \textrm{pol} \, \dagger} \right] \rrangle (z) = \epsilon^{ij} \, S_P^j \, F (x_{10}^2, z) , 
\end{align}
where the function $F (x_{10}^2, z)$ again may include logarithms of $x_{10}^2$ or perturbatively small powers of $x_{10}^2$ (e.g. $x_{10}^{\textrm{const} \, \sqrt{\as}}$); however, we assume that it cannot contain integer powers of  $x_{10}^2$. Indeed the initial condition \eqref{Fi0} was derived for a quark target, that is, for a hadron so large that the dipole interacts with a single quark in it. One can instead imagine a situation where the target is perturbatively small, even smaller than the dipole in the projectile: then the initial conditions may not be proportional to the dipole size squared, like \eq{Fi0}, and the ansatz \eqref{ansatz2} would appear to be more appropriate for the initial condition, and, therefore, for the evolved dipole amplitude as well. 

With this new ansatz \eqref{ansatz2}, the DLA regime of \eq{Fi_evol4} becomes much simpler, with only the first and the last terms in the kernel contributing. We arrive at (cf. helicity evolution in \cite{Kovchegov:2015pbl,Kovchegov:2016zex})
\begin{subequations}\label{Fi_evol6}
\begin{align}
& F (x^2_{10} , z)  = F^{(0)} (x^2_{10} , z) + \frac{\as \, N_c}{4 \pi} \int\limits_{\frac{1}{s \, x^2_{10}}}^z \frac{d z'}{z'}  \int\limits_{\frac{1}{z' s}}^{x_{10}^2} \frac{d x_{21}^2}{x_{21}^2} \, \left[ F (x_{21}^2, z')  -  3 \, \Gamma (x_{10}^2, x_{21}^2, z') \right] , \\
& \Gamma (x_{10}^2, x_{21}^2, z') = F^{(0)} (x^2_{10} , z') + \frac{\as \, N_c}{4 \pi} \int\limits_{\frac{1}{s \, x^2_{10}}}^{z'} \frac{d z''}{z''} \int\limits_{\frac{1}{z'' s}}^{\min \{x_{10}^2, \frac{z'}{z''} x_{21}^2 \}} \frac{d x_{32}^2}{x_{32}^2} \, \left[ F (x_{32}^2, z'')  - 3 \, \Gamma (x_{10}^2, x_{32}^2, z'') \right]. 
\end{align} 
\end{subequations}
Our aim now is to solve these equations, following \cite{Kovchegov:2017jxc}. 

Defining the new variables
\begin{subequations}
\begin{align}
\eta \equiv \sqrt{\frac{\as N_c}{4 \pi}} \, \ln \frac{zs}{\Lambda^2}, \ \ \ s_{10} \equiv \sqrt{\frac{\as N_c}{4 \pi}} \, \ln \frac{1}{x_{10}^2 \Lambda^2} \\
\eta' \equiv \sqrt{\frac{\as N_c}{4 \pi}} \, \ln \frac{z's}{\Lambda^2}, \ \ \ s_{21} \equiv \sqrt{\frac{\as N_c}{4 \pi}} \, \ln \frac{1}{x_{12}^2 \Lambda^2} \\
\eta'' \equiv \sqrt{\frac{\as N_c}{4 \pi}} \, \ln \frac{z''s}{\Lambda^2}, \ \ \ s_{32} \equiv \sqrt{\frac{\as N_c}{4 \pi}} \, \ln \frac{1}{x_{32}^2 \Lambda^2}
\end{align}
\end{subequations}
and putting, for simplicity, $F^{(0)} (x^2_{10} , z)=1$, we rewrite Eqs.~\eqref{Fi_evol6} as
\begin{subequations}\label{Fi_evol7}
\begin{align}
& F (s_{10} , \eta)  = 1 + \int\limits_{s_{10}}^\eta d \eta'  \int\limits_{s_{10}}^{\eta'} ds_{21} \, \left[ F (s_{21}, \eta')  - 3 \,  \Gamma (s_{10}, s_{21}, \eta') \right] , \\
& \Gamma (s_{10}, s_{21}, \eta') = 1 + \int\limits_{s_{10}}^{\eta'} d \eta'' \int\limits_{\max \{ s_{10} , s_{21} + \eta'' - \eta' \} }^{\eta''} ds_{32}   \, \left[ F (s_{32}, \eta'')  - 3 \, \Gamma (s_{10}, s_{32}, \eta'') \right]. 
\end{align} 
\end{subequations}
These equations have the same kernel as the large-$N_c$ DLA helicity equations~(3) from \cite{Kovchegov:2017jxc}. Therefore, the solution must have the same scaling property,
\begin{align}\label{scaling}
F (s_{10} , \eta)  = F (\eta - s_{10}), \ \ \ \Gamma (s_{10}, s_{21}, \eta') = \Gamma (\eta' - s_{10}, \eta' - s_{21}).
\end{align}
Using \eqref{scaling} in Eqs.~\eqref{Fi_evol7} yields
\begin{subequations}\label{Fi_evol8}
\begin{align}
& F (\zeta)  = 1 + \int\limits_{0}^\zeta d \xi  \int\limits_{0}^{\xi} d\xi' \, \left[ F (\xi')  - 3 \, \Gamma (\xi, \xi') \right] , \label{Fi_evol8a} \\
& \Gamma (\zeta, \zeta') = F (\zeta')  + \int\limits_{\zeta'}^\zeta d \xi  \int\limits_{0}^{\zeta'} d\xi' \, \left[ F (\xi')  - 3 \,  \Gamma (\xi, \xi') \right] , 
\end{align} 
\end{subequations}
confirming the scaling ansatz from \eq{scaling}. The ``neighbor" dipole amplitude is defined only for $x_{10} > x_{21}$, that is, $\zeta > \zeta'$. 

\begin{figure}[ht]
\begin{center}
\includegraphics[width= 0.75 \textwidth]{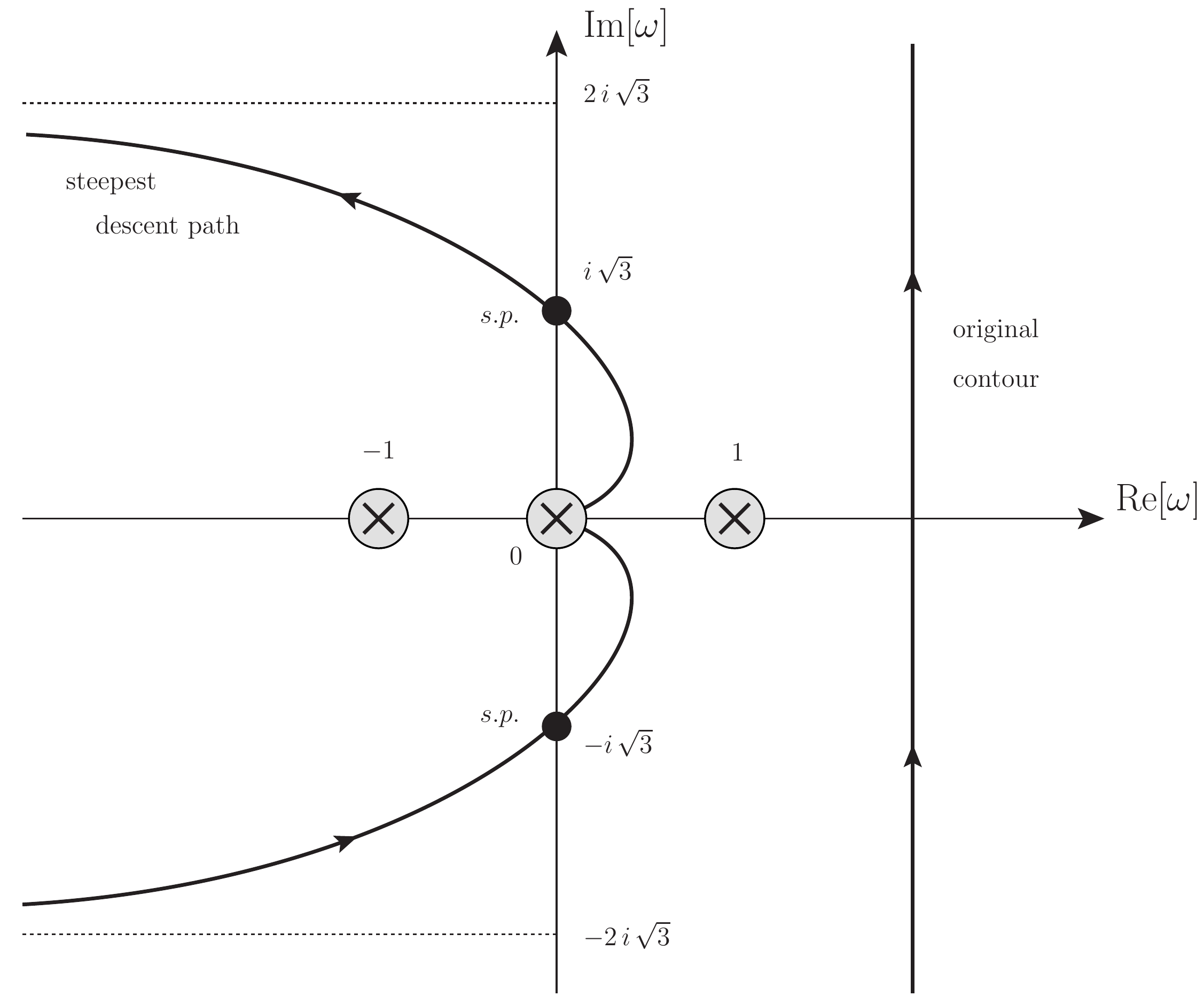} 
\caption{The complex plane structure of the $\omega$-integral in \eq{Fsol}. The singularities are denoted by circled crosses: the poles at $\omega = \pm 1$ and the essential singularity at $\omega =0$. The original contour is denoted by the vertical solid straight line to the right of all the singularities. The saddle points at $\omega_{s.p.} = \pm i \, \sqrt{3}$ are marked by the thick dots, while the steepest descent path is sketched by a curved solid line crossing the saddle points and going through the origin. The dashed horizontal lines denote the asymptotics of the steepest descent contour.}
\label{FIG:complex_plane}
\end{center}
\end{figure}

Repeating the steps from \cite{Kovchegov:2017jxc}, as detailed in \app{sec:app_sol}, one arrives at the solution of Eqs.~\eqref{Fi_evol8} in the integral form,
\begin{subequations}\label{FGsol}
\begin{align}\label{Fsol}
& F (\zeta)  = \int \frac{d \omega}{2 \pi i} \, e^{\left( \omega - \frac{3}{\omega} \right) \zeta} \frac{\omega^2 + 3}{\omega \, (\omega^2 -1)}, \\
& \Gamma (\zeta, \zeta') = \frac{2}{3}  \int \frac{d \omega}{2 \pi i} \, e^{ \omega \zeta' - \frac{3\, \zeta}{\omega} } \frac{\omega^2 + 3}{\omega \, (\omega^2 -1)} + \frac{1}{3} \int \frac{d \omega}{2 \pi i} \, e^{\left( \omega - \frac{3}{\omega} \right) \zeta'} \frac{\omega^2 + 3}{\omega \, (\omega^2 -1)}. 
\end{align} 
\end{subequations}
The leading high-energy asymptotics of $F (\zeta)$ is given by the saddle points at $\omega = \pm i \sqrt{3}$ in the exponent. This is illustrated in \fig{FIG:complex_plane}, which also shows the entire complex-plane structure of the integrand of \eq{Fsol} (cf. \cite{Kovchegov:2017way}). In distorting the original integration contour (the straight vertical line on the right of \fig{FIG:complex_plane}) to the steepest descent path we pick up the pole at $\omega=1$. However, the contribution of this pole falls of exponentially with $\zeta$, that is, it scales as $\sim e^{-2 \, \zeta}$, and can be safely discarded for $\zeta \gg 1$. We thus see that the integral in \eq{Fsol} is indeed dominated by the contribution of the steepest descent contour. (Note that this situation is the exact opposite of the helicity distribution in \cite{Kovchegov:2017jxc}, where the contribution of the right-most pole dominated over the steepest descent path.)

%%%%%%%%%%%%%%%%%%%%%%%%%%%%%%%%%%%%%%%%%%%%%%%%%%%%%%%%%%%%%%%%%%%%%%%%%%%%%%%%
%%%%%%%%%%%%%%%%%%%%%%%%%%%%%%%%%%%%%%%%%%%%%%%%%%%%%%%%%%%%%%%%%%%%%%%%%%%%%%%%

Distorting the integration contour to run along the steepest descent path, and integrating over the regions near the saddle points at $\omega_{s.p.} = \pm i \, \sqrt{3}$ by expanding $\omega \approx i \, \sqrt{3} + \rho \, e^{i 3 \pi/4} - \rho^2/(2 \sqrt{3})$ and $\omega \approx - i \, \sqrt{3} + \rho \, e^{i \pi/4} - \rho^2/(2 \sqrt{3})$, respectively, with some small real parameter $\rho$ (which is then integrated from $-\infty$ to $\infty$), one arrives at 
\begin{align}\label{asympt_F1}
F (\zeta) \approx \frac{3^{1/4}}{8 \sqrt{\pi} \ \zeta^{3/2}} \, \sin \left( 2 \, \sqrt{3} \, \zeta - \frac{\pi}{4} \right)
\end{align}
such that
\begin{align}\label{asympt_F2}
F (s_{10} , \eta) & = F (\zeta) \approx \frac{3^{1/4}}{8 \sqrt{\pi} \ \zeta^{3/2}} \, \sin \left( 2 \, \sqrt{3} \, \zeta - \frac{\pi}{4} \right) = \frac{3^{1/4}}{8 \sqrt{\pi} \ (\eta - s_{10})^{3/2}} \, \sin \left( 2 \, \sqrt{3} \, (\eta - s_{10}) - \frac{\pi}{4} \right) \notag \\
& = \frac{3^{1/4}}{8 \sqrt{\pi} \ \left[ \sqrt{\frac{\as N_c}{4 \pi}} \, \ln (z s x_{10}^2) \right]^{3/2}} \, \sin \left( 2 \, \sqrt{3} \, \sqrt{\frac{\as N_c}{4 \pi}} \, \ln (z s x_{10}^2) - \frac{\pi}{4} \right) .
\end{align}
We see that the dipole amplitude $F (s_{10} , \eta)$ is an oscillating function of its arguments: such oscillations are interesting and reminiscent of the oscillations found in \cite{Kovchegov:2020hgb} for quark helicity in the large-$N_c \& N_f$ limit (with $N_f$ the number of quark flavors).

We cross-checked the result \eqref{asympt_F1} by solving Eqs.~\eqref{Fi_evol7} numerically and found a very good agreement between the analytic and numerical solutions for $\zeta = \eta - s_{10} \gtrsim 3$. Furthermore, to test our assumption that Eqs.~\eqref{Fi_evol5} and Eqs.~\eqref{Fi_evol6} give the same small-$x$ asymptotics for the Sivers function, we solved Eqs.~\eqref{Fi_evol5} numerically, obtaining the solution which exhibited the oscillating behavior with the same period and similarly decreasing-with-$\zeta$ amplitude of the oscillations as \eq{asympt_F1}.

Employing the result from \eq{asympt_F2} in Eqs.~\eqref{ansatz2} and \eqref{Sivers_sub_eik7} we see that the $z$-integral in the latter is dominated by its lower limit, which gives a contribution independent of the center-of-mass energy squared $s$, and, consequently, of $x$.  Therefore, we conclude that the sub-eikonal small-$x$ asymptotics of the quark Sivers function at large $N_c$ and in DLA is given by a constant,
\begin{align}\label{Sivers_sub_eik8}
f_{1 \: T}^{\perp \: q} (x,{k}_T^2) \Big|_\textrm{sub-eikonal} \sim \left( \frac{1}{x} \right)^0 = \textrm{const} (x) .
\end{align}
The approach to the constant asymptotics of \eq{Sivers_sub_eik8} should be oscillatory with decreasing amplitude of such oscillations, due to the form of the amplitude in \eq{asympt_F2}. This way, in principle, some residual effects of the oscillations from \eq{asympt_F2} may be observable experimentally.

Let us point out that the result \eqref{Sivers_sub_eik8} is, in a way, similar to the case of the odderon: while, unlike the odderon case, the (DLA) evolution at the sub-eikonal order does significantly affect the dipole amplitude $F (x^2_{10} , z)$, the $x$-dependence of the Sivers function is almost unaffected by the evolution, just like it was for the eikonal odderon contribution. The same conclusion \eqref{Sivers_sub_eik8} can be obtained by using the initial amplitude \eqref{F0init} in Eqs.~\eqref{ansatz2} and \eqref{Sivers_sub_eik7}.

%%%%%%%%%%%%%%%%%%%%%%%%%%%%%%%%%%%%%%%%%%%%%%%%%%%%%%%%%%%%%%%%%%%%%%%%%%%%%%%%%%%

\subsection{Small-$x$ Asymptotics of the Quark Sivers Function: a Summary}

We conclude this Section by summarizing the results of our calculations. The quark Sivers function at small $x$ receives contributions at the eikonal and sub-eikonal order. The eikonal contribution is coming from the spin-dependent odderon and is given by $f_{1 \: T}^{\perp \: q} \sim 1/x$ with an almost non-perturbative accuracy (see Eqs.~\eqref{siv} and \eqref{oddsiv}), in agreement with \cite{Dong:2018wsp,Boer:2015pni}. The sub-eikonal contribution to quark Sivers function is calculated in this work for the first time. At large $N_c$ and in DLA it is given by \eq{Sivers_sub_eik8}. We, therefore, conclude that, at small-$x$, one can describe the small-$x$ asymptotics of the quark Sivers function as
\begin{align}\label{Sivers_small-x}
f_{1 \: T}^{\perp \: q} (x, k_T^2)  = C_O (k_T^2, x) \, \frac{1}{x} + C_1 (k_T^2) \, \left( \frac{1}{x} \right)^0 + \ldots
\end{align}
with some functions $C_O (k_T^2, x)$ and $C_1 (k_T^2)$, which can be obtained from the above results. The function $C_O (k_T^2, x)$ also depends on $x$, but in a much slower way than the powers explicitly shown in \eq{Sivers_small-x} (see, e.g., Eq.~(26) in \cite{Contreras:2020lrh}). The ellipsis in \eq{Sivers_small-x} denote the order-$x$ sub-sub-eikonal corrections along with the powers of $\ln(1/x)$-suppressed pre-asymptotic corrections to the saddle-point asymptotics shown in \eq{Sivers_small-x}. 

The situation we have found is qualitatively similar to what was suggested for the $h_1$ structure function in \cite{Kirschner:1996jj}: a sum of the odderon and DLA contributions. (Note, however, that for a related quantity, the valence quark transversity TMD, only the DLA contribution was found in \cite{Kovchegov:2018zeq}.)

%%%%%%%%%%%%%%%%%%%%%%%%%%%%%%%%%%%%%%%%%%%%%%%%%%%%%%%%%%%%%%%%%%%%%%%%%%%%%%%%%%%

\section{Conclusions and Outlook}
\label{sec:con}

In this paper we have accomplished several results. In \eq{Full} we have constructed the full sub-sub-eikonal polarized Wilson line/quark $S$-matrix operator which can also be used to obtain the small-$x$ asymptotics of a number of quark TMDs which have not been studied at small $x$ yet. We employed this operator to study the Sivers function $f_{1 \: T}^{\perp}$ up to the sub-eikonal order. In the future, one TMD that can be studied with the help of this new operator is the Worm-Gear function $g_{1T}^\perp$ coupling the proton's transverse spin to the quark helicity: such coupling is given by the $\delta_{\chi,-\chi'}$ terms which can be extracted from \eq{Full}. One can also apply the same operator treatment to study the small-$x$ asymptotics of other leading-twist quark TMDs, such as the Boer-Mulders ($h_{1}^\perp$) function, by employing the quark polarization-independent $\sim \delta_{\chi,\chi'}$ sub-eikonal corrections from the quark $S$-matrix operator constructed here. Similarly one can study Pretzelosity ($h_{1T}^\perp$) which couples transversely polarized quarks ($\sim \chi \, \delta_{\chi,-\chi'}$) to the transverse spin of the proton, with both spins orthogonal to each other. One can also study the other Worm-Gear function $h_{1L}^\perp$ coupling the transversely-polarized quark ($\sim \chi \, \delta_{\chi,\chi'}$) to the longitudinal spin of the proton.  Then, together with the known results for the unpolarized quark TMD at small $x$  \cite{Mueller:1999wm,McLerran:1998nk,Kovchegov:2015zha,Xiao:2017yya}, quark helicity \cite{Kovchegov:2015pbl,Kovchegov:2018znm,Kovchegov:2017jxc} and transversity \cite{Kovchegov:2018zeq} one would have obtained the small-$x$ asymptotics for all the leading-twist quark TMDs. Leading-twist gluon TMDs can also be analyzed in a similar way. This would allow one to make predictions for the data of future small-$x$ experiments studying the proton's spin structure, such as those at the upcoming EIC.

The standard Collins-Soper-Sterman (CSS) \cite{Collins:1981uk,Collins:1984kg} equations usually applied to TMDs evolve them in $Q^2$ and not in $x$. As a consequence, the CSS evolution cannot predict the $x$-dependence of TMDs, particularly at small $x$. Constructing small-$x$ evolution of TMDs, which is able to predict the $x$ dependence of TMDs, like what was done in this paper for the quark Sivers function, is thus vital for making predictions for the future TMD measurements at the EIC and for completing our understanding of TMDs in the important small-$x$ region of phase space probed in high energy collisions. In addition, understanding the small-$x$ asymptotics of TMDs would provide a unique angle on the proton momentum and spin structure in the regime dominated by sea quarks and gluons. 

To illustrate our method we have constructed the small-$x$ quark Sivers function using the operator formalism introduced in \cite{Kovchegov:2018znm} to study the quark helicity TMD and used again in \cite{Kovchegov:2018zeq} to construct the valence quark transversity TMD. We have reproduced the conclusion of \cite{Dong:2018wsp} that the spin-dependent odderon dominates in the small-$x$ asymptotics of the quark Sivers function. In addition, we have found the sub-eikonal correction to this odderon-dominated asymptotics. Perhaps naturally, the sub-eikonal contribution to the Sivers function comes from the gauge-covariant operator representing the sub-eikonal phase \eqref{phase}, since existence of a phase is essential for the Sivers function. The two terms, eikonal and sub-eikonal, are summarized in \eq{Sivers_small-x}.  For the STSA $A_N$ observable, if we conjecture that $A_N \sim x \, f_{1 \: T}^{\perp}$ as far as the $x$-dependence is concerned, our prediction \eqref{Sivers_small-x} implies $A_N \sim C_O + x \, C_1$. Since the data \cite{Adams:1991rw,Adams:1991cs,Abelev:2008af,Adler:2005in} appears to be closer to $A_N \sim x$ scaling than to $A_N \sim $~const, we see that the sub-eikonal correction may turn out to be more important for the description of the existing data on STSA.  One could speculate that the sub-eikonal correction somehow dominates over the spin-dependent odderon contribution in the experimentally-probed $x$-region, possibly just numerically if $C_1 \gg C_O$ for some (probably non-perturbative\footnote{In a purely perturbative approach, $C_O$ is not parametrically suppressed compared to $C_1$ (both are order-$\as^3$): while a definitive conclusion can be reached only by performing a detailed calculation, it appears puzzling that only the $C_1$ contribution is seen in the data so far (if our interpretation of the data on the $x$-dependence of $A_N$ is correct). One can suspect that non-perturbative effects may alter this perturbative conclusion of $C_1$ and $C_O$ being comparable. Indeed, the historical difficulty of detecting odderon contributions to QCD processes might suggest that terms like $C_O$ are suppressed by a mechanism which cannot be seen by perturbative calculations.}) reason. Note that the lowest-order perturbative Sivers function scales as $f_{1 \: T}^{\perp} \sim x$ at small $x$ \cite{Meissner:2007rx}, leading to $A_N \sim x^2$, which seems to disagree with the data \cite{Adams:1991rw,Adams:1991cs,Abelev:2008af,Adler:2005in}, though a more detailed analysis of the $x$-dependence of $A_N$ in the data is needed to draw firm conclusions. The sub-eikonal correction we found, once better quantified, may also provide a background for the future spin-dependent odderon searches. These results give an exciting possible new direction for such future experimental studies, particularly in light of the recent announcement by D0 and TOTEM collaborations of odderon detection through the asymmetry between $pp$ and $p\bar{p}$ collisions \cite{TOTEM:2020zzr}. Future experiments such as those to be conducted at the EIC will be able to probe transverse spin asymmetries at small-$x$ and potentially observe both the spin-dependent odderon contribution and the sub-eikonal correction derived in this work.

%||||||||||||||||||||||||||||||||||||||||||||||||||||||||||||||||||||||||||||||||||||||||||||||||||||||||||||||||||||||||
%||||||||||||||||||||||||||||||||||||||||||||||||||||||||||||||||||||||||||||||||||||||||||||||||||||||||||||||||||||||||
%||||||||||||||||||||||||||||||||||||||||||||||||||||||||||||||||||||||||||||||||||||||||||||||||||||||||||||||||||||||||

\section*{Acknowledgments}

\label{sec:acknowledgement}

The authors would like to thank Markus Diehl, Daniel Pitonyak and Jian Zhou for discussions.

This material is based upon work supported by the U.S. Department of
Energy, Office of Science, Office of Nuclear Physics under Award
Number DE-SC0004286.

%||||||||||||||||||||||||||||||||||||||||||||||||||||||||||||||||||||||||||||||||||||||||||||||||||||||||||||||||||||||||
%||||||||||||||||||||||||||||||||||||||||||||||||||||||||||||||||||||||||||||||||||||||||||||||||||||||||||||||||||||||||
%||||||||||||||||||||||||||||||||||||||||||||||||||||||||||||||||||||||||||||||||||||||||||||||||||||||||||||||||||||||||

%%%%%%%%%%%%%%%%%%%%%%%%%%%%%%%%%%%%%%%%%%%%%%%%%%%%%%%%%%%%%%%%%%%%%%%%%%%%%%%%%

\appendix
\section{Solution of the Large-$N_c$ Evolution Equations}
\label{sec:app_sol}

Here we solve Eqs.~\eqref{Fi_evol8} following the strategy presented in \cite{Kovchegov:2017jxc}. Differentiating Eqs.~\eqref{Fi_evol8} yields
\begin{subequations}\label{FG1}
\begin{align}
& \pd_{\zeta} F (\zeta) = \int\limits_0^\zeta d \xi' \left[ F (\xi') - 3 \, \Gamma (\zeta, \xi') \right], \\
& \pd_{\zeta} \Gamma (\zeta, \zeta') = \int\limits_0^{\zeta'} d \xi' \left[ F (\xi') - 3 \, \Gamma (\zeta, \xi') \right]. \label{FG1b}
\end{align}
\end{subequations}

Introducing the Laplace transforms 
\begin{align}\label{Laplace}
F (\zeta) = \int \frac{d \omega}{2 \pi i} \, e^{\omega \zeta} \, F_\omega, \ \ \ \Gamma (\zeta, \zeta') = \int \frac{d \omega}{2 \pi i} \, e^{\omega \zeta'} \, \Gamma_\omega (\zeta)
\end{align}
we reduce \eq{FG1b} to 
\begin{align}\label{G1}
\pd_{\zeta} \Gamma_\omega (\zeta) = \frac{1}{\omega} \, \left[ F_\omega - 3 \, \Gamma_\omega (\zeta) \right]. 
\end{align}
Solution of \eq{G1} is
\begin{align}\label{Gsol1}
\Gamma_\omega (\zeta) - \frac{1}{3} \, F_\omega = e^{- \frac{3}{\omega} \, \zeta} \, \left[ \Gamma_\omega (0) - \frac{1}{3} \, F_\omega \right] \equiv e^{- \frac{3}{\omega} \, \zeta} \, H_\omega.
\end{align}
Employing it in \eq{Laplace}, along with the $\Gamma (\zeta, \zeta ) = F (\zeta)$ condition, we arrive at
\begin{subequations}\label{FG2}
\begin{align}
& F (\zeta) = \frac{3}{2} \int \frac{d \omega}{2 \pi i} \, e^{\left( \omega - \frac{3}{\omega} \right) \, \zeta} \, H_\omega , \\
&  \Gamma (\zeta, \zeta') = \int \frac{d \omega}{2 \pi i} \, e^{ \omega \zeta' - \frac{3}{\omega} \, \zeta} \, H_\omega + \frac{1}{2} \int \frac{d \omega}{2 \pi i} \, e^{\left( \omega - \frac{3}{\omega} \right) \, \zeta'} \, H_\omega. \label{FG2b}
\end{align}
\end{subequations}

Substituting Eqs.~\eqref{FG2} into Eqs.~\eqref{FG1} results in two constraints, 
\begin{align}\label{Hconst}
\int \frac{d \omega}{2 \pi i} \, e^{ - \frac{3}{\omega} \, \zeta} \, \frac{1}{\omega} \,  H_\omega = 0 , \ \ \  \int \frac{d \omega}{2 \pi i} \, e^{\left( \omega - \frac{3}{\omega} \right) \, \zeta} \, \left( \omega - \frac{1}{\omega} \right) \, H_\omega =0. 
\end{align}
To satisfy these, we write
\begin{align}
H_\omega = \frac{\omega}{\omega^2 -1} \, f_\omega
\end{align}
and look for the unknown function $f_\omega$ in the form
\begin{align}
f_\omega = \sum_{n=-\infty}^\infty d_n \, \omega^n .
\end{align}
After a calculation similar to  \cite{Kovchegov:2017jxc} and involving a series of Bessel functions $J_n (2 \sqrt{3} \, \zeta)$ we arrive at
\begin{align}
f_\omega = d_0 \, \left( 1 + \frac{3}{\omega^2} \right)
\end{align}
with some unknown constant $d_0$. The constant can be fixed by requiring that that $F(0) =1$, as follows from \eq{Fi_evol8a}. This gives $d_0 = 2/3$. We thus obtain
\begin{align}\label{Hsol}
H_\omega = \frac{2}{3} \, \frac{\omega^2 + 3}{\omega \, (\omega^2 -1)}.
\end{align}
Using \eq{Hsol} in Eqs.~\eqref{FG2} yields the solution \eqref{FGsol} in the main text.

%%%%%%%%%%%%%%%%%%%%%%%%%%%%%%%%%%%%%%%%%%%%%%%%%%%%%%%%%%%%%%%%%%%%%%%%%%%%%%%

%

%%%%%%%%%%%%%%%%%%%%%%%%%%%%%%%%%%%%%%%%%%%%%%%%%%%%%%%%%%%%%%%%%%%%%%%

\end{document}